\pdfoutput=1
\documentclass[11pt,a4paper]{article}
\usepackage{jheppub}
\usepackage[caption=false]{subfig}
\usepackage{graphicx}
\usepackage{hyperref}
\usepackage{multirow}
\usepackage{color}

\def\Lag{\mathcal{L}}
\def\p{\partial}

\def\=:{=\hspace{-.7em}\raisebox{1.1ex}{.}\hspace{.1em}\raisebox{-0.2ex}{.}}

\def\cC{\mathcal{C}}
\def\cI{\mathcal{I}}
\newcommand{\ab}[1]{\langle#1\rangle}

\newcommand{\beq}{\begin{eqnarray}}
\newcommand{\eeq}{\end{eqnarray}}
\newcommand{\non}{\nonumber\\}

\newcommand{\Tr}{\mathop{\mbox{\rm Tr}}}

\begin{document}

\title{Higher-order Skyrme hair of black holes}

\author{Sven Bjarke Gudnason${}^1$ and}
\author{Muneto Nitta${}^2$}
\affiliation{${}^1$Institute of Modern Physics, Chinese Academy of
  Sciences, Lanzhou 730000, China}
\affiliation{${}^2$Department of Physics, and Research and Education
  Center for Natural Sciences, Keio University, Hiyoshi 4-1-1,
  Yokohama, Kanagawa 223-8521, Japan}
\emailAdd{bjarke(at)impcas.ac.cn}
\emailAdd{nitta(at)phys-h.keio.ac.jp}

\abstract{
Higher-order derivative terms are considered as replacement for the
Skyrme term in an Einstein-Skyrme-like model in order to pinpoint
which properties are necessary for a black hole to possess stable
static scalar hair. 
We find two new models able to support stable black hole hair in the
limit of the Skyrme term being turned off.
They contain 8 and 12 derivatives, respectively, and are roughly the
Skyrme-term squared and the so-called BPS-Skyrme-term squared.
In the twelfth-order model we find that the lower branches, which are
normally unstable, become stable in the limit where the Skyrme term is
turned off.
We check this claim with a linear stability analysis.
Finally, we find for a certain range of the gravitational coupling and
horizon radius, that the twelfth-order model contains 4 solutions as
opposed to 2. More surprisingly, the lowest part of the would-be
unstable branch turns out to be the stable one of the 4 solutions. 
}


\maketitle

\section{Introduction}

Black holes pose interesting questions about fundamental physics, such
as quantum information and whether quantum theory is truly unitary at
all scales.
A simpler question is whether a black hole (BH) is characterized just
by global charges at infinity or it has the ability to possess hair.
That is, can the BH stably couple to fields of the standard model
yielding a non-vacuum configuration that surrounds said BH.
An attempt at answering this question in the negatory is made by the
weak no-hair conjecture: for a scalar field coupled to Einstein
gravity, all BHs fall in the Kerr-Newman family of solutions,
i.e.~being characterized by their mass, charge and spin.

The first stable counterexample was provided in the Einstein-Skyrme
model \cite{Luckock:1986tr,Glendenning:1988qy,Droz:1991cx,Heusler:1991xx,Heusler:1992av,Bizon:1992gb,Volkov:1998cc,Shiiki:2005pb}.
In some sense a counterexample in a specific scalar field theory is
not too exciting.
However, the Skyrme model \cite{Skyrme:1961vq,Skyrme:1962vh} is
believed to be an effective field theory describing QCD at low
energies, at least in the limit of a large number of
colors \cite{Witten:1983tw,Witten:1983tx}, 
where baryons are identified with solitons -- called the Skyrmions.
This means that at low energies, the presence of effective operators
induced by QCD would potentially be able to give rise to stable BH
hair. 

Recently, a program of trying to understand in more detail whether the
Skyrme term plays a special role in stabilizing BH hair has been
pursued.
It started with a simple observation stating that the BPS-Skyrme
model, which consists of a sextic derivative term and a potential,
cannot sustain stable BH hair \cite{Gudnason:2015dca}.
This was later proved analytically in Ref.~\cite{Adam:2016vzf}. 
The sixth-order derivative term which we shall call the BPS-Skyrme term,
was known for long time as it is induced by integrating out the
$\omega$ meson \cite{Adkins:1983nw,Jackson:1985yz}.
This term later caught much attention due to the existence of a
submodel with an energy bound that is saturable, hence the name
BPS-Skyrme model \cite{Adam:2010fg,Adam:2010ds}.\footnote{The name may
  suggest that the model can easily be supersymmetrized, but that is
  not the case due to the target space being non-K\"ahler and odd
  dimensional. So far, the only available supersymmetric extension of
  Skyrme-like models consists solely of a four-derivative term, with
  the target space complexified
  \cite{Gudnason:2015ryh,Gudnason:2016iex}. 
  }
The BPS-Skyrme model consists of the sextic BPS-Skyrme term and an
appropriate potential term.
The standard Skyrme model, in comparison, contains a kinetic term and
the fourth-order Skyrme term \cite{Skyrme:1961vq,Skyrme:1962vh}; the
latter term can be viewed as a curvature term.
Finally, a model sometimes called the generalized Skyrme model,
contains the standard Skyrme model as well as the BPS-Skyrme term 
(see,
e.g.~Refs.~\cite{Gudnason:2014hsa,Gudnason:2015nxa,Adam:2016lir,Adam:2016drk,Gudnason:2016tiz,Adam:2017czk}
for recent papers). 
The latter two models can have potentials as well.
In Refs.~\cite{Gudnason:2016kuu,Adam:2016vzf} it was shown numerically 
that in the generalized Skyrme model, the BH hair ceases to exist in
the limit where the Skyrme term is turned off.
This was somewhat surprising because in the gravitating case and in
the flat-space limit of the model, the BPS-Skyrme term is able to
stabilize Skyrmions.

In Ref.~\cite{Gudnason:2017opo} we constructed a family of
higher-order derivative terms, which could be used exactly for testing
the BH hair stability.
These terms are not the most generic terms possible, but are
constructed with the same concept of minimality that underlies the
Skyrme term and the BPS-Skyrme term.
First of all, all the terms are constructed out of the strain tensor,
which implies that only one derivative acts on each field component in
the Lagrangian density.
This guarantees a second order equation of motion, which however is
highly nonlinear in single derivatives.
The above mentioned minimality for the kinetic, the Skyrme and the
BPS-Skyrme term, can be described as each term possessing at most 2
derivatives in the same spatial direction.
This is achieved by means of antisymmetrization.
The same minimality is then applied to the terms with eight, ten and
twelve derivatives.
Now the minimal number of derivatives in one spatial direction cannot
be lower than 4.
By means of antisymmetrization, all terms with 6 or more derivatives
in one spatial direction are eliminated and the outcome is what we
denoted the minimal Lagrangians.

Another attempt at constructing higher-order Lagrangians was done by
Marleau, who found a recipe for constructing terms that for the
spherically symmetric hedgehog gave very simple Lagrangian
densities \cite{Marleau:1989fh,Marleau:1990nh,Marleau:1991jk}. 
The construction yields Lagrangian densities for the hedgehog with
exactly 2 radial derivatives.
This implies that there must be $2(n-1)$ angular derivatives in a term
with $2n$ derivatives.
Unfortunately, for $n>3$ that yielded Lagrangian densities with
non-positive definite static energies.
In particular, in Ref.~\cite{Gudnason:2017opo} we showed that although
the latter construction gives stable radial equations of motion, the
angular directions contain instabilities which can be triggered by a
baby-Skyrme string in the baryon number 0 sector.
For a BH hair configuration it is even worse, as the expansion of the
Einstein equation yields a first derivative of the profile function at
the horizon which is positive (from $f_h\lesssim \pi$); hence no
solutions exist.
Thus the instabilities present in the construction are immediately
seen by the gravitational background.
This obstacle lead us to construct the above-mentioned minimal
Lagrangians. 

Let us briefly mention some further activities in the field of
Skyrme-type BH hair.
The BH Skyrme hair was extended from the Schwarzschild case to
spacetimes which are asymptotically anti-de
Sitter (AdS) \cite{Shiiki:2005aq,Shiiki:2005xn,Perapechka:2016cof} and
de Sitter (dS) \cite{Brihaye:2005an}. 
In particular, Ref.~\cite{Perapechka:2016cof} extended the result of
Refs.~\cite{Gudnason:2016kuu,Adam:2016vzf} to AdS spacetime, i.e.~that
the sextic BPS-Skyrme term is not able to support stable BH hair.

Gravitating Skyrmions have also been considered in the literature, see
e.g.~\cite{Heusler:1991xx,Bizon:1992gb,Volkov:1998cc,Zajac:2009am,Zajac:2010mu}.
Collective quantization of their zero modes have also been
studied \cite{Shiiki:2004jj}.
The spherically symmetric case of the gravitating Skyrmions has been
extended to a higher charged axially symmetric
case \cite{Sawado:2003at,Sawado:2004yq} and it was again
quantized \cite{Sato:2006xy}. 
Axially symmetric BH hair was also constructed in
Ref.~\cite{Sawado:2004yq}.
Spinning gravitating Skyrmions were
considered \cite{Ioannidou:2006nn,Perapechka:2017bsb} and it was
further found that the BPS-Skyrme model does not possess spinning
Skyrmions \cite{Perapechka:2017bsb} -- neither in a curved nor in a
flat background. 
More exotic configurations like the gravitating axisymmetric
sphalerons in the Einstein-Skyrme model have also been
studied \cite{Shnir:2015aba}.
A lower-dimensional example was considered, where exactly solvable
gravitating baby-Skyrmions were found in $2+1$
dimensions \cite{Adam:2018bot}, see Ref.~\cite{Wachla:2018pdx} for a
gauged version thereof.
In 5 dimensions, a generalization of the Skyrme model to O(5) also
possesses solitons and in particular, solitonic hair of
BHs \cite{Brihaye:2017wqa} with and without spin.
An exotic study in this direction involves nonstandard boundary
conditions for the metric, identifies 3-space with $\mathbb{R}P^3$
without the point at infinity and contemplates a $\pi_2$-valued
Skyrmion that can give rise to a negative gravitational mass, thus
antigravity \cite{Klinkhamer:2018dta}. 
Neutron stars and in particular the equation of state, which is of 
crucial importance in that subject, have been studied in the
BPS-Skyrme model coupled to gravity \cite{Adam:2014dqa,Adam:2015lpa}.
Due to the simplicity of the model, the solutions could be compared to
the mean field approximation \cite{Adam:2015lpa}, which interestingly
showed some deviations.
Unfortunately, since neutron stars are expected to possess some amount
of spin, which is acquired during a star's core collapse, the
BPS-Skyrme model cannot quite be a good approximation due to the
instability found in Ref.~\cite{Perapechka:2017bsb}.
In some limit it may give reasonable answers, nevertheless.

Another interesting idea was proposed by Dvali and Gu\ss mann, stating
that the baryon number may actually be conserved by BHs and become
Skyrmionic hair once swallowed by the
BH \cite{Dvali:2016mur,Dvali:2016sac}.
They also proposed that there may be ways for an observer outside the 
horizon to measure the number of swallowed baryons, although they rely
on certain assumptions.

In this paper, we will focus on what terms are able to sustain stable
static hair on a Schwarzschild-type BH.
We will only consider the above mentioned minimal Lagrangians as
components in the full Lagrangian and leave other possibilities for
future work.
Specifically, the new models that we consider have a kinetic term, the
Skyrme term as well as a $2n$-th order derivative term, with
$n=4,5,6$.
For brevity, we shall call them the $2+4+2n$ models.
The idea is then to slowly turn off the Skyrme term, the limit will be
denoted the $2+2n$ model.
We find that only the $2+8$ and the $2+12$ models are stable.
In the case of the $2+8$ model, there is a one-parameter family of
models (which we will denote $\gamma\in[0,1]$) and only at the
endpoint ($\gamma=0$), BH hair solutions cease to exist.
The $2+8$ model (with $\gamma=\frac13$) can be thought of as the
kinetic term and the Skyrme-term squared, while the $2+12$ model is
the kinetic term and the BPS-Skyrme-term squared.

As known in the literature, usually for the Skyrme-type BH hair, there
are two branches of solutions; one upper branch (in the value of the
profile function at the horizon) and one lower branch.
The lower branch in the standard Skyrme model is found to be
unstable; it has a higher Arnowitt-Deser-Misner (ADM) mass and this is
backed up by a linear perturbation analysis, which shows that the
lower branch has one negative eigenvalue in the perturbation
spectrum \cite{Heusler:1991xx,Heusler:1992av,Shiiki:2005pb}.
In the $2+12$ model, we find a new behavior, where the ADM mass of the
lower branch crosses over that of the upper branch and thus becomes
the stable one.
To this end, we carry out a linear perturbation analysis to check the
eigenvalue spectrum, and indeed the lower branches have regions where
they contain only positive eigenvalues. 
Further studies in this direction, however, is needed in order to
determine full nonlinear stability.
Finally, in the $2+12$ model, we find a range of the gravitational
coupling where not 2, but 4 solutions exist.
That is, the lower branch ceases to be single valued in the
gravitational coupling.
The surprising result is that the lower part of the lower branch --
farthest away from the upper branch -- turns out to be the one with
the lowest ADM mass.

The paper is organized as follows.
In the next section, we will set up the notation of the
higher-derivative terms and construct the component Lagrangians that
we will use for the model mentioned above.
Sec.~\ref{sec:BHhair} then couples our generic model Lagrangian with 
Einstein gravity and in Sec.~\ref{sec:m242n} the explicit Einstein
equations, equations of motion and boundary conditions are written
down for each model and finally, the numerical results will be
presented. 
Linear stability of the $2+4+12$ model will then be analyzed in
Sec.~\ref{sec:linstability}.
Due to the surprising spectrum of eigenvalues of the perturbations, we
check the Skyrme model limit in Sec.~\ref{sec:Skyrme_model_limit}.
The models with BH hair surviving without the presence of the Skyrme
term are then studied as functions of the gravitational coupling in
Sec.~\ref{sec:m22n}.
Finally, Sec.~\ref{sec:discussion} concludes with a summary and
discussion of our results, as well as an outlook on future work.

\section{Lagrangian components of a higher-order Skyrme model}\label{sec:hoterms}

The class of models we will study in this paper is built out of
several component Lagrangians with different numbers of derivatives.
In this section, we will set up the framework for the component
Lagrangians and their corresponding energy-momentum tensors.
The $2n$-th order Lagrangians that we consider in this paper
are those proposed in Ref.~\cite{Gudnason:2017opo} and were termed
minimal Lagrangians.
They are minimal in the sense that they contain the smallest possible
number of derivatives in the $i$-th direction.
There are of course many more possibilities for terms with $2n$
derivatives, which we will not consider here.

The minimal Lagrangians of order $2n$ are \cite{Gudnason:2017opo},
\begin{align}
\Lag_2 &= -c_2 \cC_1,\label{eq:L2}\\
\Lag_4 &= -\frac{c_4}{2} \cC_2,\\
\Lag_6 &= -\frac{c_6}{3} \cC_3,\label{eq:L6}\\
\Lag_8 &= \frac{c_{8|4,4}}{2}\left(\cC_4 - \frac12\cC_2^2\right)
  - \frac{c_{8|4,2,2}}{4} \cC_4
  + \frac{3a_{3,1}}{4}\cI_4,\\
\Lag_{10} &= -\frac{c_{10|4,4,2}}{6} \cC_2 \cC_3
  - a_{4,1} \cC_1\cI_4
  + a_5 \cI_5, \\
\Lag_{12} &= -\frac{c_{12|4,4,4}}{9} \cC_3^2
  - a_{4,1,1} \cC_1^2\cI_4
  + a_{4,2} (\cC_2 - \cC_1^2)\cI_4
  + a_{5,1} \cC_1\cI_5
  - a_6 \cI_6,\label{eq:L12}
\end{align}
where we have defined the following building blocks
\begin{align}
\mathcal{C}_1 &\equiv \ab{1},\label{eq:C1}\\
\mathcal{C}_2 &\equiv -\ab{2} + \ab{1}^2,\\
\mathcal{C}_3 &\equiv \ab{3} - \frac32\ab{2}\ab{1} + \frac12\ab{1}^3,\label{eq:C3}\\
\mathcal{C}_4 &\equiv \ab{4} - \frac12\ab{2}^2 - \ab{2}\ab{1}^2 + \frac12\ab{1}^4,\\
\mathcal{C}_5 &\equiv -\ab{5} + \frac53\ab{3}\ab{1}^2 + \frac54\ab{2}^2\ab{1}
  - \frac52\ab{2}\ab{1}^3 + \frac{7}{12}\ab{1}^5, \\
\mathcal{C}_6 &\equiv \ab{6} - 2\ab{3}\ab{2}\ab{1} - \frac14\ab{2}^3
  + \frac32\ab{2}^2\ab{1}^2  - \frac14\ab{2}\ab{1}^4,\label{eq:C6}
\end{align}
as well as the quantities
\begin{align}
\cI_4 &\equiv \cC_4 - \frac43\cC_1\cC_3,\label{eq:I4}\\
\cI_5 &\equiv \cC_5 - \frac56\cC_2\cC_3,\\
\cI_6 &\equiv \cC_6 - \frac13\cC_3^2,\label{eq:I6}
\end{align}
and the O(4) invariants are written neatly as
\beq
\ab{r} \equiv \prod_{p=1}^r
g^{\mu_{p+1|r}\nu_p} \mathbf{n}_{\mu_p}\cdot\mathbf{n}_{\nu_p}.
\label{eq:rinv}
\eeq
Here, $r=1,\ldots,6$,  
$\mathbf{n}_{\mu} \equiv \partial_{\mu}  \mathbf{n}$,
and the modulo function in the first index, $p+1|r$ (meaning $p+1$ 
mod $r$), simply ensures that the index $\mu_{r+1}$ is just $\mu_1$. 
The coefficients $c_{2n}\geq 0$ in the
Lagrangians in Eqs.~\eqref{eq:L2}-\eqref{eq:L12} all have to be
positive semi-definite, whereas the coefficients $a$ can take any 
real values. In fact, we will show shortly that the coefficients $a$
are irrelevant for the Lagrangian formulation of the theory. 

The scalar fields $\mathbf{n}(x)=(n^0,n^1,n^2,n^3)$ are related to the chiral 
Lagrangian field $U(x) \in$ SU(2) as
\beq
U = n^0 \mathbf{1}_2 + i n^a \tau^a,
\eeq
with $\tau^a$ being the Pauli matrices and $a=1,2,3$ is summed over. 
The field $U$ transforms as $U \to g_LUg_R^\dag$ with
$g_{L,R}\in$ SU(2)$_{L,R}$ and hence the symmetry group is
SU(2)$_L\times$SU(2)$_{R}$, which is spontaneously broken to the
diagonal SU(2)$_{L+R}$ (or it is also simultaneously broken explicitly
by a pion mass term, which however we will not consider in this
paper). 

The first three Lagrangians (Eqs.~\eqref{eq:L2}-\eqref{eq:L6}) are
well known. The first, $\mathcal{L}_2$, is the standard kinetic term,
the second, $\mathcal{L}_4$, is the Skyrme term and the third,
$\mathcal{L}_6$, is the BPS-Skyrme term \cite{Adam:2010fg}.
The physical interpretation of the three terms is that the kinetic or
Dirichlet term accounts for the kinetic energy, the Skyrme term
measures the curvature on the O(4) target space and finally, the
BPS-Skyrme term acts like a perfect fluid term \cite{Adam:2014nba}.

We will now show that $\mathcal{I}_{4,5,6}$ vanish identically.
In order to do this, we will utilize the eigenvalues of the
four-dimensional strain tensor, defined by
using the left invariant form $L_{\mu} \equiv U^\dagger \partial_{\mu} U$,  
as
\beq
\widetilde{D}_{\mu\nu} \equiv -\frac{1}{2}\Tr[L_\mu L_\nu]
= \mathbf{n}_\mu\cdot\mathbf{n}_\nu
= \left[\widetilde{V}
\begin{pmatrix}
0\\
&\lambda_1^2\\
&&\lambda_2^2\\
&&&\lambda_3^2
\end{pmatrix}
\widetilde{V}^{\rm T}
\right],
\label{eq:straintensor}
\eeq
for which the invariants \eqref{eq:rinv} simply read
\beq
\ab{r} = \lambda_1^{2r} + \lambda_2^{2r} + \lambda_3^{2r}.
\label{eq:rinvlambda}
\eeq
Here, $\lambda_{1,2,3}$ are the eigenvalues of the strain tensor 
and $\widetilde{V}$ is the corresponding diagonalization matrix. 

Inserting the above relation into $\mathcal{I}_{4,5,6}$ of
Eqs.~\eqref{eq:I4}-\eqref{eq:I6} and using the
definitions in Eqs.~\eqref{eq:C1}-\eqref{eq:C6}, it follows that they vanish
identically
\beq
\mathcal{I}_4 = \mathcal{I}_5 = \mathcal{I}_6 = 0.
\eeq
Using this, we can simplify the Lagrangians $\Lag_{4,5,6}$ to
\begin{align}
\Lag_8 &= \frac{c_{8|4,4}}{2}\left(\cC_4 - \frac12\cC_2^2\right)
  - \frac{c_{8|4,2,2}}{4} \cC_4,\label{eq:L8simpl}\\
\Lag_{10} &= -\frac{c_{10|4,4,2}}{6} \cC_2 \cC_3, \\
\Lag_{12} &= -\frac{c_{12|4,4,4}}{9} \cC_3^2.\label{eq:L12simpl}
\end{align}

Since we are interested in black holes, we need the stress-energy
tensor corresponding to the above Lagrangian densities
\begin{align}
T_{\mu\nu} &= -2\sum_r\frac{\delta\ab{r}}{\delta g^{\mu\nu}}
  \frac{\p\Lag}{\p\ab{r}}
+ g_{\mu\nu} \Lag \non
&=-2\sum_{r,s}\frac{\delta\ab{r}}{\delta g^{\mu\nu}}
  \frac{\p\cC_s}{\p\ab{r}} \frac{\p\Lag}{\p\cC_s}
+ g_{\mu\nu} \Lag \non
&= -\sum_s \cC_{\mu\nu}^{(s)} \frac{\p\Lag}{\p\cC_s} + g_{\mu\nu} \Lag,
\end{align}
where we used the chain rule to express the stress-energy tensor in
terms of
\beq
\cC_{\mu\nu}^{(s)} \equiv 2\sum_r\frac{\delta\ab{r}}{\delta g^{\mu\nu}}
  \frac{\p\cC_s}{\p\ab{r}},
\eeq
for which we get
\begin{align}
\cC_{\mu\nu}^{(1)} &= 2\ab{1}_{\mu\nu},\\
\cC_{\mu\nu}^{(2)} &= -4\ab{2}_{\mu\nu} + 4\ab{1}\ab{1}_{\mu\nu},\\
\cC_{\mu\nu}^{(3)} &= 6\ab{3}_{\mu\nu} - 6\ab{1}\ab{2}_{\mu\nu}
  - 3\ab{2}\ab{1}_{\mu\nu} + 3\ab{1}^2\ab{1}_{\mu\nu},\\
\cC_{\mu\nu}^{(4)} &= 8\ab{4}_{\mu\nu} - 4\ab{2}\ab{2}_{\mu\nu}
  - 4\ab{1}^2\ab{2}_{\mu\nu} - 4\ab{2}\ab{1}\ab{1}_{\mu\nu}
  + 4\ab{1}^3\ab{1}_{\mu\nu},\\
\cC_{\mu\nu}^{(5)} &= -10\ab{5}_{\mu\nu} + 10\ab{1}^2\ab{3}_{\mu\nu}
  + \frac{20}{3}\ab{3}\ab{1}\ab{1}_{\mu\nu} + 10\ab{2}\ab{1}\ab{2}_{\mu\nu}
  + \frac52\ab{2}^2\ab{1}_{\mu\nu} \non
&\phantom{=\ }
  - 10\ab{1}^3\ab{2}_{\mu\nu} 
  - 15\ab{2}\ab{1}^2\ab{1}_{\mu\nu} + \frac{35}{6}\ab{1}^4\ab{1}_{\mu\nu},\\
\cC_{\mu\nu}^{(6)} &= 12\ab{6}_{\mu\nu} - 12\ab{2}\ab{1}\ab{3}_{\mu\nu}
  - 8\ab{3}\ab{1}\ab{2}_{\mu\nu} - 4\ab{3}\ab{2}\ab{1}_{\mu\nu}
  - 3\ab{2}^2\ab{2}_{\mu\nu} \non
&\phantom{=\ }
  + 12\ab{2}\ab{1}^2\ab{2}_{\mu\nu} 
  + 6\ab{2}^2\ab{1}\ab{1}_{\mu\nu} - \ab{1}^4\ab{2}_{\mu\nu}
  - 2\ab{2}\ab{1}^3\ab{1}_{\mu\nu},
\end{align}
where we have used
\beq
\frac{\delta\ab{r}}{\delta g^{\mu\nu}} = r \ab{r}_{\mu\nu},
\eeq
and defined
\begin{align}
\ab{1}_{\mu\nu} &\equiv \mathbf{n}_\mu\cdot\mathbf{n}_{\nu}\\
\ab{2}_{\mu\nu} &\equiv
  g^{\mu_2\nu_1}(\mathbf{n}_\mu\cdot\mathbf{n}_{\nu_1})
  (\mathbf{n}_{\mu_2}\cdot\mathbf{n}_{\nu})\\
\ab{r}_{\mu\nu} &\equiv
g^{\mu_2\nu_1}
  (\mathbf{n}_\mu\cdot\mathbf{n}_{\nu_1})
  (\mathbf{n}_{\mu_r}\cdot\mathbf{n}_\nu)
\prod_{p=2}^{r-1}
g^{\mu_{p+1}\nu_p} \mathbf{n}_{\mu_p}\cdot\mathbf{n}_{\nu_p}, \qquad
\textrm{for}\ r>2.
\end{align}
Finally, we can write down the stress-energy tensor
$T_{\mu\nu}^{(2n)}$ for each component Lagrangian, $\Lag_{2n}$, as
\begin{align}
T_{\mu\nu}^{(2)} &= c_2 \cC_{\mu\nu}^{(1)} + g_{\mu\nu} \Lag_2,\\
T_{\mu\nu}^{(4)} &= \frac{c_4}{2} \cC_{\mu\nu}^{(2)}
  + g_{\mu\nu} \Lag_4,\\
T_{\mu\nu}^{(6)} &= \frac{c_6}{3} \cC_{\mu\nu}^{(3)}
  + g_{\mu\nu} \Lag_6,\\
T_{\mu\nu}^{(8)} &= -\frac{c_{8|4,4}}{2}\left(\cC_{\mu\nu}^{(4)}
  - \cC_2 \cC_{\mu\nu}^{(2)}\right)
  + \frac{c_{8|4,2,2}}{4} \cC_{\mu\nu}^{(4)}
  + g_{\mu\nu}\Lag_8,\\
T_{\mu\nu}^{(10)} &= \frac{c_{10|4,4,2}}{6}\left(\cC_3 \cC_{\mu\nu}^{(2)}
  + \cC_2\cC_{\mu\nu}^{(3)}\right)
  + g_{\mu\nu}\Lag_{10},\\
T_{\mu\nu}^{(12)} &= \frac{2c_{12|4,4,4}}{9}\cC_3 \cC_{\mu\nu}^{(3)}
  + g_{\mu\nu}\Lag_{12}.
\end{align}

We will now specialize to the case with spherical symmetry, which is
relevant for the 1-Skyrmion sector. Thus we can assume the hedgehog
Ansatz for the Skyrme field
\beq
U = \mathbf{1}_2\cos f(r) + \frac{i x^a\tau^a}{r}\sin f(r),
\label{eq:Uhedgehog}
\eeq
with a profile function $f(r)$ satisfying the boundary condition 
$f(r\to \infty)\to 0$, 
which in terms of the four-vector $\mathbf{n}$ reads
\beq
\mathbf{n} =
\big(\sin f(r)\sin\theta\cos\phi,
 \sin f(r)\sin\theta\sin\phi,
 \sin f(r)\cos\theta,
 \cos f(r)\big).
\eeq
In this paper, we will choose a metric tensor of the form
\beq
ds^2 = -N(r)^2C(r) dt^2 + \frac{1}{C(r)} dr^2 + r^2d\theta^2
  + r^2\sin^2\theta d\phi^2,
\label{eq:metric}
\eeq
which is compatible with a spherically symmetric Schwarzschild black
hole. 

We can now plug in the hedgehog Ansatz and metric to the component
Lagrangians and corresponding stress-energy tensors.
First we note that the invariants have an astonishingly simple form
for the hedgehog
\beq
\ab{s} = C^s f_r^{2s} + \frac{2\sin^{2s}(f)}{r^{2s}},
\eeq
where $f_r\equiv \p_r f$ is the radial derivative of $f$.
We will use this notation for derivatives throughout the paper. 

We will first evaluate the building blocks $\cC_s$ with the hedgehog
and Schwarzschild metric
\begin{align}
\cC_1 &= C f_r^2 + \frac{2\sin^2(f)}{r^2},\\
\cC_2 &= \frac{4\sin^2(f)}{r^2} C f_r^2 + \frac{2\sin^4(f)}{r^4},\\
\cC_3 &= \frac{3\sin^4(f)}{r^4} C f_r^2,\\
\cC_4 &= \frac{4\sin^4(f)}{r^4} C^2 f_r^4 + \frac{8\sin^6(f)}{r^6} C f_r^2,\\
\cC_5 &= \frac{10\sin^6(f)}{r^6} C^2 f_r^4 + \frac{5\sin^8(f)}{r^8} C f_r^2,\\
\cC_6 &= \frac{3\sin^8(f)}{r^8} C^2 f_r^4,
\end{align}
and the identities $\cI_s$ are readily checked to vanish
\beq
\cI_4 = \cI_5 = \cI_6 = 0.
\eeq
It is now easy to write down the Lagrangian densities
\begin{align}
-\Lag_2 &= c_2\left(C f_r^2 + \frac{2\sin^2(f)}{r^2}\right),\\
-\Lag_4 &= c_4\frac{\sin^2(f)}{r^2}\left(2C f_r^2
  + \frac{\sin^2(f)}{r^2}\right),\\
-\Lag_6 &= c_6 \frac{\sin^4(f)}{r^4} C f_r^2,\\
-\Lag_8 &= c_{8|4,4}\frac{\sin^4(f)}{r^4}\left(2C^2 f_r^4
    + \frac{\sin^4(f)}{r^4}\right)
  + c_{8|4,2,2}\frac{\sin^4(f)}{r^4}\left(C^2 f_r^4
    + \frac{2\sin^2(f)}{r^2} C f_r^2\right),\\
-\Lag_{10} &= c_{10|4,4,2} \frac{\sin^6(f)}{r^6}\left(2 C^2 f_r^4
    + \frac{\sin^2(f)}{r^2} C f_r^2\right),\\
-\Lag_{12} &= c_{12|4,4,4} \frac{\sin^8(f)}{r^8} C^2 f_r^4.
\end{align}
It is convenient to note that the invariant derived with respect
to the inverse metric, has a very simple form once we plug in the
hedgehog Ansatz and Schwarzschild metric. Due to spherical symmetry
and diagonal metric, they remain diagonal and their components read
\begin{align}
\ab{s}_{tt} &= 0,\\
\ab{s}_{rr} &= C^{s-1} f_r^{2s},\\
\ab{s}_{\theta\theta} &= \frac{\sin^{2s}(f)}{r^{2(s-1)}},\\
\ab{s}_{\phi\phi} &= \frac{\sin^2(\theta)\sin^{2s}(f)}{r^{2(s-1)}}.
\end{align}
Let us now evaluate all the once-derived building blocks with respect
to the inverse metric, namely, $\cC_{\mu\nu}^{(s)}$. 
\begin{align}
\cC_{rr}^{(1)} &= 2f_r^2,\quad&
\cC_{\theta\theta}^{(1)} &= 2\sin^2(f),\\
\cC_{rr}^{(2)} &= \frac{8\sin^2(f)}{r^2} f_r^2,\quad&
\cC_{\theta\theta}^{(2)} &= 4\sin^2(f)\left(C f_r^2
  + \frac{\sin^2(f)}{r^2}\right),\\
\cC_{rr}^{(3)} &= \frac{6\sin^4(f)}{r^4} f_r^2,\quad&
\cC_{\theta\theta}^{(3)} &= \frac{6\sin^4(f)}{r^2} C f_r^2,\\
\cC_{rr}^{(4)} &= \frac{16\sin^4(f)}{r^4}\left(C f_r^4
  + \frac{\sin^2(f)}{r^2} f_r^2\right),\quad&
\cC_{\theta\theta}^{(4)} &= \frac{8\sin^4(f)}{r^2} \left(C^2 f_r^4
  + \frac{3\sin^2(f)}{r^2} C f_r^2\right),\\
\cC_{rr}^{(5)} &= \frac{10\sin^6(f)}{r^6}\left(4C f_r^4
+ \frac{\sin^2(f)}{r^2} f_r^2\right),\quad&
\cC_{\theta\theta}^{(5)} &= \frac{10\sin^6(f)}{r^4}\left(3C^2 f_r^4
  + \frac{2\sin^2(f)}{r^2} C f_r^2\right),\\
\cC_{rr}^{(6)} &= \frac{12\sin^8(f)}{r^8} C f_r^4,\quad&
\cC_{\theta\theta}^{(6)} &= \frac{12\sin^8(f)}{r^6} C^2 f_r^4,
\end{align}
and all $\cC_{tt}^{(s)}=0$ due to the static Ansatz
while all
$\cC_{\phi\phi}^{(s)}=\sin^2(\theta)\cC_{\theta\theta}^{(s)}$ due to
spherical symmetry.

Finally, we can evaluate the stress-energy tensors with the hedgehog
Ansatz and Schwarzschild metric giving
\begin{align}
T_{tt}^{(2)} &= c_2 N^2C \left(C f_r^2
  + \frac{2\sin^2(f)}{r^2}\right),\label{eq:Ttt2}\\
T_{rr}^{(2)} &= c_2 C^{-1}\left(C f_r^2
  - \frac{2\sin^2(f)}{r^2}\right),\\
T_{\theta\theta}^{(2)} &= -c_2 r^2 C f_r^2,
\end{align}
\begin{align}
T_{tt}^{(4)} &= c_4 N^2C \frac{\sin^2(f)}{r^2}\left(2C f_r^2
  + \frac{\sin^2(f)}{r^2}\right),\\
T_{rr}^{(4)} &= c_4 C^{-1}\frac{\sin^2(f)}{r^2} \left(2C f_r^2
  - \frac{\sin^2(f)}{r^2}\right),\\
T_{\theta\theta}^{(4)} &= c_4 \frac{\sin^4(f)}{r^2},
\end{align}
\begin{align}
T_{tt}^{(6)} &= c_6 N^2C^2 \frac{\sin^4(f)}{r^4} f_r^2,\\
T_{rr}^{(6)} &= c_6 \frac{\sin^4(f)}{r^4} f_r^2,\\
T_{\theta\theta}^{(6)} &= c_6 \frac{\sin^4(f)}{r^2} C f_r^2,
\end{align}
\begin{align}
T_{tt}^{(8)} &= (2c_{8|4,4} + c_{8|4,2,2}) N^2C^3 \frac{\sin^4(f)}{r^4} f_r^4
  + 2c_{8|4,2,2} N^2C^2 \frac{\sin^6(f)}{r^6} f_r^2
  + c_{8|4,4} N^2C \frac{\sin^8(f)}{r^8},\\
T_{rr}^{(8)} &= 3(2c_{8|4,4} + c_{8|4,2,2}) \frac{\sin^4(f)}{r^4} C f_r^4
  + 2c_{8|4,2,2} \frac{\sin^6(f)}{r^6} f_r^2
  - c_{8|4,4}\frac{\sin^8(f)}{C r^8},\\
T_{\theta\theta}^{(8)} &=
  (2c_{8|4,4} + c_{8|4,2,2}) \frac{\sin^4(f)}{r^2} C^2 f_r^4
  + 4c_{8|4,2,2}\frac{\sin^6(f)}{r^4} C f_r^2
  + 3c_{8|4,4}\frac{\sin^8(f)}{r^6},
\end{align}
\begin{align}
T_{tt}^{(10)} &= c_{10|4,4,2} N^2C \frac{\sin^6(f)}{r^6}\left(2C^2 f_r^4
  + \frac{\sin^2(f)}{r^2} C f_r^2\right),\\
T_{rr}^{(10)} &= c_{10|4,4,2} \frac{\sin^6(f)}{r^6}\left(6 C f_r^4
  + \frac{\sin^2(f)}{r^2} f_r^2\right),\\
T_{\theta\theta}^{(10)} &=
  c_{10|4,4,2} \frac{\sin^6(f)}{r^4}\left(4C^2 f_r^4
  + \frac{3\sin^2(f)}{r^2} C f_r^2\right),
\end{align}
\begin{align}
T_{tt}^{(12)} &= c_{12|4,4,4} N^2C^3 \frac{\sin^8(f)}{r^8} f_r^4,\label{eq:Ttt12}\\
T_{rr}^{(12)} &= 3c_{12|4,4,4} \frac{\sin^8(f)}{r^8} C f_r^4,\\
T_{\theta\theta}^{(12)} &= 3c_{12|4,4,4} \frac{\sin^8(f)}{r^6} C^2 f_r^4,
\end{align}
and due to spherical symmetry, we have that
$T_{\phi\phi} = \sin^2(\theta)T_{\theta\theta}$.

\section{Black hole Skyrme hair}\label{sec:BHhair}

Once we have chosen the Lagrangian, $\Lag$, out of our set of
Lagrangians and fixed the constants, we are ready to solve the
Einstein equation 
\beq
G_{\mu\nu}
= R_{\mu\nu} - \frac{1}{2}g_{\mu\nu} R
= 8\pi G T_{\mu\nu},
\eeq
fleshed out as
\begin{align}
- \frac{1}{r}N^2CC_r + \frac{1}{r^2}N^2C - \frac{1}{r^2}N^2C^2 &=
  8\pi G T_{tt},\\
\frac{1}{r^2}\left(1 - \frac{1}{C}\right)
  + \frac{C_r}{r C} + \frac{2N_r}{r N} &= 8\pi G T_{rr},\\
\frac{1}{2}r^2 C_{rr}
  + r C_r
  + \frac{r C N_r}{N}
  + \frac{3r^2 C_r N_r}{2N}
  + \frac{r^2C N_{rr}}{N} &= 8\pi G T_{\theta\theta},
\end{align}
for the metric in Eq.~\eqref{eq:metric}.
After taking suitable linear combinations, we can write them as
\begin{align}
-C_r + \frac{1}{r} - \frac{C}{r} &= 8\pi G\frac{r T_{tt}}{N^2C},\label{eq:EEQ1}\\
\frac{N_r}{N} &= 4\pi G \frac{r T_{tt}}{N^2C^2} + 4\pi G r T_{rr},
\label{eq:EEQ2}
\end{align}
which with two boundary conditions determine $N,C$ and hence the
metric. 

Finally, we also need the equation of motion coming from varying the
matter Lagrangian
\beq
-\frac{1}{N r^2}
\p_r\left(N r^2 \frac{\p\Lag}{\p f_r}\right)
+\frac{\p\Lag}{\p f}
= 0.
\eeq
The black hole horizon is defined as $C(r_h)=0$, where $r_h$ is the
horizon radius.
The other boundary conditions we need to impose, is that the profile
function goes to zero at spatial infinity, $f(\infty)=0$, and the
correct value of the first derivative at the horizon radius,
$f_r(r_h)$. 
The latter can be derived by taking the $r\to r_h$ limit of the first 
Einstein equation and the equation of motion, yielding
\begin{align}
C_r - \frac{1}{r_h} = -8\pi G \frac{r_h}{N(r_h)}
  \lim_{r\to r_h} \frac{T_{tt}(r)}{C(r)}, \label{eq:Cr_rh_to_zero}\\
\lim_{r\to r_h}
\left[-\frac{1}{N r^2}
\p_r\left(N r^2 \frac{\p\Lag}{\p f_r}\right)
+\frac{\p\Lag}{\p f}\right] = 0, \label{eq:EOM_rh_to_zero1}
\end{align}
where $\lim_{r\to r_h}C(r)=0$. 
Since the metric function $C$ accompanies the radial derivative
squared, the limit is relatively simple.
Indeed, by looking at the time-time component of the energy-momentum
tensors \eqref{eq:Ttt2}-\eqref{eq:Ttt12}, only the kinetic term, the
Skyrme term and the eighth-order term have a nonvanishing contribution
to the right-hand side of Eq.~\eqref{eq:Cr_rh_to_zero}.

We can rewrite Eq.~\eqref{eq:EOM_rh_to_zero1} by using the fact that
$f_r$ is necessarily accompanied by at least a factor of $C$.
Therefore, the first term can only give a nonzero contribution 
when the radial derivative hits a factor of $C$.\footnote{One may
naively think that a contribution can come from $N_\rho/N$, but the
combination $T_{\rho\rho}+T_{tt}/(N^2C^2)$ does not contain negative
powers of $C$ even though each term separately does. Thus the Einstein
equation is not singular at the horizon. }
Thus we can write
\beq
\lim_{r\to r_h}
\left[- C_r \frac{\p^2\!\Lag}{\p C\p f_r}
  + \frac{\p\Lag}{\p f}\right] = 0, \label{eq:EOM_rh_to_zero2}
\eeq
where we have used Eq.~\eqref{eq:EEQ2}.

In our calculations, we will use a variable simply related to $C$,
namely
\beq
C = 1 - \frac{2m}{r},
\eeq
i.e.~$m$ and it has the physical interpretation as a gravitational
mass function; more precisely,
$\lim_{r\to\infty}m=m(\infty)=m_{\rm ADM}$ is the ADM mass.
It is a good observable to distinguish stable and unstable branches
of solutions by (without turning to a linear stability analysis).

As the inverse metric that accompanies the double derivative of the
profile function, $f_{rr}$, is $g^{rr}=C$, the equation of motion is
singular exactly at the horizon. 
For this reason, we will start all calculations a tiny step $r_\delta$
from the horizon: i.e.~at $r=r_h+r_\delta$ and extrapolate the values
of the fields linearly using the derivatives of $f$ and $\mu$.

It is well known that the topological charge or topological degree is
only a full integer when there is no BH horizon.
That is, part of the topological charge is swallowed by the BH.
In this paper we only consider the spherically symmetric hedgehog, for
which the topological charge outside the horizon is less than unity.
In particular, we have
\begin{align}
B &= \frac{1}{2\pi^2} \int_{\tiny\begin{array}{c}\textrm{outside}\\\textrm{horizon}\end{array}} d^3x\; \sqrt{-g}
  \sqrt{-g^{tt}} \sqrt{\frac{\cC_3}{3}} 
  = \frac{2}{\pi} \int_{r_h}^{\infty} dr \; \sin^2(f) f_r
  = \frac{2f_h - \sin 2f_h}{2\pi},
\end{align}
where $f_h\equiv f(r_h)$ is the value of the profile function at
the horizon.
$B<1$ for $f_h<\pi$, which is the case for all the BH hair solutions
we find in this paper.
Note that since the first derivative of $B$ with respect to $f_h$ is
zero at $f_h=\pi$, the charge is close to one for a range of
$f_h\lesssim\pi$.

\section{The \texorpdfstring{$2+4+2n$}{2+4+2n} model}\label{sec:m242n}

This model is based on the standard black hole Skyrme hair with an
added higher-order term, which is higher order than the Skyrme term.
The purpose of this model is to construct a stable modified black hole
hair, and then slowly turn off the Skyrme term to see if the hair
persists.

\subsection{General setup}
The model is defined as
\beq
\Lag = \Lag_2 + \Lag_4 + \Lag_{2n} + \frac{R}{16\pi G},
\eeq
where $n=3,4,5,6$ specifies the higher-order term added to the normal
black hole Skyrme hair, $R$ is the Ricci scalar and $G$ is Newton's
constant. 

We will now switch to dimensionless units
\beq
\rho \equiv a r,
\eeq
where $a$ sets the (inverse) length scale and has mass dimension 1. 
Hence, we can write the Lagrangian as
\beq
\Lag = a^3 M_0 \left(\frac{1}{a M_0}\Lag_2 + \frac{a}{M_0}\Lag_4
+ \frac{a^{2n-3}}{M_0}\Lag_{2n}
+ \frac{R}{4\alpha}\right),
\eeq
where we have defined
\beq
\alpha = 4\pi G a M_0,
\label{eq:alpha_def}
\eeq
i.e.~the effective (dimensionless) gravitational coupling, while $M_0$
will turn out to set the mass scale for the soliton hair, as we will
show shortly. 

We will use the freedom of choosing the scales to remove $c_2$ and
$c_{2n}$ from the dynamical equations:
\beq
a M_0 = c_2, \qquad
\frac{M_0}{a^{2n-3}} = c_{2n},
\eeq
which we can invert to
\beq
a = \left(\frac{c_2}{c_{2n}}\right)^{\frac{1}{2n-2}}, \qquad
M_0 = c_2\left(\frac{c_{2n}}{c_2}\right)^{\frac{1}{2n-2}}.
\eeq
The mass units of the sub-Lagrangian constants are $[c_{2n}]=4-2n$,
and thus it can readily be verified that both $M_0$ and $a$ have mass
dimension one: $[M_0]=[a]=1$ (as they should). 

Hence, we can now write the rescaled Lagrangian as
\beq
\Lag = a^3 M_0\left(\frac{1}{c_2}\Lag_2 + \frac{\beta}{c_4}\Lag_4
+ \frac{1}{c_{2n}}\Lag_{2n}
+ \frac{R}{4\alpha}\right),
\label{eq:Lrescaled}
\eeq
where we have defined
\beq
\beta \equiv
\frac{c_4}{c_2}\left(\frac{c_2}{c_{2n}}\right)^{\frac{1}{n-1}},
\eeq
which can be verified to be a dimensionless parameter of the model. 
The dynamical equations now only depend on the effective gravitational
coupling $\alpha$, and the coupling $\beta$.

Finally, we will show that the soliton hair in the rescaled units has a
mass given by $M_0$ times a dimensionless integral (number):
\begin{align}
M &= -4\pi \int_{r_h}^\infty dr\; r^2 N
  \left(\Lag_2(r) + \Lag_4(r) + \Lag_{2n}(r)\right) \non
  &= -4\pi M_0\int_{\rho_h}^\infty d\rho\; \rho^2 N
  \left(\frac{1}{c_2}\Lag_2(\rho) + \frac{\beta}{c_4}\Lag_4(\rho)
    + \frac{1}{c_{2n}}\Lag_{2n}(\rho)\right),
\end{align}
and as a check, we can see that the Einstein equations are now
dimensionless 
\beq
G_{\mu\nu} =
2\alpha
\left(\frac{1}{c_2}T_{\mu\nu}^{(2)}(\rho)
  + \frac{\beta}{c_4}T_{\mu\nu}^{(4)}(\rho)
  + \frac{1}{c_{2n}}T_{\mu\nu}^{(2n)}(\rho)\right),
\eeq
and the suitable linear combinations in Eqs.~\eqref{eq:EEQ1}-\eqref{eq:EEQ2}
read
\begin{align}
-C_\rho + \frac{1}{\rho} - \frac{C}{\rho} &=
2\alpha\left[
\rho C f_\rho^2
+\frac{2\sin^2f}{\rho}
+\frac{2\beta C\sin^2(f)f_\rho^2}{\rho}
+\frac{\beta\sin^4f}{\rho^3}
+\frac{\rho T_{tt}^{(2n)}}{c_{2n}N^2C}
\right],
\label{eq:EEQ1_242n}\\
\frac{1}{\alpha}\frac{N_\rho}{N} &=
2\rho f_\rho^2
+\frac{4\beta \sin^2(f)f_\rho^2}{\rho}
+\frac{\rho T_{tt}^{(2n)}}{c_{2n}N^2C^2}
+\frac{\rho T_{\rho\rho}^{(2n)}}{c_{2n}},
\label{eq:EEQ2_242n}
\end{align}
for $n=3,4,5,6$.

It will be useful in the next subsections, to have dimensionless
expressions for the boundary conditions.
Let us start with Eq.~\eqref{eq:Cr_rh_to_zero}:
\beq
C_\rho - \frac{1}{\rho_h} =
-2\alpha\left[
\frac{2\sin^2f}{\rho}
+\frac{\beta\sin^4f}{\rho^3}
+\frac{\rho_h}{c_{2n}N^2(\rho_h)}\lim_{\rho\to\rho_h}\frac{T_{tt}^{(2n)}}{C}
\right],
\label{eq:C_rho_at_rh}
\eeq
whereas the matter equation of motion is slightly more involved.
We will use the expression \eqref{eq:EOM_rh_to_zero2} and write it in
dimensionless units as
\begin{align}
2C_\rho f_\rho
+\frac{4\beta\sin^2(f)C_\rho f_\rho}{\rho^2}
-\frac{C_\rho(\rho_h)}{c_{2n}}\lim_{\rho\to\rho_h}
  \frac{\p^2\Lag_{2n}}{\p C\p f_\rho} &\non
\mathop-\frac{2\sin 2f}{\rho^2}
-\frac{2\beta\sin^2(f)\sin 2f}{\rho^4}
+\frac{1}{c_{2n}}\lim_{\rho\to\rho_h}\frac{\p\Lag_{2n}}{\p f}
&= 0.
\label{eq:f_rho_at_rh}
\end{align}
For the numerical calculations, we will use the dimensionless mass
function, $\mu$, given via
\beq
C = 1 - \frac{2\mu}{\rho}.
\eeq
The boundary conditions at the black hole horizon can now be written
as the following expansion
\begin{align}
\mu &= \frac{\rho_h}{2} + (\rho - \rho_h)\mu_\rho(\rho_h)
  + \mathcal{O}\left((\rho-\rho_h)^2\right),\\
f &= f_h + (\rho - \rho_h)f_\rho(\rho_h)
  + \mathcal{O}\left((\rho-\rho_h)^2\right),
\end{align}
where $f_h\equiv f(\rho_h)$ is the shooting parameter and provides
nonperturbative information about the soliton hair. 
$\mu_\rho(\rho_h)$ is extracted from Eq.~\eqref{eq:C_rho_at_rh} by
using that $C_\rho=\frac{2\mu}{\rho^2}-\frac{2\mu_\rho}{\rho}$.
$f_\rho(\rho_h)$, i.e.~the derivative of the profile function of the
Skyrme hair at the horizon, is isolated in Eq.~\eqref{eq:f_rho_at_rh}
and $C_\rho$ is eliminated by insertion of Eq.~\eqref{eq:C_rho_at_rh}.

The last observable that we will use here, is the Hawking temperature
\beq
T_H = \frac{N(\rho_h) C_\rho(\rho_h)}{4\pi}.
\eeq
$C_\rho(\rho_h)$ is almost a local quantity; it can be found by
just knowing the value of the profile at the horizon, $f_h$.
This is not so with $N(\rho_h)$; this is a nonlocal quantity and it
must be obtained by integrating $N$ from infinity down to the horizon
radius, $\rho_h$, using Eq.~\eqref{eq:EEQ2_242n}, where we have used
the boundary condition $N(\infty)=1$.
The Hawking temperature defined above is dimensionless; to convert to
the dimensionful temperature one should multiply by $a$ yielding
$aT_H$.

In the following subsections, we will study each case of $n=3,4,5,6$
in turn.

\subsection{The \texorpdfstring{$2+4+6$}{2+4+6} model}

This model was already studied in Ref.~\cite{Gudnason:2016kuu}, and
the result was that the sixth-order BPS-Skyrme term cannot stabilize
solitonic black hole hair.
For completeness, we will review the results here and provide a few
new figures. 
This will help facilitate a comparison between the characteristics of
this model and the other ones.

Let us first complete the Einstein
equations \eqref{eq:EEQ1_242n}-\eqref{eq:EEQ2_242n},
\begin{align}
-C_\rho + \frac{1}{\rho} - \frac{C}{\rho} &=
2\alpha\left[
\rho C f_\rho^2
+\frac{2\sin^2f}{\rho}
+\frac{2\beta C\sin^2(f)f_\rho^2}{\rho}
+\frac{\beta\sin^4f}{\rho^3}
+\frac{\sin^4(f)C f_\rho^2}{\rho^3}
\right],\non
\frac{1}{\alpha}\frac{N_\rho}{N} &=
2\rho f_\rho^2
+\frac{4\beta \sin^2(f)f_\rho^2}{\rho}
+\frac{2\sin^4(f)f_\rho^2}{\rho^3},
\end{align}
while the equation of motion for the profile function is
\begin{align}
C f_{\rho\rho}
+\frac{2Cf_\rho}{\rho}
+C_\rho f_\rho
+\frac{N_\rho C f_\rho}{N}
-\frac{\sin 2f}{\rho^2} \non
+\frac{2\beta\sin^2f}{\rho^2}\left(
  C f_{\rho\rho}
  +C_\rho f_\rho
  +\frac{N_\rho C f_\rho}{N}
  -\frac{\sin 2f}{2\rho^2}
 \right)
+\frac{\beta\sin(2f) C f_\rho^2}{\rho^2}
\non
+\frac{\sin^4f}{\rho^4}\left(
  C f_{\rho\rho}
  -\frac{2C f_\rho}{\rho}
  +C_\rho f_\rho
  +\frac{N_\rho C f_\rho}{N}
  \right)
+\frac{\sin^2(f)\sin(2f) C f_\rho^2}{\rho^4}
= 0.
\end{align}
Finally, the boundary
conditions \eqref{eq:C_rho_at_rh}-\eqref{eq:f_rho_at_rh} can be
written as
\begin{align}
\mu &= \frac{\rho_h}{2}
+\alpha\sin^2(f_h)\left(2
  + \frac{\beta\sin^2f_h}{\rho_h^2}\right)(\rho-\rho_h)
  + \mathcal{O}\left((\rho-\rho_h)^2\right),\\
f &= f_h
+\rho_h^3 \sin(2f_h)
\frac{\rho_h^2 + \beta\sin^2(f_h)}
{\left(\rho_h^4 + 2\beta\rho_h^2\sin^2f_h + \sin^4 f_h\right)\Xi_6}
(\rho-\rho_h)
+ \mathcal{O}\left((\rho-\rho_h)^2\right),\\
\Xi_6 &\equiv
\rho_h^2 - 4\alpha\rho_h^2\sin^2f_h - 2\alpha\beta\sin^4 f_h.
\end{align}

\begin{figure}[!thp]
\begin{center}
\mbox{\subfloat[]{\includegraphics[width=0.49\linewidth]{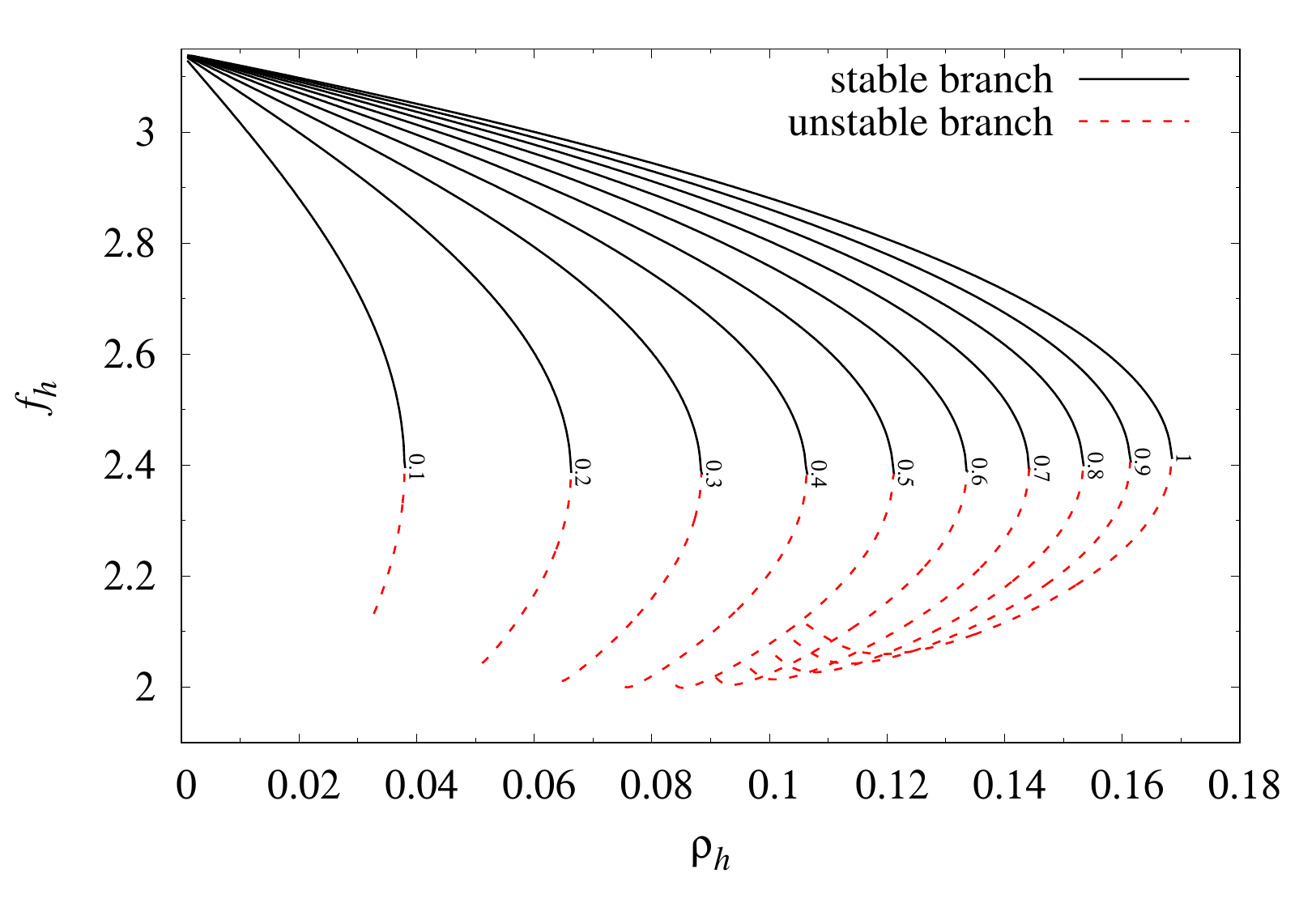}}
\subfloat[]{\includegraphics[width=0.49\linewidth]{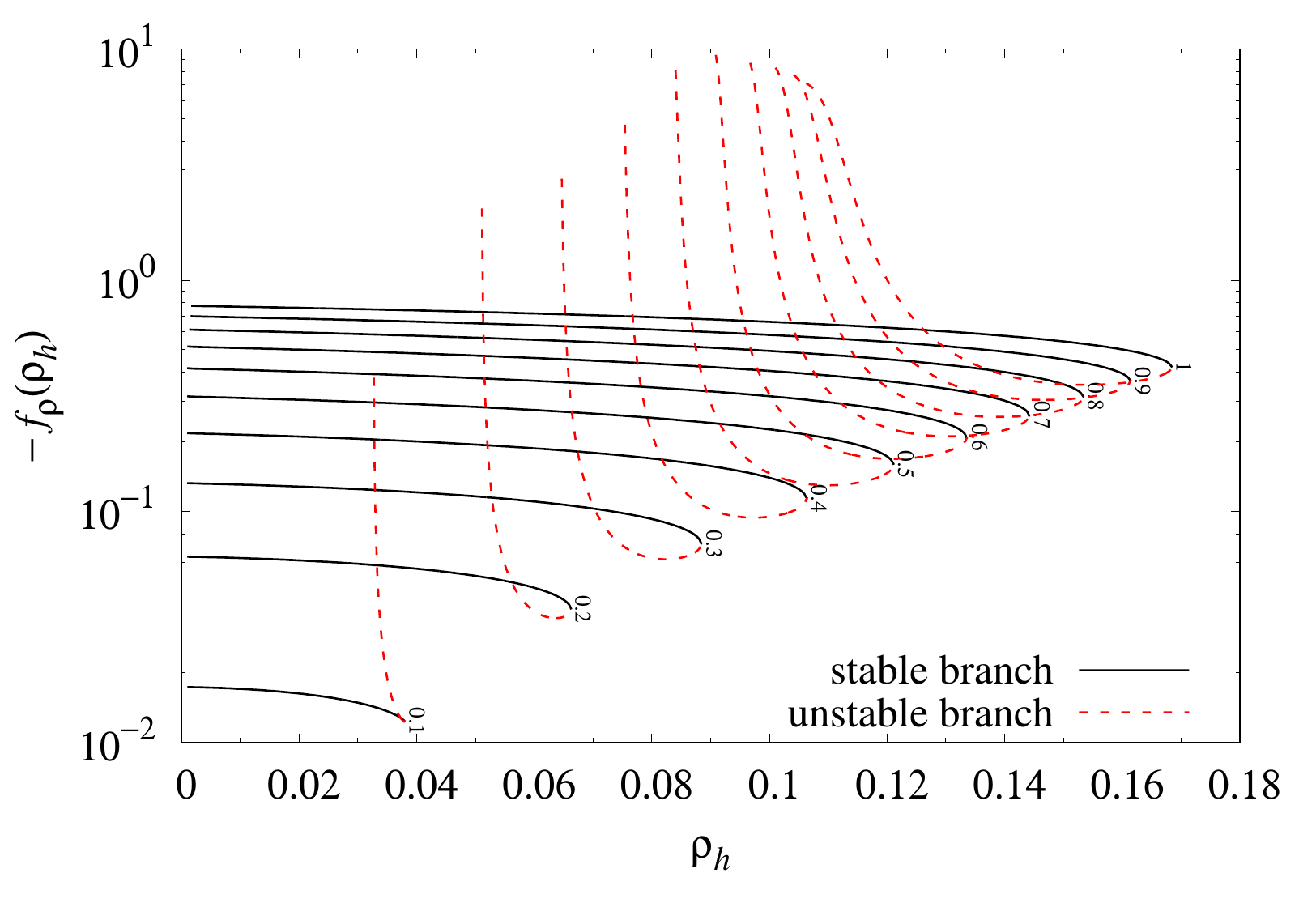}}}
\mbox{\subfloat[]{\includegraphics[width=0.49\linewidth]{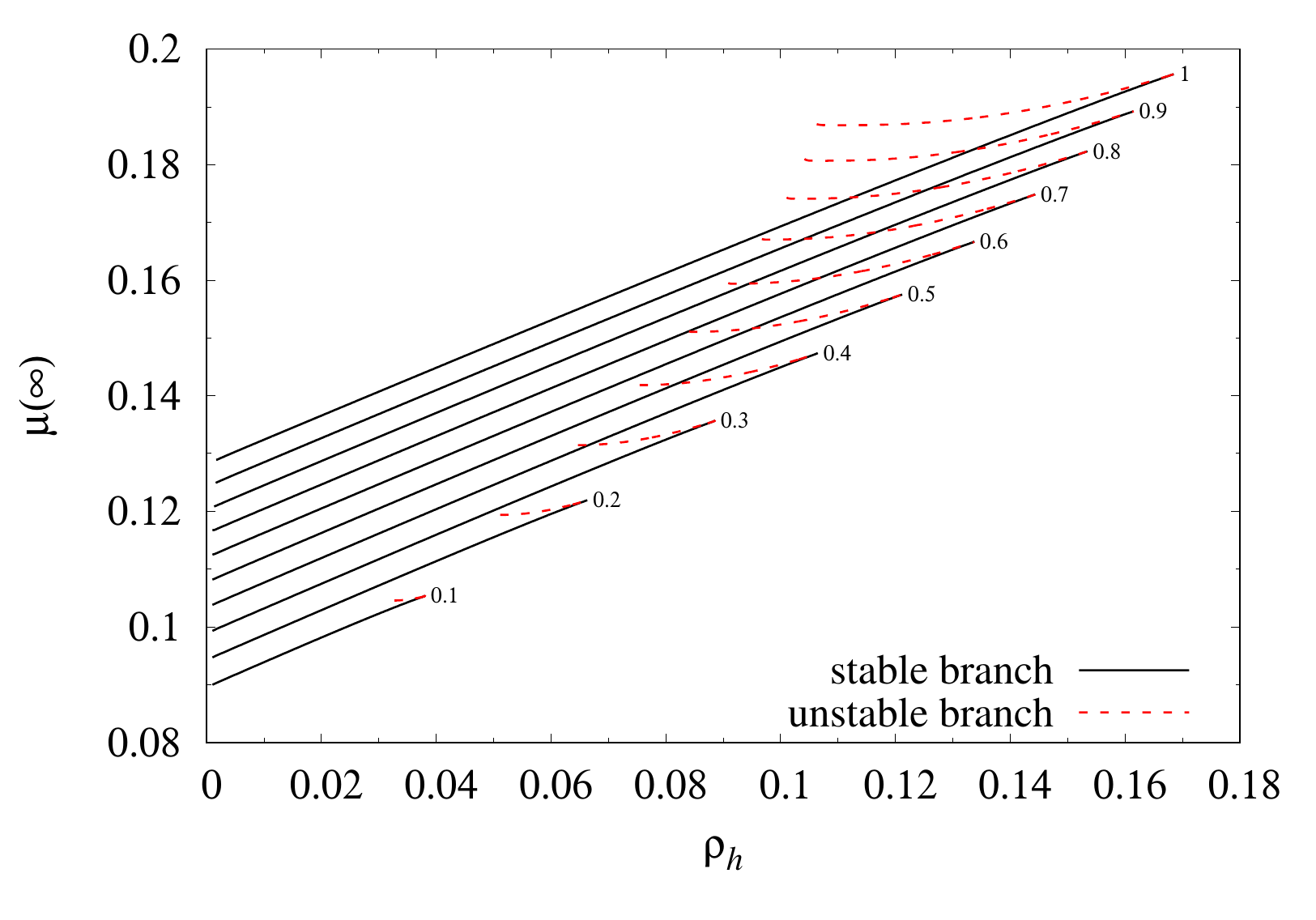}}
\subfloat[]{\includegraphics[width=0.49\linewidth]{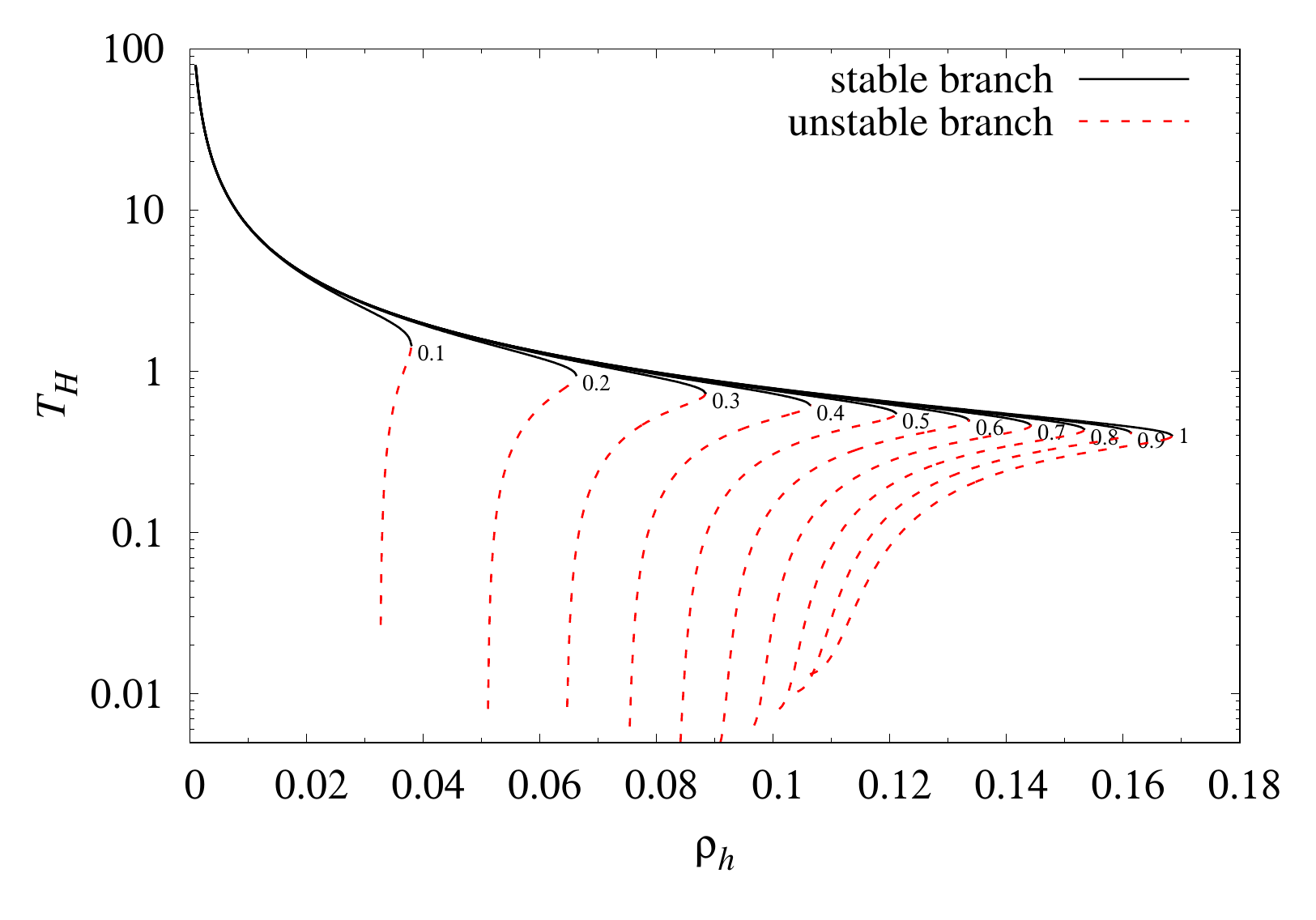}}}
\caption{Stable (black solid lines) and unstable (red dashed lines)
  branches of solutions in the $2+4+6$ model: (a) the value of the
  profile function at the horizon, $f_h$; (b) the derivative of the
  profile function at the horizon, $f_\rho(\rho_h)$; 
  (c) the ADM mass, $\mu(\infty)$;  
  (d) the Hawking temperature $T_H$, all as functions of the size of the
  black hole, i.e.~the horizon radius, $\rho_h$.
  The numbers on the figures indicate the different values of
  $\beta=0.1,0.2,\ldots,1$. 
}
\label{fig:m246_rh}
\end{center}
\end{figure}

\begin{figure}[!thp]
\begin{center}
\mbox{\subfloat[]{\includegraphics[width=0.49\linewidth]{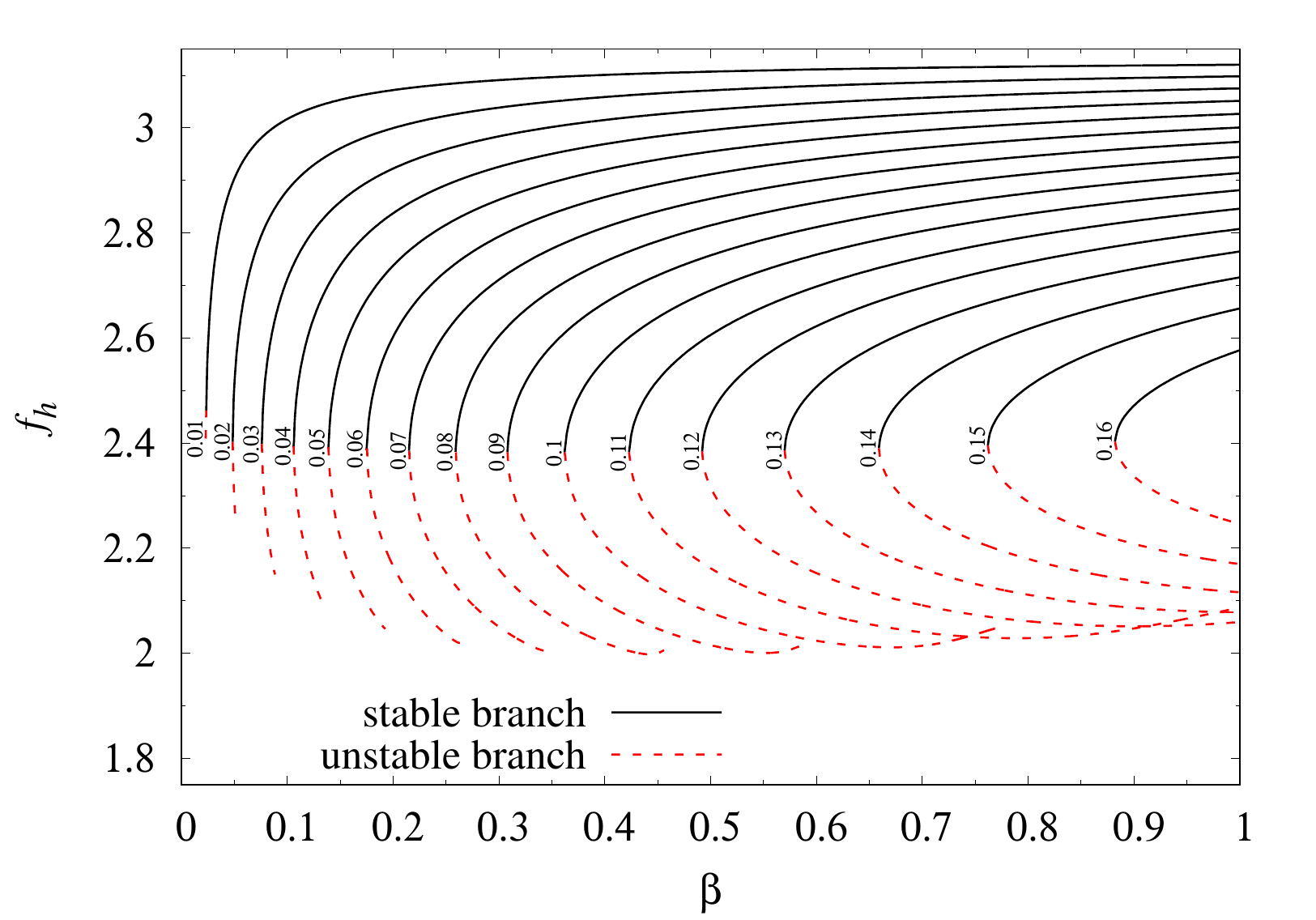}}
\subfloat[]{\includegraphics[width=0.49\linewidth]{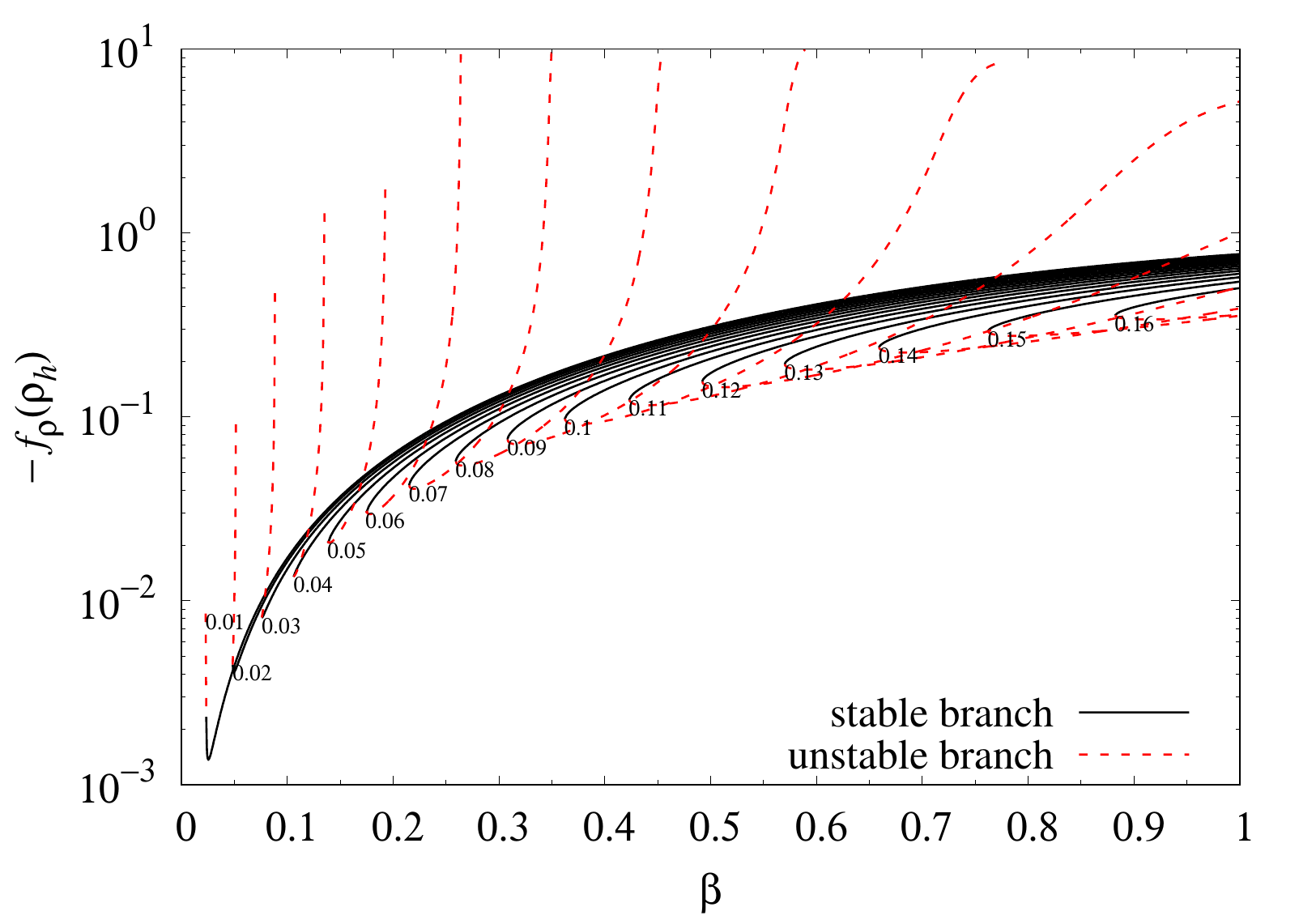}}}
\mbox{\subfloat[]{\includegraphics[width=0.49\linewidth]{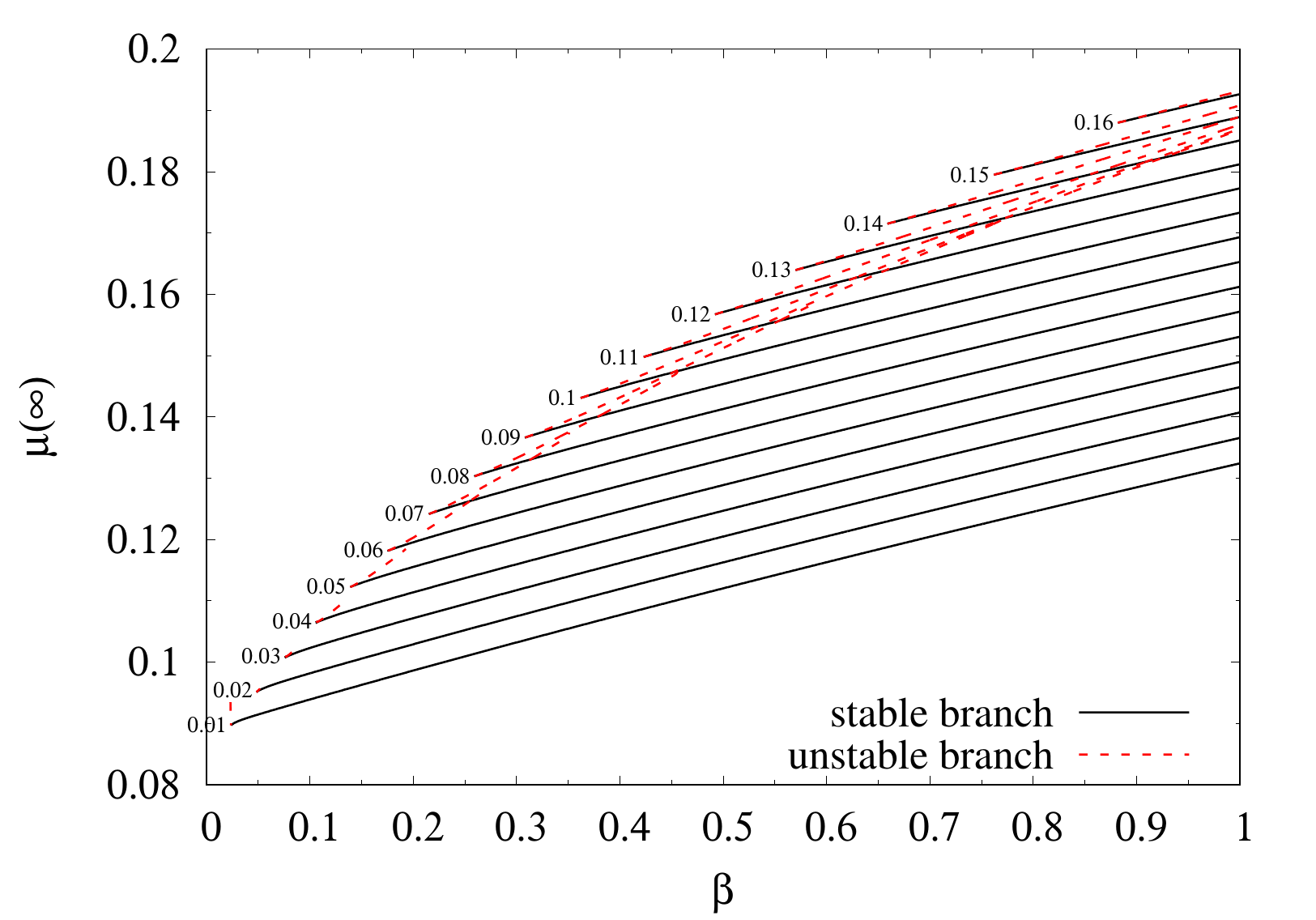}}
\subfloat[]{\includegraphics[width=0.49\linewidth]{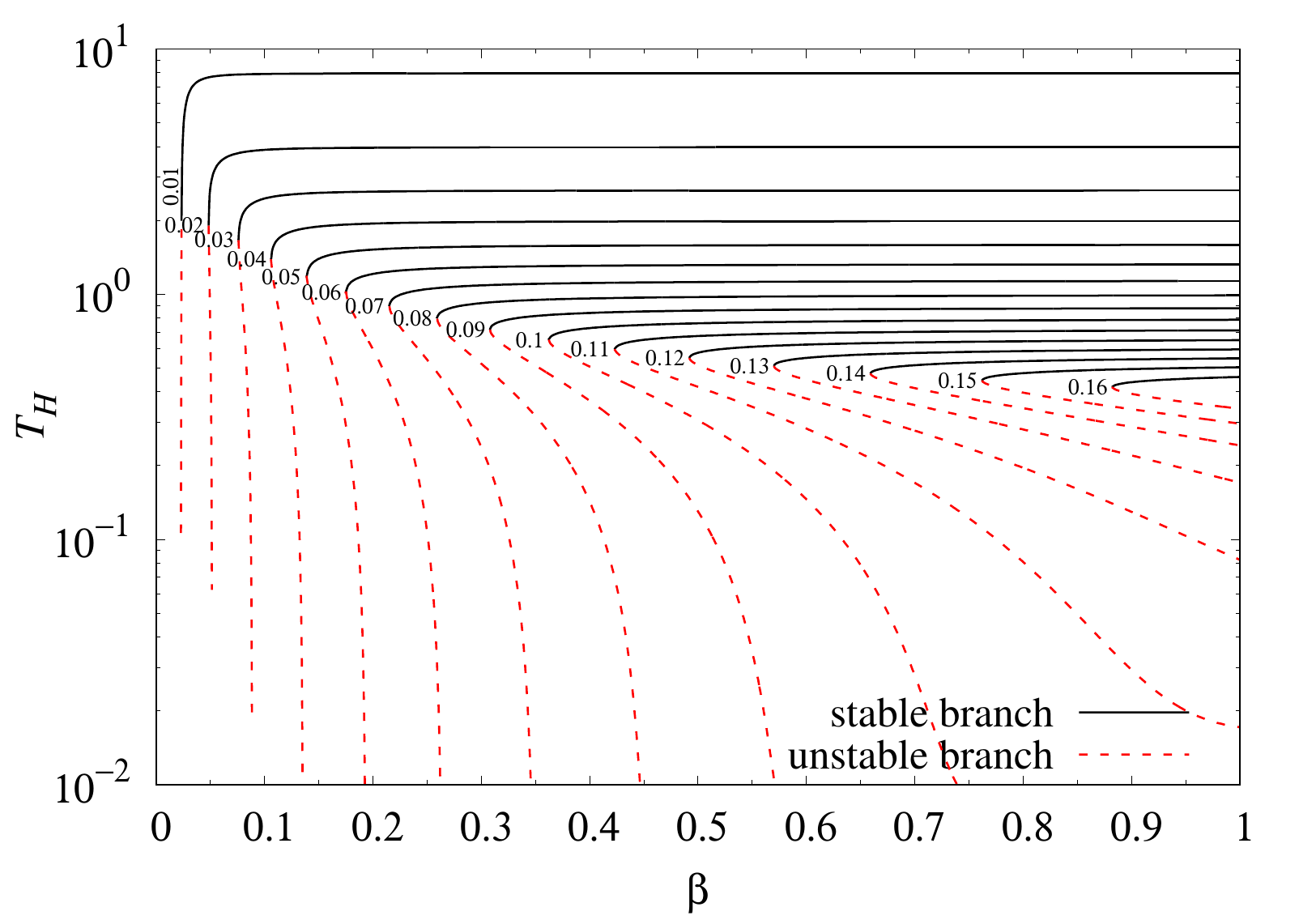}}}
\caption{Stable (black solid lines) and unstable (red dashed lines)
  branches of solutions in the $2+4+6$ model: (a) the value of the
  profile function at the horizon, $f_h$; (b) the derivative of the
  profile function at the horizon, $f_\rho(\rho_h)$; 
  (c) the ADM mass, $\mu(\infty)$;  
  (d) the Hawking temperature $T_H$, all as functions of the
  Skyrme-term coefficient, $\beta$. 
  The numbers on the figures indicate the different values of
  $\rho_h=0.01,0.02,\ldots,0.16$. }
\label{fig:m246_beta}
\end{center}
\end{figure}

We are now ready to present the numerical results for this model.
Fig.~\ref{fig:m246_rh} shows the following quantities: the profile
function at the horizon (the shooting parameter) $f_h$, the derivative
of the profile function at the horizon $f_\rho(\rho_h)$, the ADM mass
$\mu(\infty)$ and finally, the Hawking temperature $T_H$.
They are all plotted as functions of the horizon radius, $\rho_h$.
These quantities are used in many figures in this paper and we will
henceforth refer to them as the standard quantities. 
The stable branches are indicated with black solid lines and the red
dashed lines display the unstable branches. The two branches meet in
all the figures at the bifurcation point, which determines the largest
possible black hole size that can support the black hole hair for the
given parameters.
The numbers in the figures indicate the values of $\beta$ for the
different branches.
From Fig.~\ref{fig:m246_rh}(c) it is easy to verify that the unstable
branches all have larger ADM mass, for all values of the horizon
radius, $\rho_h$, than their corresponding stable branches do.

A feature of the $2+4+6$ model, which is not present in the standard
Skyrme model (the $2+4$ model), is that the unstable branches do not
continue all the way back up to the flat-space limit ($\rho_h\to 0$)
solution \cite{Gudnason:2016kuu,Adam:2016vzf}.
We can see two things that happen right about where the unstable
branches cease to exist; Fig.~\ref{fig:m246_rh}(b) shows that (minus)
the derivative of the profile function at the horizon,
$-f_\rho(\rho_h)$, increases drastically and Fig.~\ref{fig:m246_rh}(d)
shows that the temperature drops drastically, right before the
branches cease to exist.

The most important point of Fig.~\ref{fig:m246_rh} is that as $\beta$
is turned off (sent to zero), even the stable branches tend to not
exist.
In order to show it more explicitly, we provide the same figures, but
plotted as functions of $\beta$ in Fig.~\ref{fig:m246_beta}.
It is clear from Fig.~\ref{fig:m246_beta}(a), that as $\beta$ tends to
zero, only parts of the branches with very small horizon radii,
$\rho_h$ remain. In the limit of $\beta\to 0$, no branch remains and
the Skyrme hair ceases to exist.
Fig.~\ref{fig:m246_beta}(b) shows what happens to the derivative of
the profile function at the horizon as $\beta$ is turned off.
A bit counter-intuitive, perhaps, the derivative actually tends to
zero.
The ADM mass gets smaller as $\beta\to 0$, which signals that the
amount of black hole hair decreases, see Fig.~\ref{fig:m246_beta}(c). 
It is not completely clear-cut what happens to the Hawking temperature
in the limit of $\beta\to 0$, Fig.~\ref{fig:m246_beta}(d).
It is clear that only the branches with very small horizon radii,
$\rho_h$, persist for small $\beta$.
It seems that the bifurcation point moves slightly up as the horizon
radius is decreased.
In any case, for small $\rho_h$, the unstable branches possess
temperature curves that go drastically fast towards zero, until they
cease to exist. 

The Figs.~\ref{fig:m246_rh} and \ref{fig:m246_beta} provide solid
evidence for the fact that the BH hair does not exist in the
$\beta\to 0$ limit.

\subsection{The \texorpdfstring{$2+4+8$}{2+4+8} model}

This model contains the first term in increasing order of derivatives 
after the sextic term, also known as the BPS-Skyrme term -- that is,
an eighth-order derivative term.
The minimal construction of Ref.~\cite{Gudnason:2017opo} limits the
terms such that there are no powers of a derivative in the $i$-th
direction higher than four. 
This leaves us with two independent terms, as we will see and the
coefficients in Ref.~\cite{Gudnason:2017opo} are called $c_{8|4,4}$
and $c_{8|4,2,2}$ due to their composition in terms of eigenvalues,
see Eq.~\eqref{eq:rinvlambda}.
We can intuitively think of the two terms as the quadratic part of the
Skyrme term squared and the cross terms, respectively, see
Ref.~\cite{Gudnason:2017opo} for details.
The term with coefficient $c_{8|4,2,2}$ also has the interpretation as
being the BPS-Skyrme term multiplied by the kinetic (Dirichlet) term.
Furthermore, for $c_{8|4,2,2}=2c_{8|4,4}$ the two terms combine to be
the Skyrme term squared. 

The rescaled Lagrangian density \eqref{eq:Lrescaled} divides the term
$\Lag_8$ by a factor of $c_8$.
Since we have two independent coefficients, let us define
\beq
c_{8|4,4} \equiv \gamma c_8, \qquad
c_{8|4,2,2} \equiv (1-\gamma) c_8.
\eeq
After the rescaling, $c_8$ drops out and $\gamma\in[0,1]$ interpolates
between the two terms.
Here we will be interested in the following three possibilities.
The two terms separately is one way to disentangle their effect,
therefore we will consider $\gamma=0$ and $\gamma=1$.
Furthermore, we will consider $\gamma=\frac{1}{3}$ because it
corresponds to the Skyrme term squared.

The $\gamma=0$ term is composed by one eigenvalue of the strain
tensor in Eq.~\eqref{eq:straintensor} to the fourth power, multiplied by the
two other (nonzero) eigenvalues squared,
i.e.~$\lambda_1^4\lambda_2^2\lambda_3^2$ and then the product is
symmetrized, yielding
\beq
\lambda_1^4\lambda_2^2\lambda_3^2
+\lambda_1^2\lambda_2^4\lambda_3^2
+\lambda_1^2\lambda_2^2\lambda_3^4
= \lambda_1^2\lambda_2^2\lambda_3^2
(\lambda_1^2 + \lambda_2^2 + \lambda_3^2).
\eeq
The $\gamma=1$ term, on the other hand, is composed by only two
eigenvalues of the strain tensor in Eq.~\eqref{eq:straintensor}, both to the
fourth power, i.e.~$\lambda_1^4\lambda_2^4$ and then the product is
again symmetrized over all 3 eigenvalues, yielding
\beq
\lambda_1^4\lambda_2^4
+\lambda_1^4\lambda_3^4
+\lambda_2^4\lambda_3^4.
\eeq
Finally, the $\gamma=\frac{1}{3}$ term, which also corresponds to the
Skyrme term squared, reads
\beq
(\lambda_1^2\lambda_2^2
+\lambda_1^2\lambda_3^2
+\lambda_2^2\lambda_3^2)^2.
\eeq
These three cases should be enough to determine which parts of the
eighth-order term are essential for stabilizing the BH hair.

We are now ready to complete the Einstein
equations \eqref{eq:EEQ1_242n}-\eqref{eq:EEQ2_242n},
\begin{align}
-C_\rho + \frac{1}{\rho} - \frac{C}{\rho} &=
2\alpha\bigg[
\rho C f_\rho^2
+\frac{2\sin^2f}{\rho}
+\frac{2\beta\sin^2(f)C f_\rho^2}{\rho}
+\frac{\beta\sin^4f}{\rho^3} \\
&\phantom{=2\alpha\bigg[\ }
+\frac{\gamma\sin^8f}{\rho^7} 
+\frac{2(1-\gamma)\sin^6(f) C f_\rho^2}{\rho^5}
+\frac{(1+\gamma)\sin^4(f) C^2 f_\rho^4}{\rho^3}
\bigg],\non
\frac{1}{\alpha}\frac{N_\rho}{N} &=
2\rho f_\rho^2
+\frac{4\beta\sin^2(f)f_\rho^2}{\rho}
+\frac{4(1-\gamma)\sin^6(f) f_\rho^2}{\rho^5} 
+\frac{4(1+\gamma)\sin^4(f) C f_\rho^4}{\rho^3}, \nonumber
\end{align}
and the equation of motion for the profile function reads
\begin{align}
C f_{\rho\rho}
+\frac{2C f_\rho}{\rho}
+C_\rho f_\rho
+\frac{N_\rho C f_\rho}{N}
-\frac{\sin 2f}{\rho^2} \non
+\frac{2\beta\sin^2f}{\rho^2}\left(
  C f_{\rho\rho}
  +C_\rho f_\rho
  +\frac{N_\rho C f_\rho}{N}
  -\frac{\sin 2f}{2\rho^2}
  \right)
+\frac{\beta\sin(2f) C f_\rho^2}{\rho^2}
\non
+\frac{2(1-\gamma)\sin^6f}{\rho^6}\left(
  C f_{\rho\rho}
  +\frac{4C f_\rho}{\rho}
  +C_\rho f_\rho
  +\frac{N_\rho C f_\rho}{N}
  \right)
+\frac{3(1-\gamma)\sin^4(f)\sin(2f) C f_\rho^2}{\rho^6}
\non
+\frac{2(1+\gamma)\sin^4(f) C f_\rho^2}{\rho^4}\left(
  3C f_{\rho\rho}
  -\frac{2C f_\rho}{\rho}
  +2 C_\rho f_\rho
  +\frac{N_\rho C f_\rho}{N}
  \right) \non
+\frac{3(1+\gamma)\sin^2(f)\sin(2f) C^2 f_\rho^4}{\rho^4}
-\frac{2\gamma\sin^6(f)\sin 2f}{\rho^8} = 0.
\end{align}
Finally, we need the boundary
conditions \eqref{eq:C_rho_at_rh}-\eqref{eq:f_rho_at_rh}, which can be
written as
\begin{align}
\mu &= \frac{\rho_h}{2}
+\alpha\sin^2(f_h)\left(2
  +\frac{\beta\sin^2f_h}{\rho_h^2}
  +\frac{\gamma\sin^6f_h}{\rho_h^6}\right)(\rho-\rho_h)
+ \mathcal{O}\left((\rho-\rho_h)^2\right),\\
f &= f_h
+\rho_h^5\sin(2f_h)\frac{\rho_h^6 + \beta\rho_h^4\sin^2f_h + 2\gamma\sin^6f_h}
{\left(\rho_h^6 + 2\beta\rho_h^4\sin^2f_h + 2(1-\gamma)\sin^6f_h\right)
\Xi_8}(\rho-\rho_h)
+ \mathcal{O}\left((\rho-\rho_h)^2\right),\\
\Xi_8 &\equiv
\rho_h^6 - 4\alpha\rho_h^6\sin^2f_h
  - 2\alpha\beta\rho_h^4\sin^4f_h - 2\alpha\gamma\sin^8f_h.
\end{align}

\begin{figure}[!thp]
\begin{center}
\mbox{\subfloat[]{\includegraphics[width=0.49\linewidth]{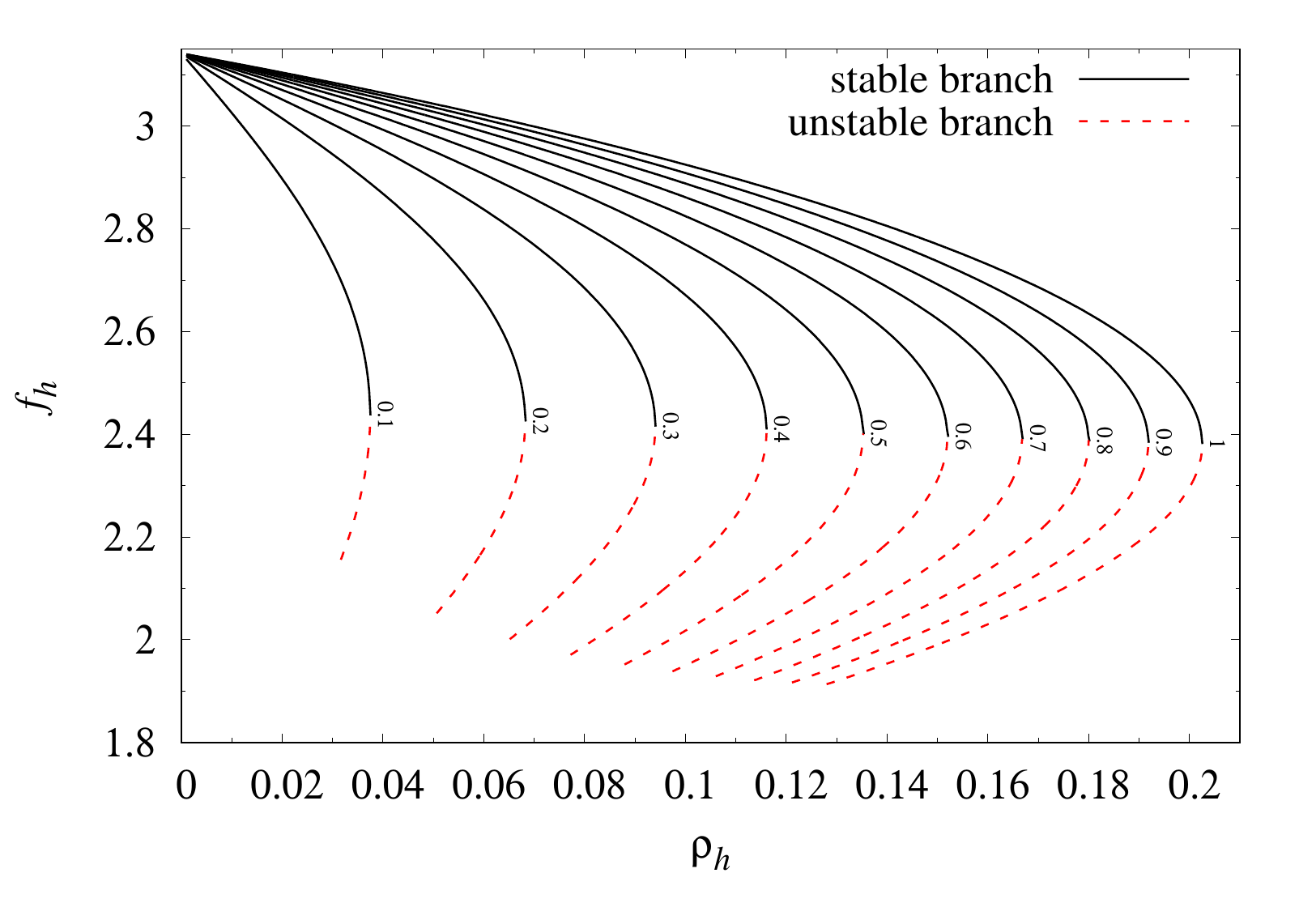}}
\subfloat[]{\includegraphics[width=0.49\linewidth]{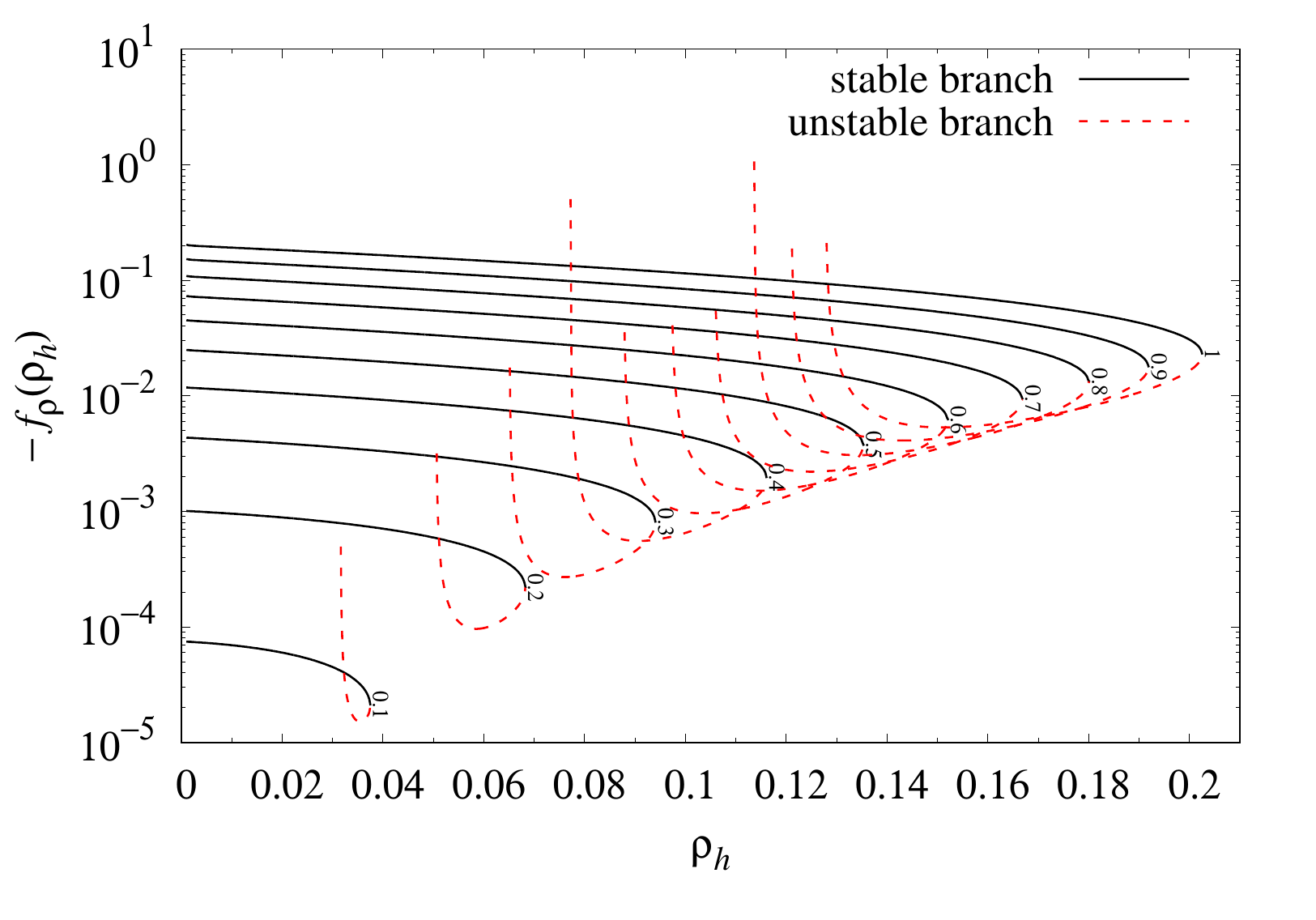}}}
\mbox{\subfloat[]{\includegraphics[width=0.49\linewidth]{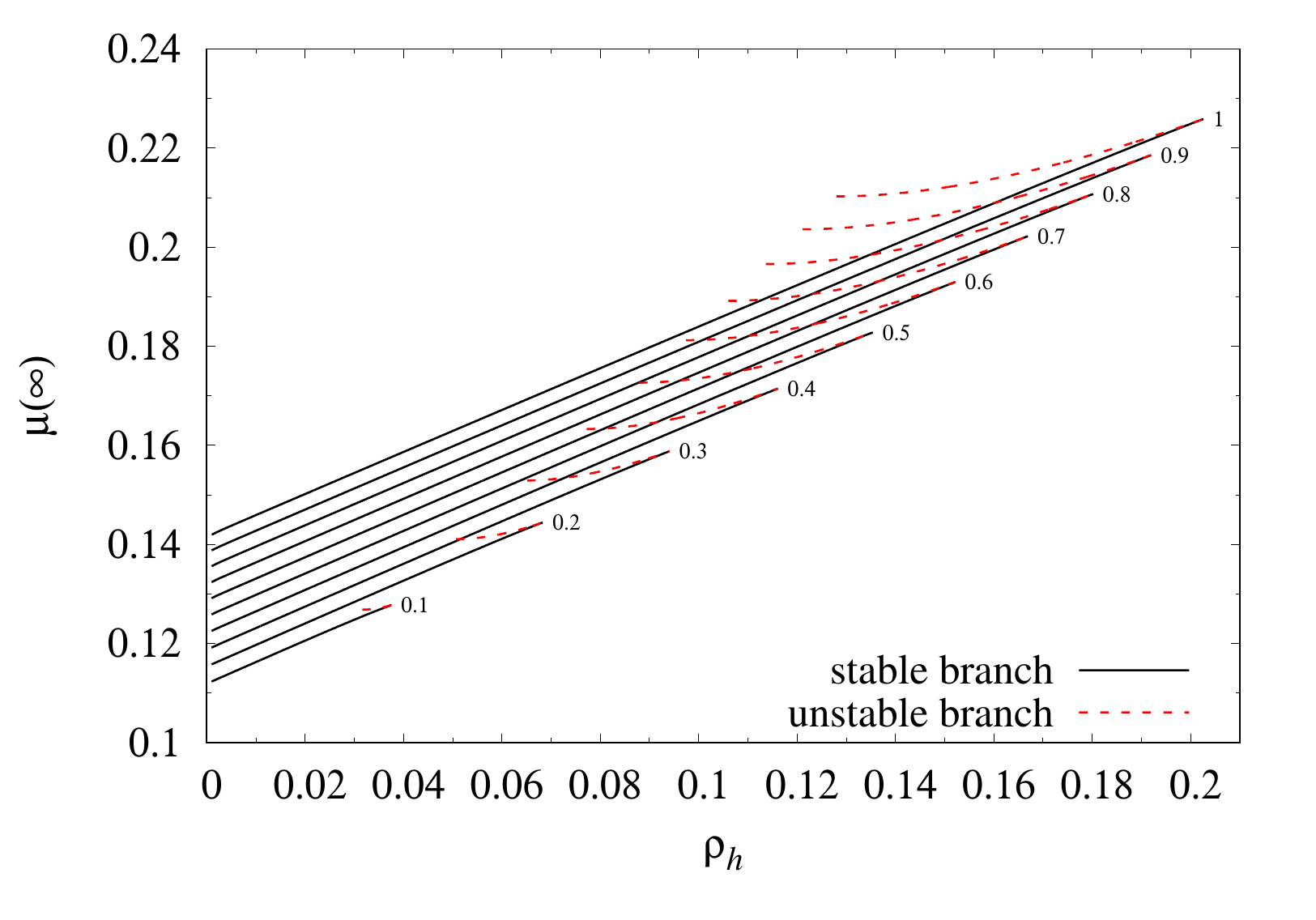}}
\subfloat[]{\includegraphics[width=0.49\linewidth]{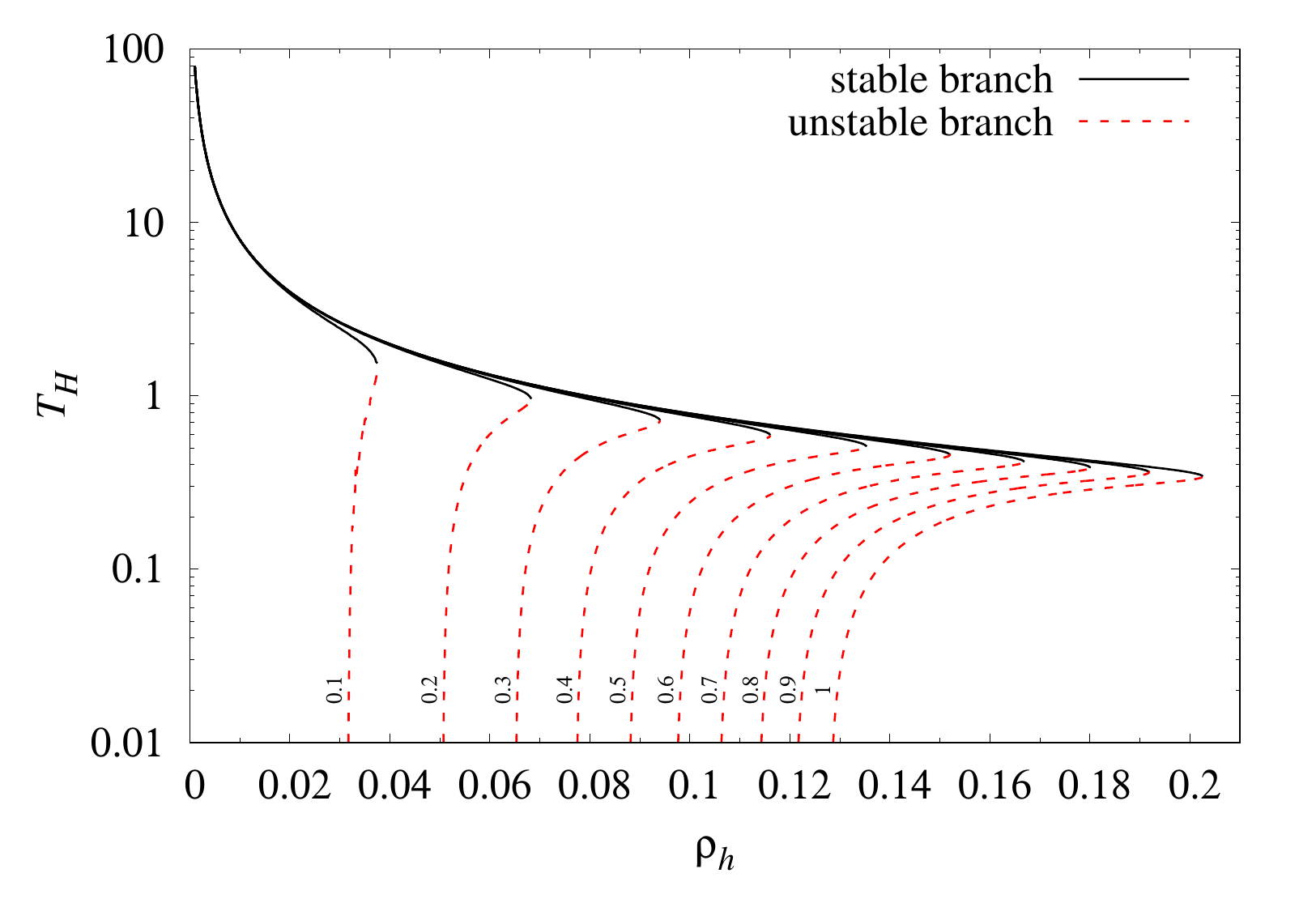}}}
\caption{Stable (black solid lines) and unstable (red dashed lines)
  branches of solutions in the $2+4+8$ model with $\gamma=0$: (a) the
  value of the profile function at the horizon, $f_h$; (b) the
  derivative of the profile function at the horizon, $f_\rho(\rho_h)$;
  (c) the ADM mass, $\mu(\infty)$; (d) the Hawking temperature $T_H$,
  all as functions of the size of the black hole, i.e.~the horizon
  radius, $\rho_h$.
  The numbers on the figures indicate the different values of
  $\beta=0.1,0.2,\ldots,1$.
}
\label{fig:m248_rh_gamma0}
\end{center}
\end{figure}

\begin{figure}[!thp]
\begin{center}
\mbox{\subfloat[]{\includegraphics[width=0.49\linewidth]{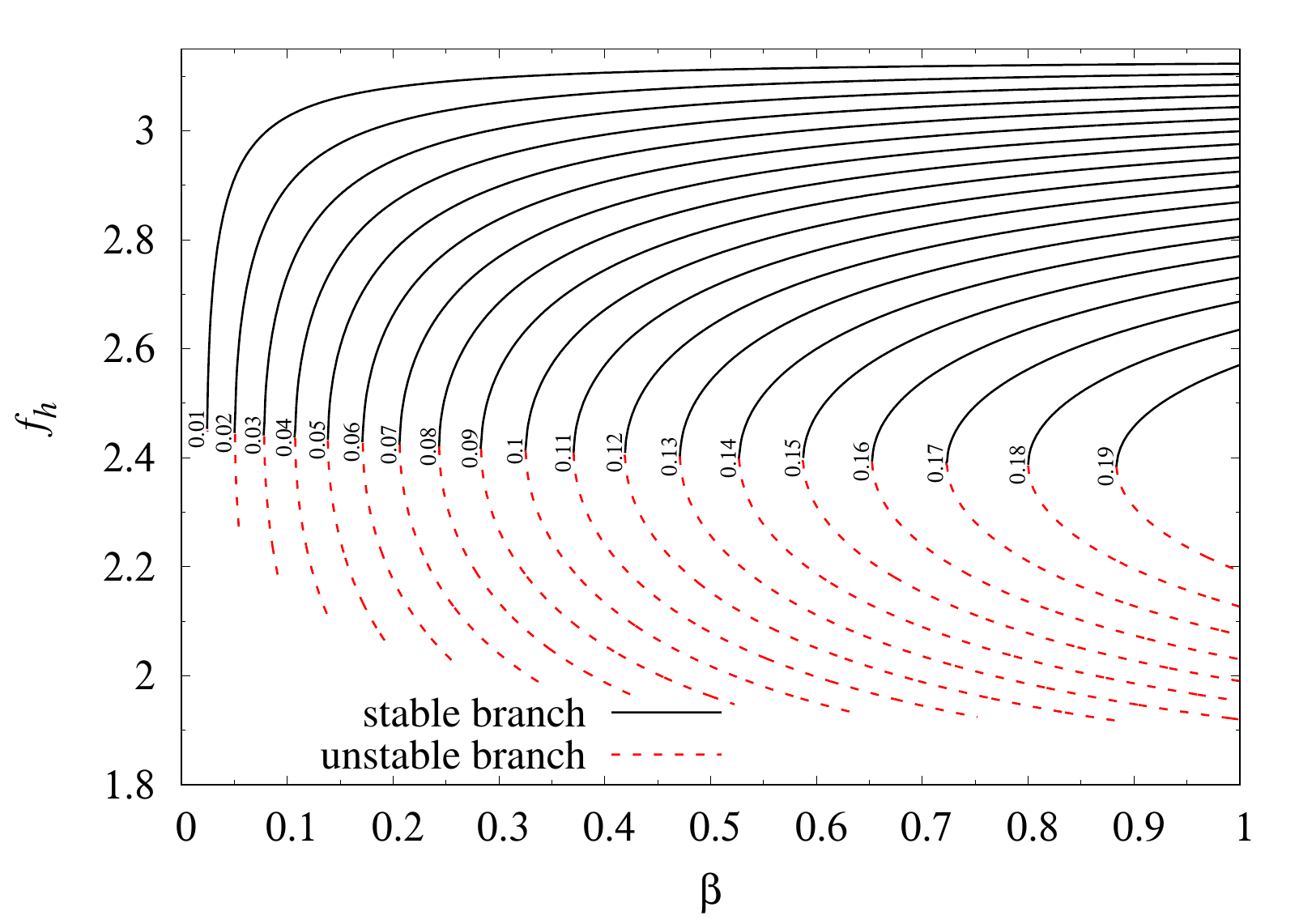}}
\subfloat[]{\includegraphics[width=0.49\linewidth]{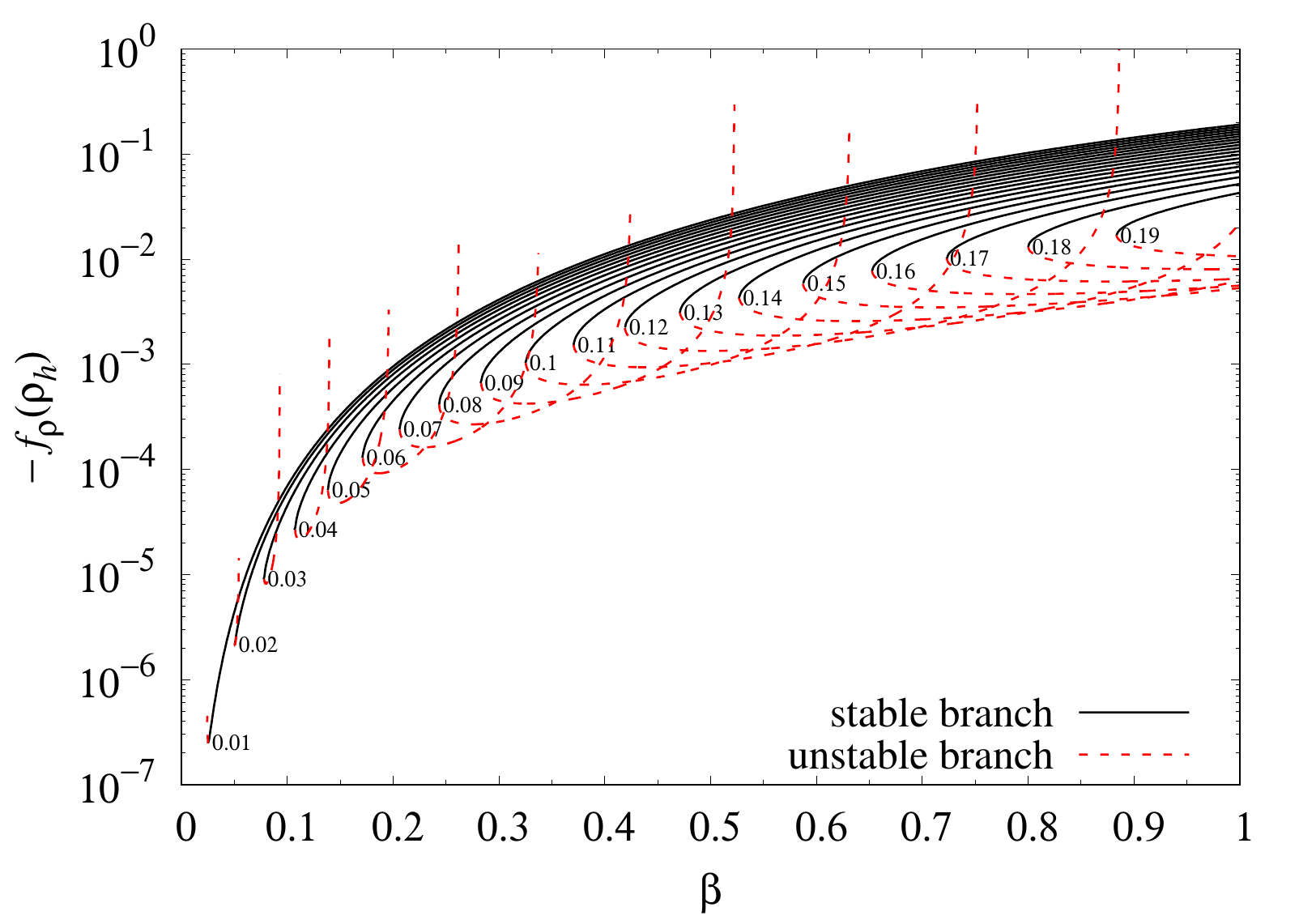}}}
\mbox{\subfloat[]{\includegraphics[width=0.49\linewidth]{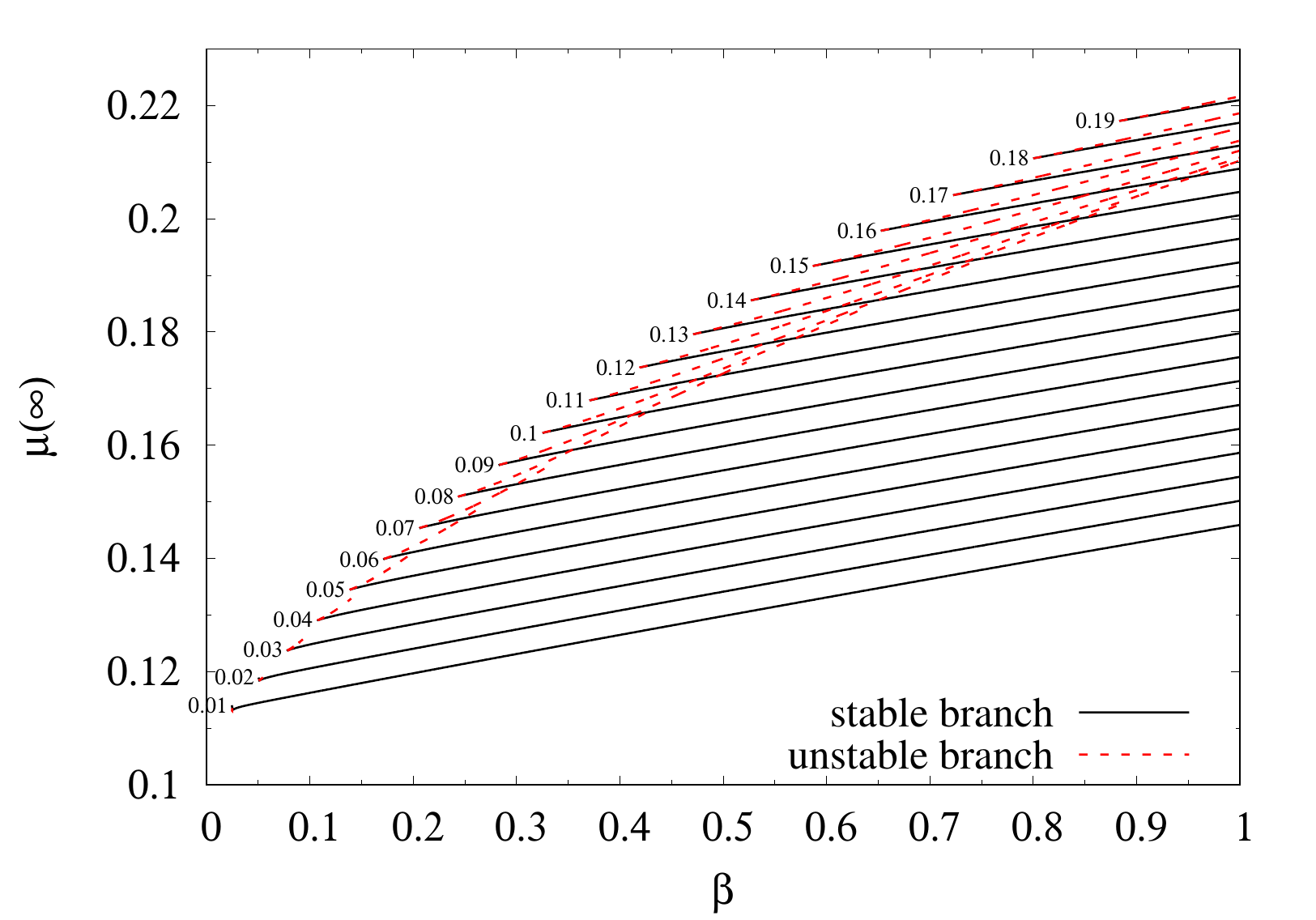}}
\subfloat[]{\includegraphics[width=0.49\linewidth]{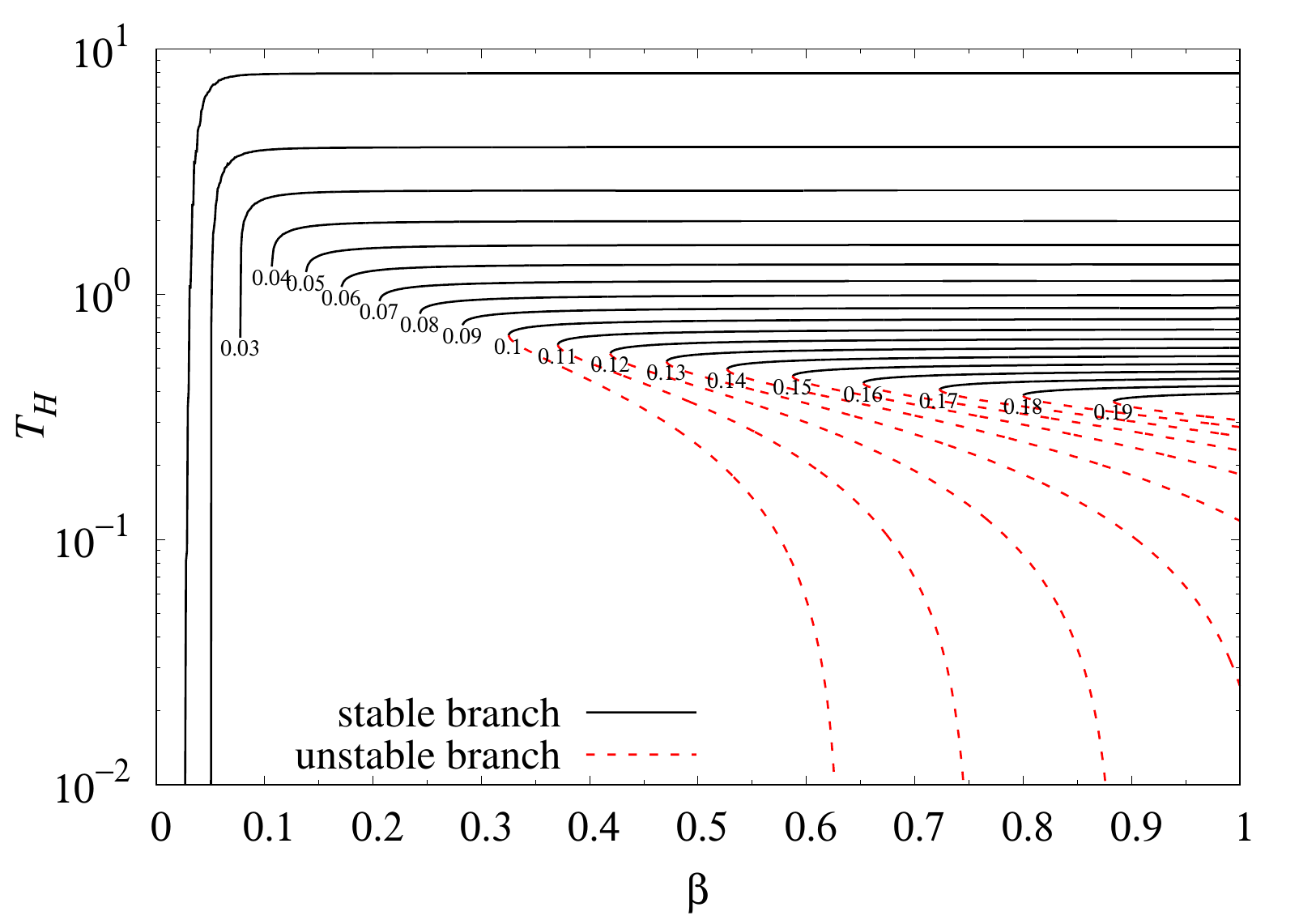}}}
\caption{Stable (black solid lines) and unstable (red dashed lines)
  branches of solutions in the $2+4+8$ model with $\gamma=0$: (a) the
  value of the profile function at the horizon, $f_h$; (b) the
  derivative of the profile function at the horizon, $f_\rho(\rho_h)$;
  (c) the ADM mass, $\mu(\infty)$; (d) the Hawking temperature $T_H$,
  all as functions of the Skyrme-term coefficient, $\beta$.
  The numbers on the figures indicate the different values of
  $\rho_h=0.01,0.02,\ldots,0.19$.
}
\label{fig:m248_beta_gamma0}
\end{center}
\end{figure}

\begin{figure}[!thp]
\begin{center}
\mbox{\subfloat[]{\includegraphics[width=0.49\linewidth]{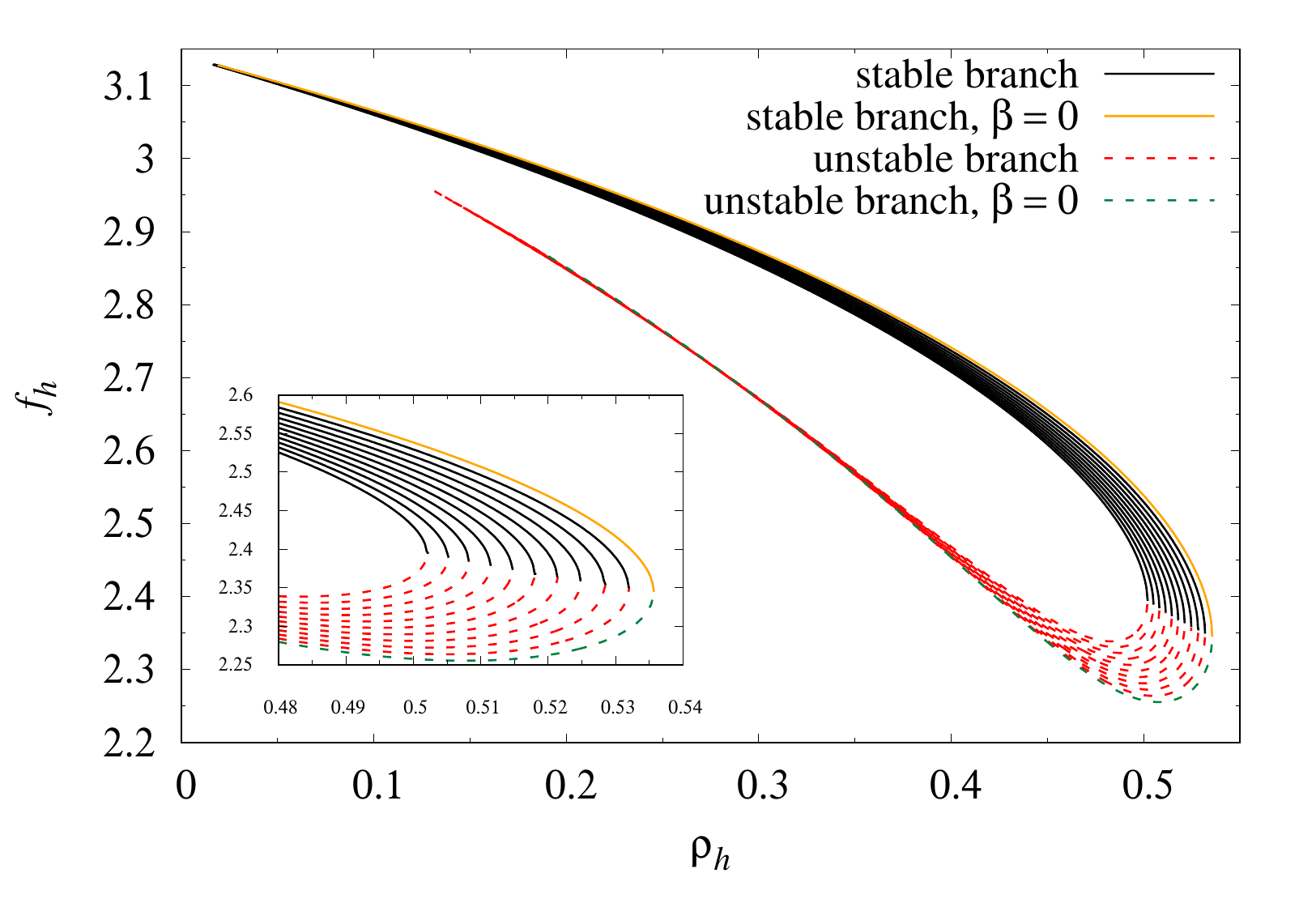}}
\subfloat[]{\includegraphics[width=0.49\linewidth]{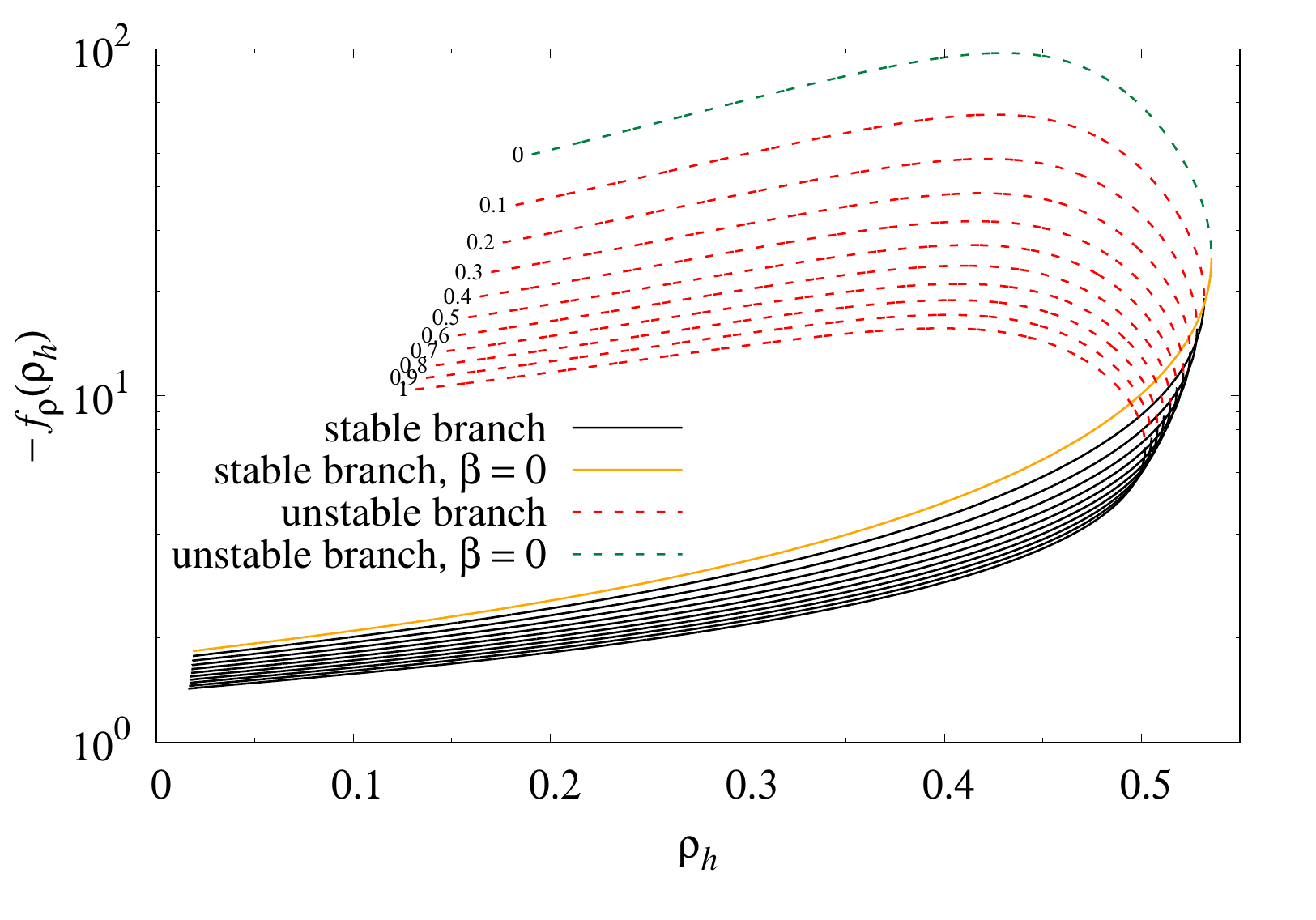}}}
\mbox{\subfloat[]{\includegraphics[width=0.49\linewidth]{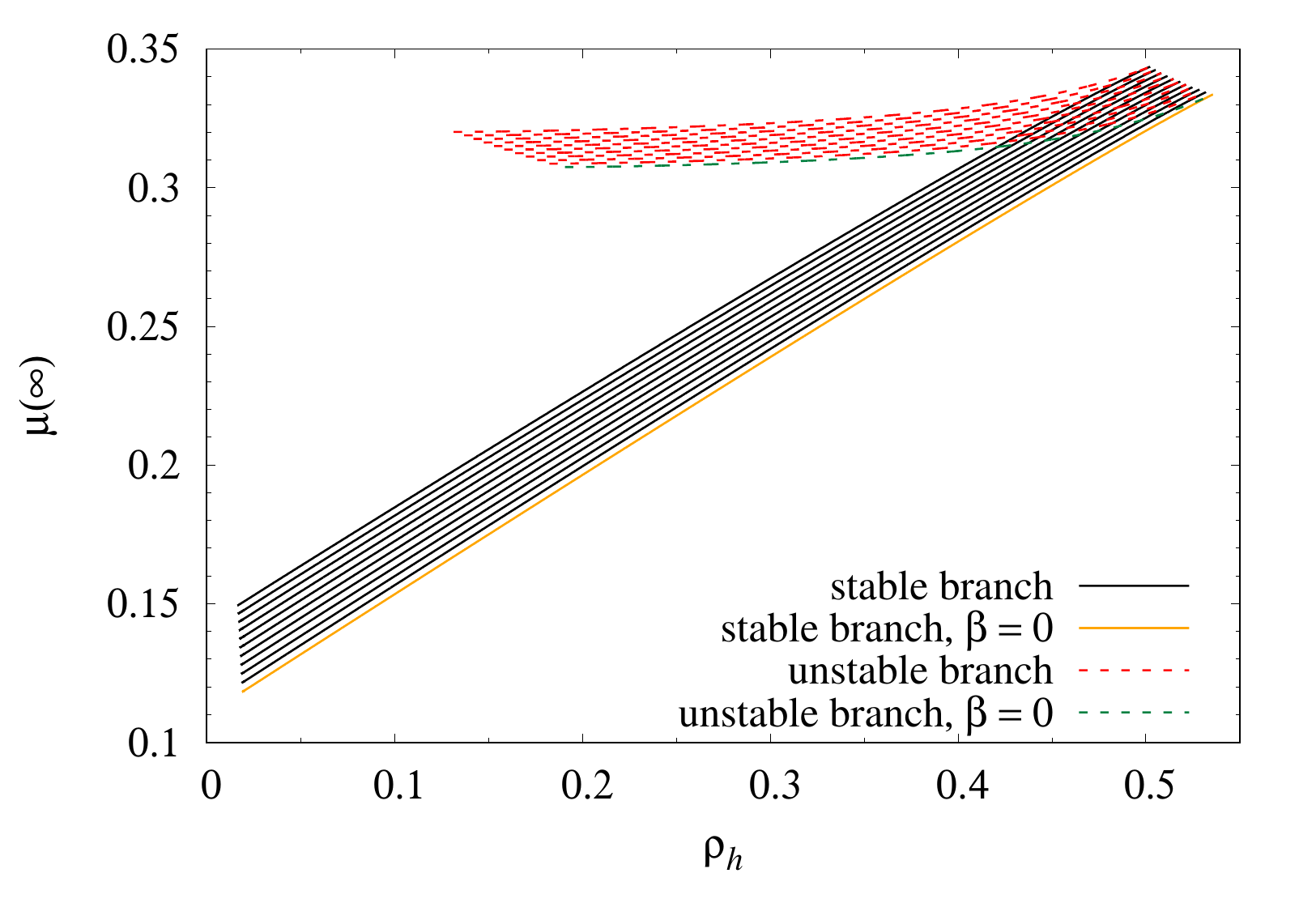}}
\subfloat[]{\includegraphics[width=0.49\linewidth]{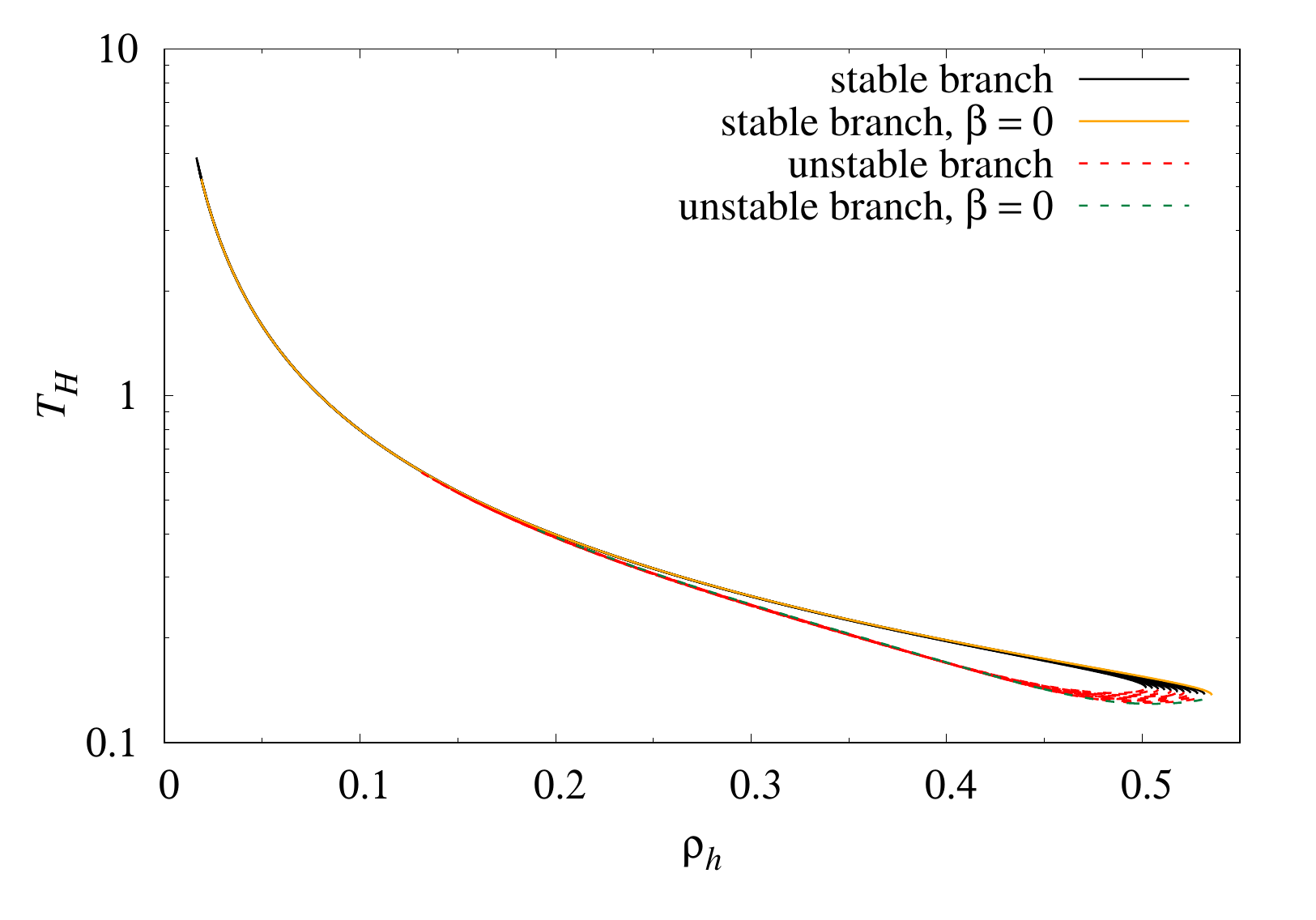}}}
\caption{Stable (black solid lines) and unstable (red dashed lines)
  branches of solutions in the $2+4+8$ model with $\gamma=1$: (a) the
  value of the profile function at the horizon, $f_h$; (b) the
  derivative of the profile function at the horizon, $f_\rho(\rho_h)$;
  (c) the ADM mass, $\mu(\infty)$; (d) the Hawking temperature $T_H$,
  all as functions of the size of the black hole, i.e.~the horizon
  radius, $\rho_h$.
  The numbers in (b) indicate the different values of
  $\beta=0,0.1,0.2,\ldots,1$.
  In (a) $\beta=1$ for the innermost branch (both stable and
  unstable), and decreases to $\beta=0$ for the outermost branch.
  In (c) the topmost branches (both stable and unstable) correspond to
  $\beta=1$ and the branches move downward as $\beta\to 0$. 
  The stable and unstable $\beta=0$ branches are shown in orange and
  dark green colors, respectively. 
}
\label{fig:m248_rh_gamma1}
\end{center}
\end{figure}

\begin{figure}[!thp]
\begin{center}
\mbox{\subfloat[]{\includegraphics[width=0.49\linewidth]{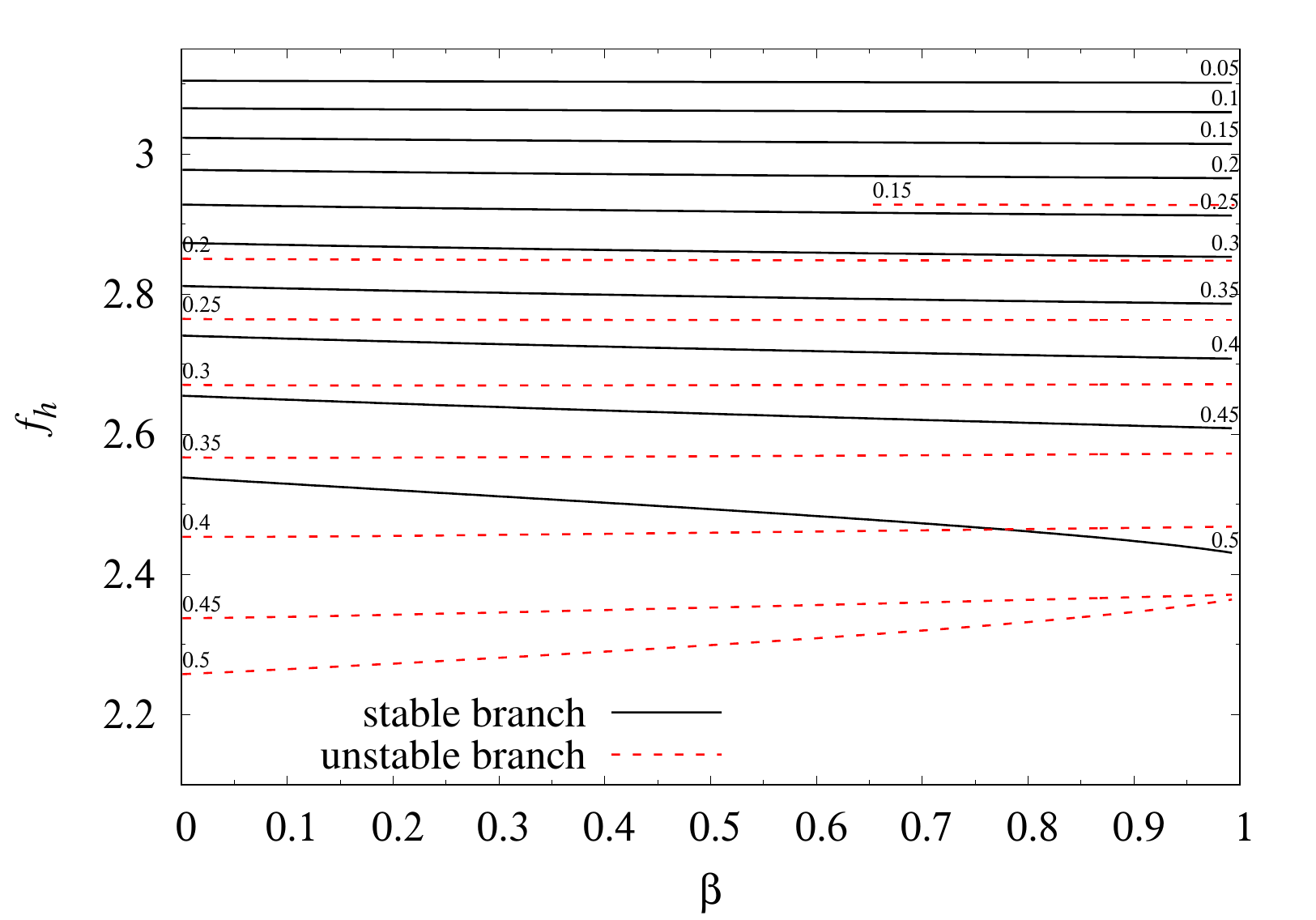}}
\subfloat[]{\includegraphics[width=0.49\linewidth]{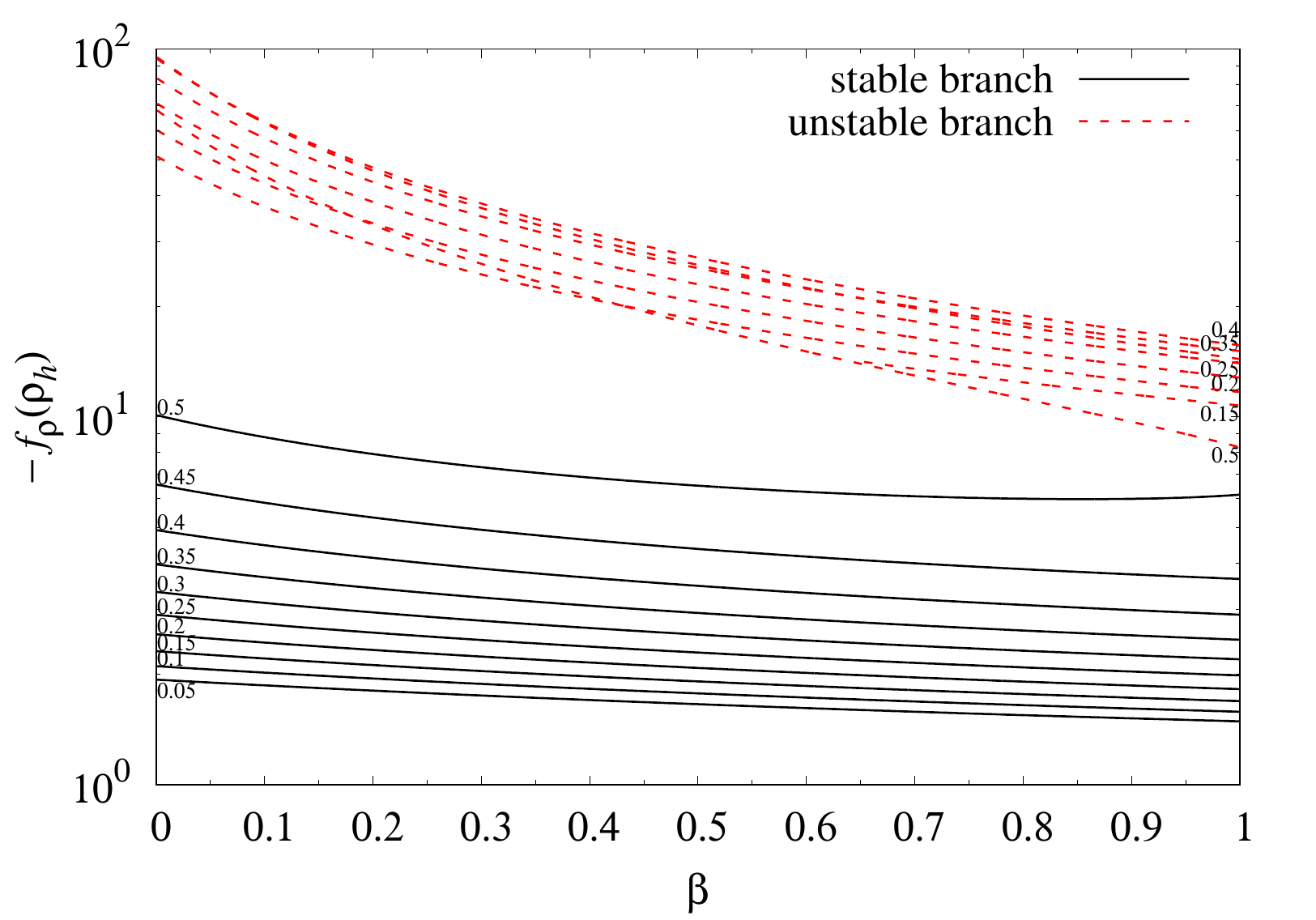}}}
\mbox{\subfloat[]{\includegraphics[width=0.49\linewidth]{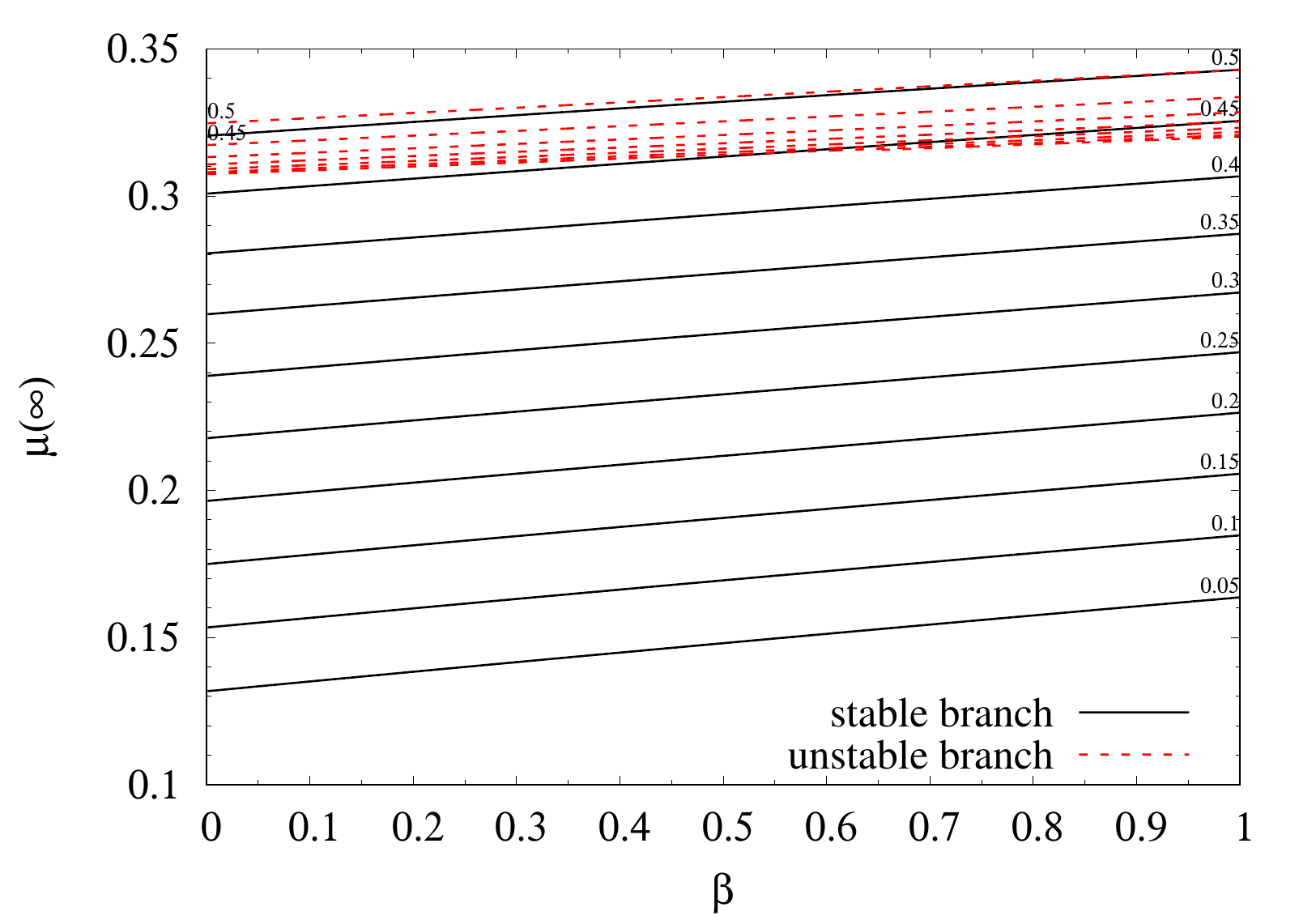}}
\subfloat[]{\includegraphics[width=0.49\linewidth]{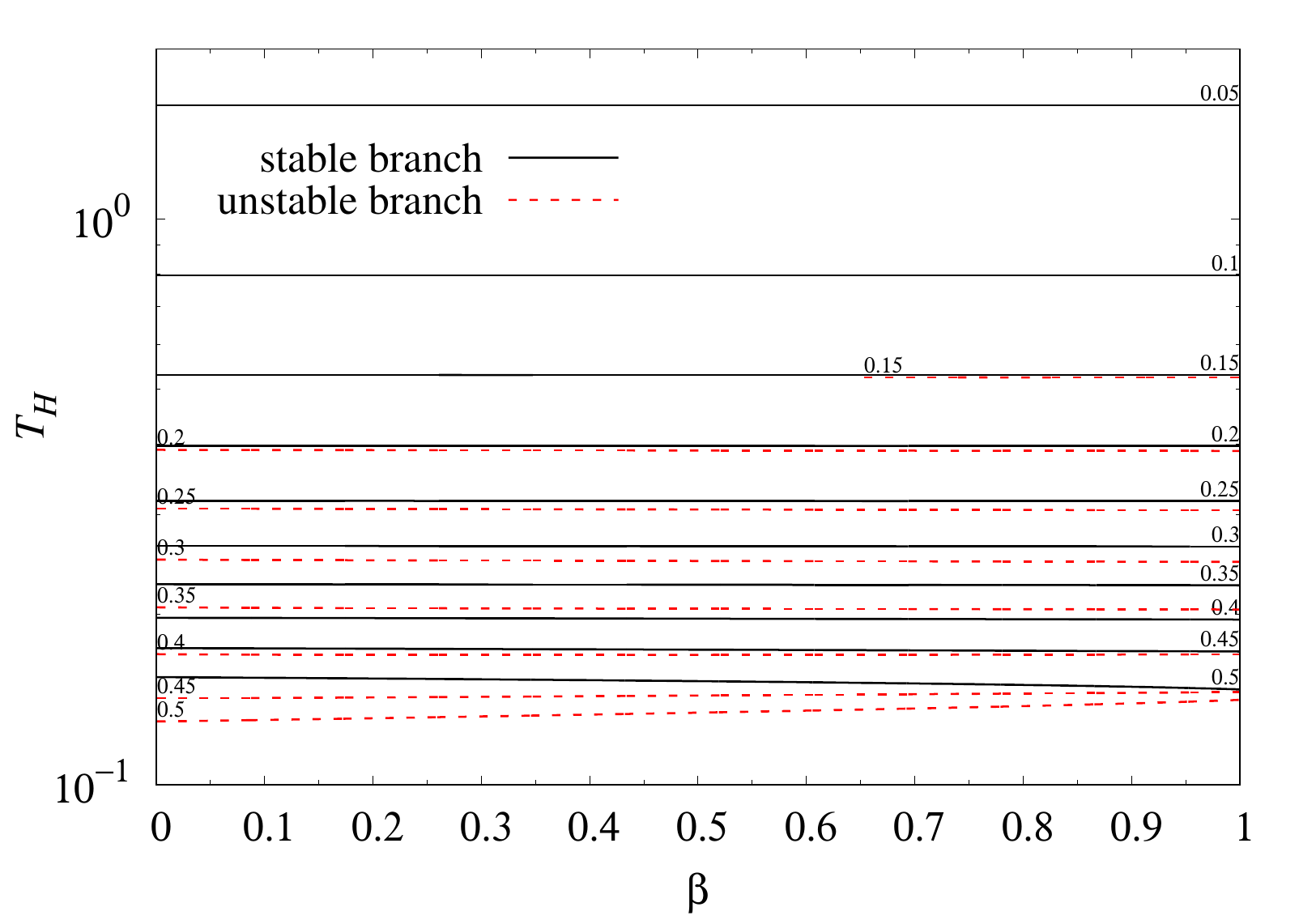}}}
\caption{Stable (black solid lines) and unstable (red dashed lines)
  branches of solutions in the $2+4+8$ model with $\gamma=1$: (a) the
  value of the profile function at the horizon, $f_h$; (b) the
  derivative of the profile function at the horizon, $f_\rho(\rho_h)$;
  (c) the ADM mass, $\mu(\infty)$; (d) the Hawking temperature $T_H$,
  all as functions of the Skyrme-term coefficient, $\beta$.
  The numbers on the figures indicate the different values of
  $\rho_h=0.05,0.1,\ldots,0.5$.
}
\label{fig:m248_beta_gamma1}
\end{center}
\end{figure}

\begin{figure}[!thp]
\begin{center}
\mbox{\subfloat[]{\includegraphics[width=0.49\linewidth]{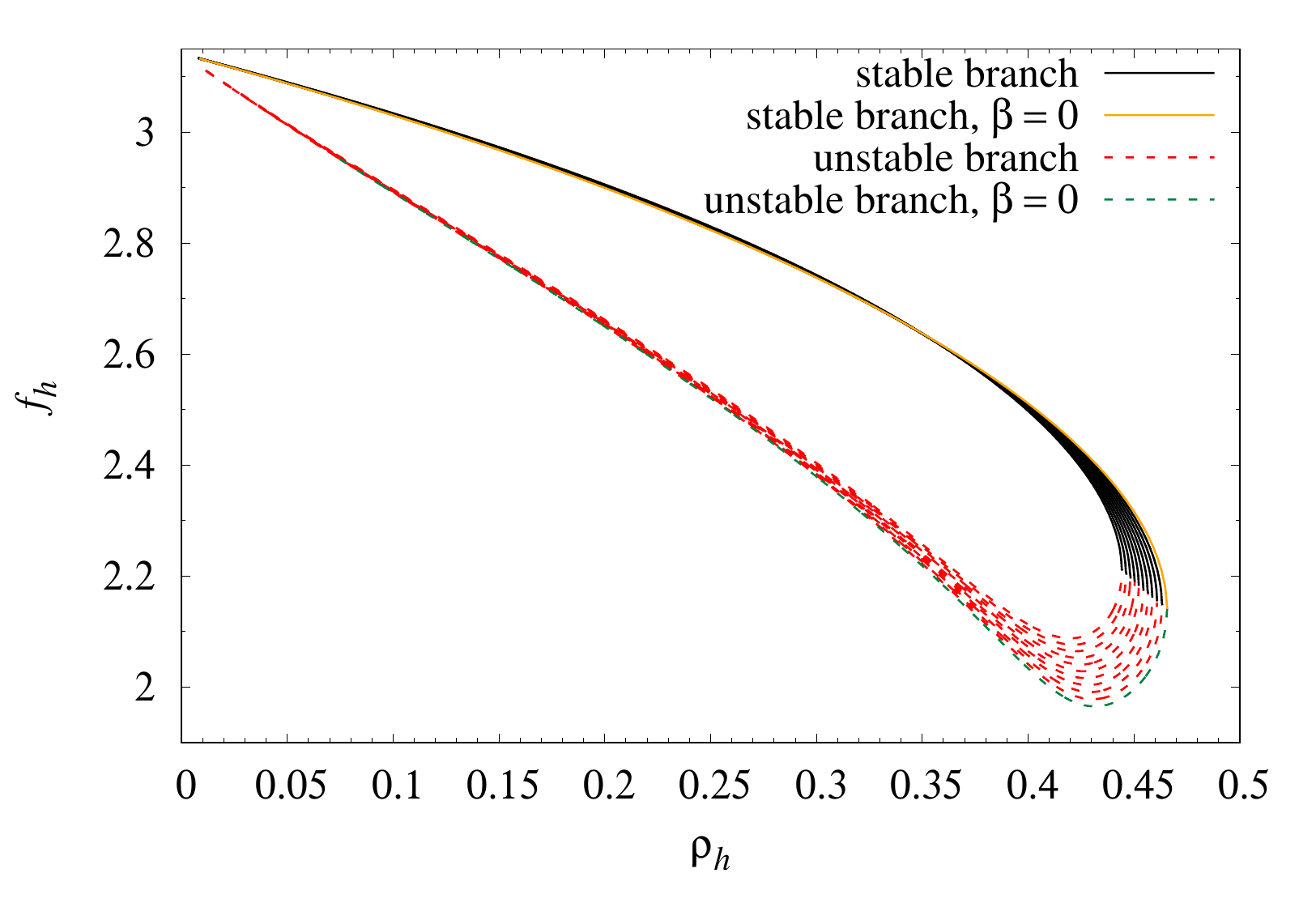}}
\subfloat[]{\includegraphics[width=0.49\linewidth]{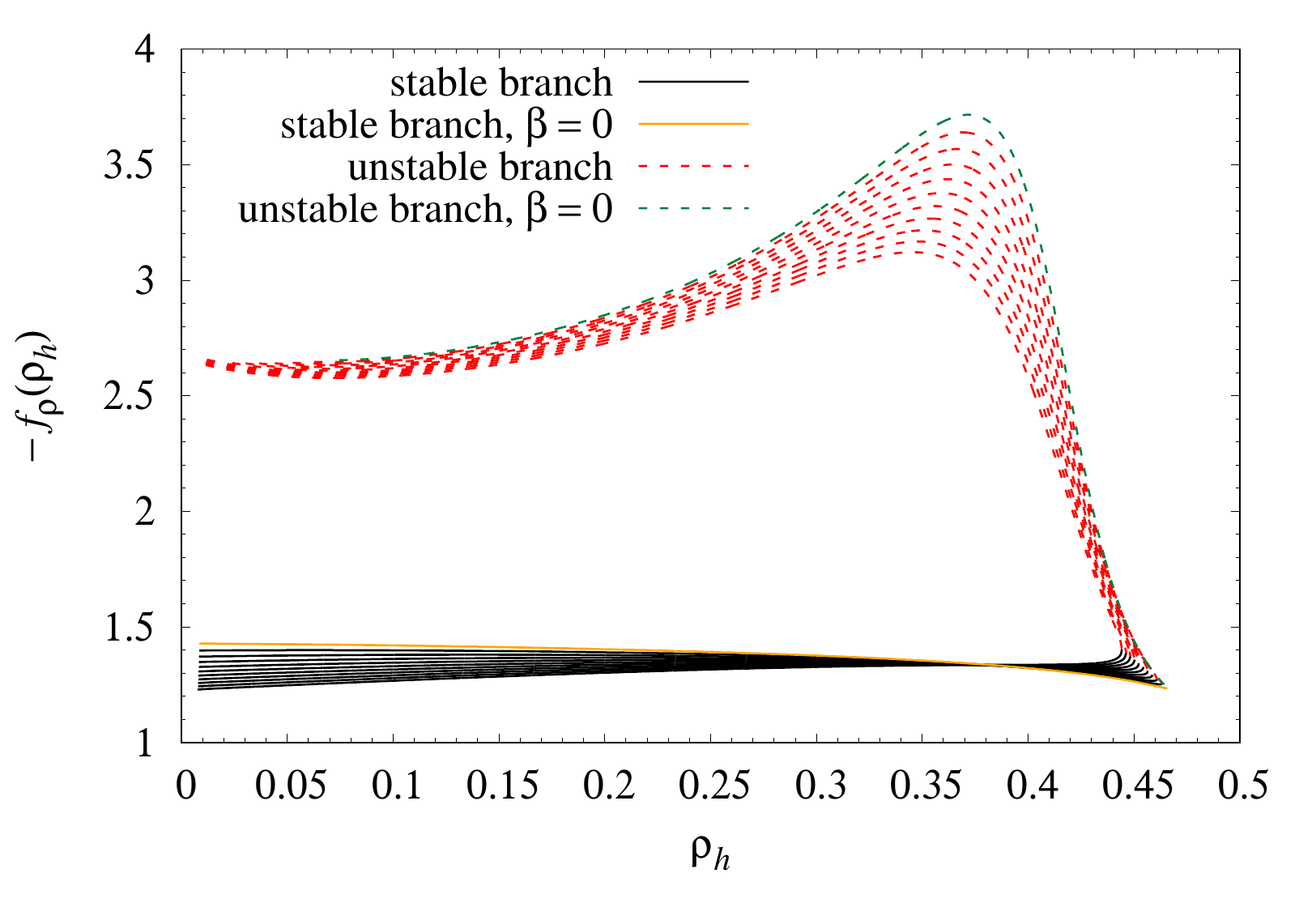}}}
\mbox{\subfloat[]{\includegraphics[width=0.49\linewidth]{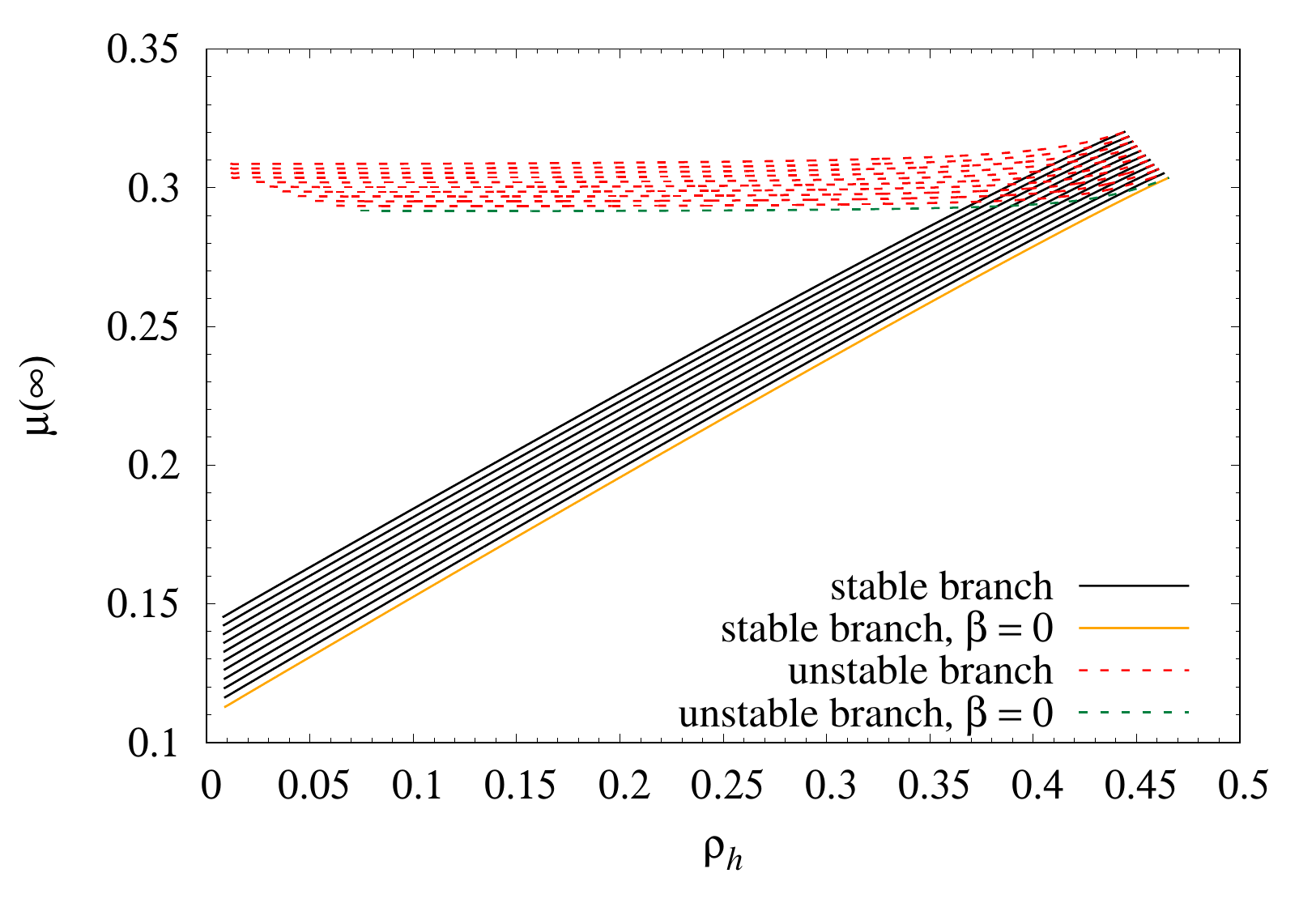}}
\subfloat[]{\includegraphics[width=0.49\linewidth]{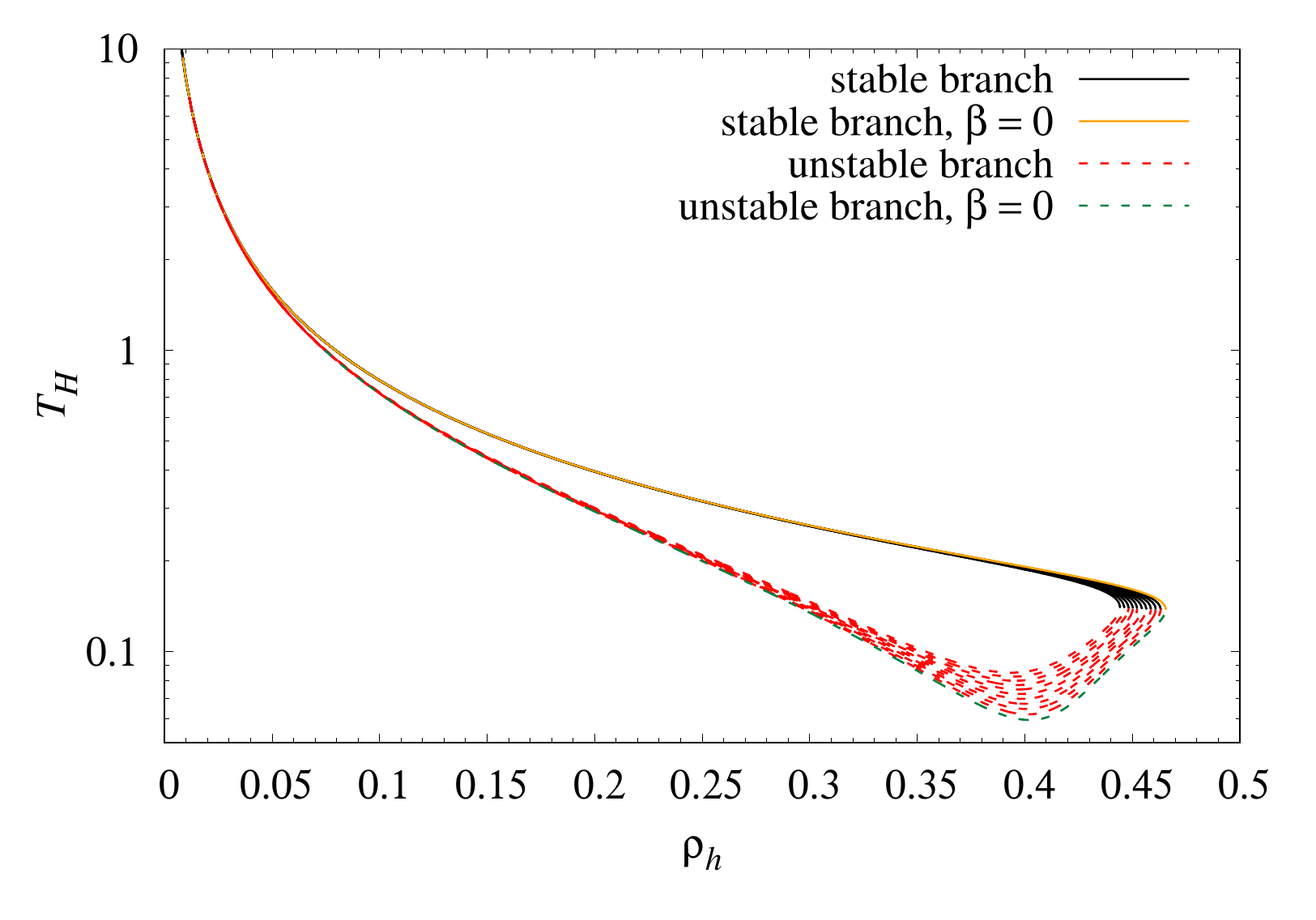}}}
\caption{Stable (black solid lines) and unstable (red dashed lines)
  branches of solutions in the $2+4+8$ model with
  $\gamma=\frac{1}{3}$: (a) the value of the profile function at the
  horizon, $f_h$; (b) the derivative of the profile function at the
  horizon, $f_\rho(\rho_h)$; (c) the ADM mass, $\mu(\infty)$; (d) the
  Hawking temperature $T_H$, all as functions of the size of the black
  hole, i.e.~the horizon radius, $\rho_h$.
  In (a) and (d) $\beta=1$ for the innermost branch (both stable and
  unstable), and decreases to $\beta=0$ for the outermost branch.
  In (b) $\beta=1$ starts as the lowest branch and crosses over to the
  higher branch as $\rho_h$ is increased, for the stable branch.
  For the unstable branch, $\beta=1$ corresponds to the lowest
  branch. 
  In (c) the topmost branches (both stable and unstable) correspond to
  $\beta=1$ and the branches move downward as $\beta\to 0$. 
  The stable and unstable $\beta=0$ branches are shown in orange and
  dark green colors, respectively. 
}
\label{fig:m248_rh_gamma1third}
\end{center}
\end{figure}

We will start with the case of $\gamma=0$.
The numerical results for this case are shown in
Fig.~\ref{fig:m248_rh_gamma0} and Fig.~\ref{fig:m248_beta_gamma0}, 
where the standard quantities are shown as functions of the horizon
radius, $\rho_h$, and as functions of the Skyrme-term coefficient,
$\beta$, respectively. 

Qualitatively, everything is very similar to the $2+4+6$ model. That
is, the branches shrink as $\beta$ tends to zero; in particular, the
bifurcation point, which represents the largest possible black hole
possessing BH hair, moves to smaller and smaller values of the horizon
radius, $\rho_h$, as $\beta$ decreases.
Fig.~\ref{fig:m248_rh_gamma0}(c) clearly shows that the unstable
branches have higher ADM masses than their respective stable ones.
Like in the case of the $2+4+6$ model, also in this model, the
unstable branches end before reaching the flat space limit at
$f_h=\pi$ for $\rho_h\to 0$.
Just before the unstable branches end, (minus) the derivative of the
profile function at the horizon, $-f_\rho(\rho_h)$ increases
drastically, see Fig.~\ref{fig:m248_rh_gamma0}(b), and the temperature 
drops drastically, see Fig.~\ref{fig:m248_rh_gamma0}(d).

Fig.~\ref{fig:m248_beta_gamma0} shows the same standard quantities as
functions of $\beta$.
Like in the $2+4+6$ model, it is clear that all branches cease to
exist in the limit of $\beta\to 0$.
In particular, we can see from Fig.~\ref{fig:m248_beta_gamma0}(b) and
(c) that (minus) the derivative of the profile function,
$-f_\rho(\rho_h)$, goes towards zero and ADM mass decreases, for
$\beta\to 0$. 
A difference with respect to the $2+4+6$ model, is that the Hawking
temperature really drops drastically to zero for $\beta\to 0$, see
Fig.~\ref{fig:m248_beta_gamma0}(d). 
More precisely, the stable branch with $\rho_h=0.01$ drops below the
temperatures of the other branches, whereas in the $2+4+6$ model, the
stable branch only drops to about $2$ and then turns into an unstable
branch which decreases to values of about $10^{-1}$ and then
terminates. 
The reason why we are interested in knowing the limiting behavior of 
these observables as functions of $\beta$ in the limit of
$\beta\to 0$, is to better understand the cause of the collapse of the
BH hair.

We will now turn to the case of the $2+4+8$ model with $\gamma=1$.
The behavior of the BH hair in this model is vastly different from the
$2+4+6$ model and the $2+4+8$ model with $\gamma=0$.
In particular, it is the first known Skyrme-type model other than the
standard Skyrme model, which possesses stable BH hair (without the
Skyrme term to stabilize it).
This can easily be seen from Fig.~\ref{fig:m248_rh_gamma1}(a) as the
branches do not collapse to small horizon radii, $\rho_h$, as $\beta$
is turned off.
In fact, a peculiarity of the model, is that the BH hair can support
larger BHs \emph{without} the Skyrme term than with it.
This can be seen in Fig.~\ref{fig:m248_rh_gamma1}(a) as the branches
are extending to larger $\rho_h$ for smaller $\beta$.
The largest branch corresponds to $\beta=0$. 
We can confirm for all $\beta$, that the unstable branches have larger
ADM masses than their corresponding stable ones, see
Fig.~\ref{fig:m248_rh_gamma1}(c).
The stable and unstable branches meet in all figures at the
bifurcation point, which also corresponds to the largest possible BH
that can support the BH hair for the given parameters.
We can also see that the derivatives of the profile function are
steeper for the unstable branches than for their corresponding stable
ones, see Fig.~\ref{fig:m248_rh_gamma1}(b).
It is observed from Fig.~\ref{fig:m248_rh_gamma1}(d) that the
temperature dependence on $\beta$ is quite mild and that the unstable
branches have lower temperatures than their corresponding stable
ones. 

For completeness, we will present the standard quantities as functions 
of $\beta$ as well, see Fig.~\ref{fig:m248_beta_gamma1}, in order to
show clearly the $\beta\to 0$ limit.
The detailed information in the figure is not so interesting, except
for the fact that all stable branches have well-defined $\beta\to 0$
limits.
The unstable branches survive the $\beta\to 0$ limit for
$\rho_h\gtrsim 0.2$.
We can confirm that all unstable branches have higher ADM masses than
their corresponding stable ones, see
Fig.~\ref{fig:m248_beta_gamma1}(c).
The unstable branches also have steeper derivatives of the profile
function and lower temperatures than the corresponding stable ones,
see Figs.~\ref{fig:m248_beta_gamma1}(b) and (d), respectively.

Finally, we will consider the last case of $\gamma=\frac{1}{3}$ which
corresponds to the Skyrme term squared.
The standard quantities are shown in
Fig.~\ref{fig:m248_rh_gamma1third} as functions of the horizon radius,
$\rho_h$.
Qualitatively, the behavior of this model is very similar to that of
$\gamma=1$.
Both cases yield a stable BH hair in the limit of vanishing $\beta$;
that is, the Skyrme term is not needed for stabilization of the
solitonic hair.
One difference with respect to the $\gamma=1$ case, is that the
unstable branches reach down to smaller BH sizes (smaller $\rho_h$).
The branch with $\beta=0$ -- which means without the Skyrme term -- is
the largest branch, see Fig.~\ref{fig:m248_rh_gamma1third}(a) and also
the branch with the smallest ADM mass, see
Fig.~\ref{fig:m248_rh_gamma1third}(c). 
As this model (with $\gamma=\frac{1}{3}$) is stable in the
$\beta\to 0$ limit, we will not show the corresponding figure with the
standard quantities as functions of $\beta$.
As can be seen from Fig.~\ref{fig:m248_rh_gamma1third}, the $\beta$
dependence is mild.

\subsection{The \texorpdfstring{$2+4+10$}{2+4+10} model}

The model we consider here is made of the standard Skyrme model with
an added higher-order derivative term that consists of the Skyrme term
multiplied by the BPS-Skyrme term. 
This model is simpler than the $2+4+8$ model as it only has one
overall coefficient $c_{10}$ which we scale away in the units, see
Eq.~\eqref{eq:Lrescaled}.

Completing the Einstein
equations \eqref{eq:EEQ1_242n}-\eqref{eq:EEQ2_242n}, we have
\begin{align}
-C_\rho + \frac{1}{\rho} - \frac{C}{\rho} &=
2\alpha\bigg[
\rho C f_\rho^2
+\frac{2\sin^2f}{\rho}
+\frac{2\beta\sin^2(f) C f_\rho^2}{\rho}
+\frac{\beta\sin^4f}{\rho^3} \non
&\phantom{=2\alpha\bigg[}
+\frac{\sin^8(f) C f_\rho^2}{\rho^7}
+\frac{2\sin^6(f) C^2 f_\rho^4}{\rho^5}
\bigg],\non
\frac{1}{\alpha}\frac{N_\rho}{N} &=
2\rho f_\rho^2
+\frac{4\beta \sin^2(f)f_\rho^2}{\rho}
+\frac{2\sin^8(f) f_\rho^2}{\rho^7}
+\frac{8\sin^6(f) C f_\rho^4}{\rho^5},
\end{align}
while the equation of motion for the profile function reads
\begin{align}
C f_{\rho\rho}
+\frac{2C f_\rho}{\rho}
+C_\rho f_\rho
+\frac{N_\rho C f_\rho}{N}
-\frac{\sin 2f}{\rho^2} \non
+\frac{2\beta\sin^2f}{\rho^2}\left(
  C f_{\rho\rho}
  +C_\rho f_\rho
  +\frac{N_\rho C f_\rho}{N}
  -\frac{\sin 2f}{2\rho^2}
  \right)
+\frac{\beta\sin(2f) C f_\rho^2}{\rho^2}
\non
+\frac{\sin^8f}{\rho^8}\left(
  C f_{\rho\rho}
  -\frac{6C f_\rho}{\rho}
  +C_\rho f_\rho
  +\frac{N_\rho C f_\rho}{N}
  \right)
+\frac{9\sin^4(f)\sin(2f) C^2 f_\rho^4}{\rho^6} \non
+\frac{2\sin^6(f) C f_\rho^2}{\rho^6}\left(
  6C f_{\rho\rho}
  -\frac{8C f_\rho}{\rho}
  +4C_\rho f_\rho
  +\frac{4N_\rho C f_\rho}{N}
  +\frac{\sin 2f}{\rho^2}
  \right)
 = 0.
\end{align}
The boundary conditions \eqref{eq:C_rho_at_rh}-\eqref{eq:f_rho_at_rh}
can be written as
\begin{align}
\mu &= \frac{\rho_h}{2}
+\alpha\sin^2(f_h)\left(2
  +\frac{\beta\sin^2f_h}{\rho_h^2}\right)(\rho-\rho_h)
+ \mathcal{O}\left((\rho-\rho_h)^2\right),\\
f &= f_h
+\rho_h^7\sin(2f_h)\frac{\rho_h^2 + \beta\sin^2(f_h)}
{\left(\rho_h^8 + 2\beta\rho_h^6\sin^2f_h + \sin^8 f_h\right)\Xi_{10}}
(\rho-\rho_h)
+ \mathcal{O}\left((\rho-\rho_h)^2\right),\\
\Xi_{10} &\equiv
\rho_h^2 - 4\alpha\rho_h^2\sin^2f_h - 2\alpha\beta\sin^4 f_h.
\end{align}

\begin{figure}[!thp]
\begin{center}
\mbox{\subfloat[]{\includegraphics[width=0.49\linewidth]{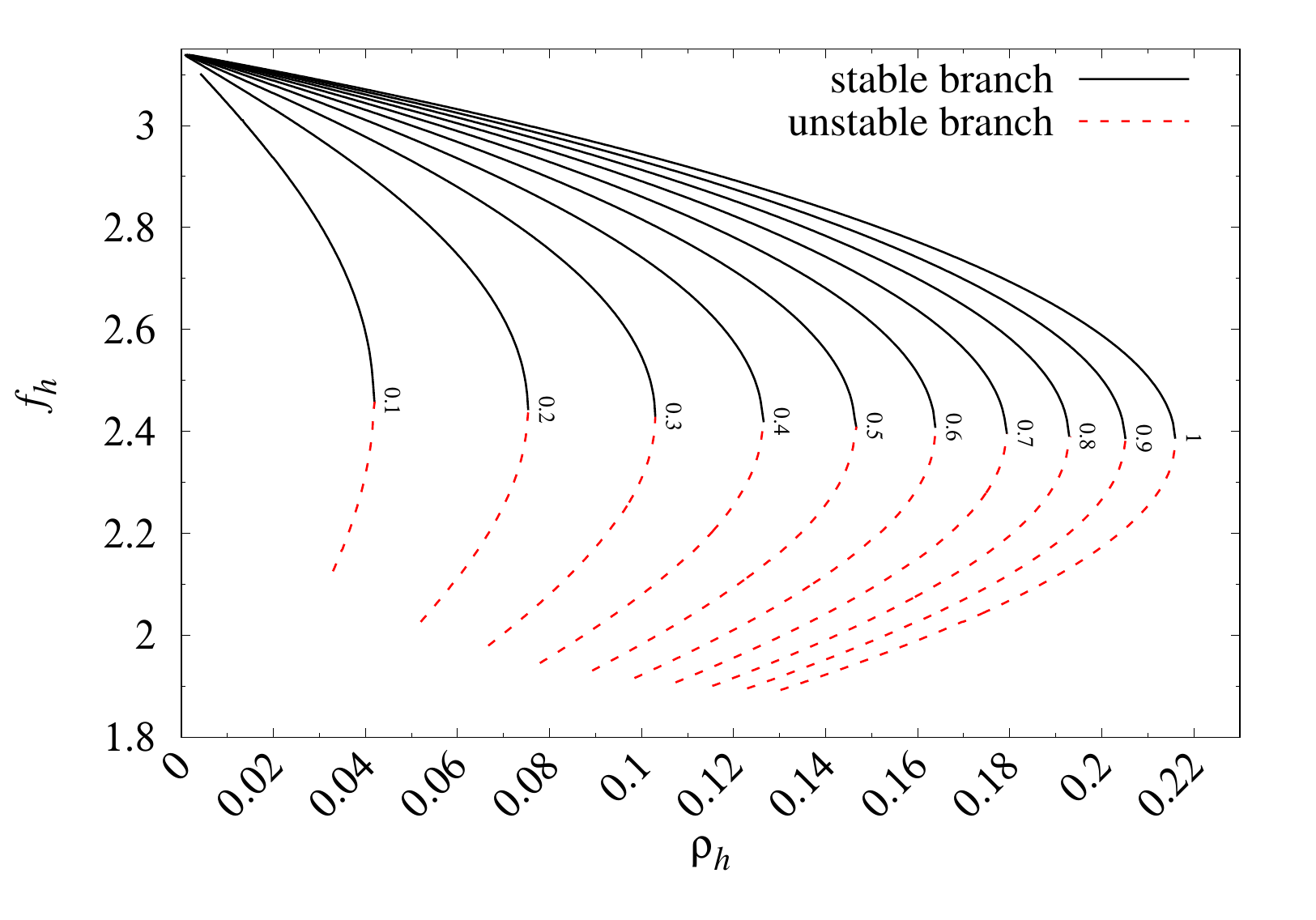}}
\subfloat[]{\includegraphics[width=0.49\linewidth]{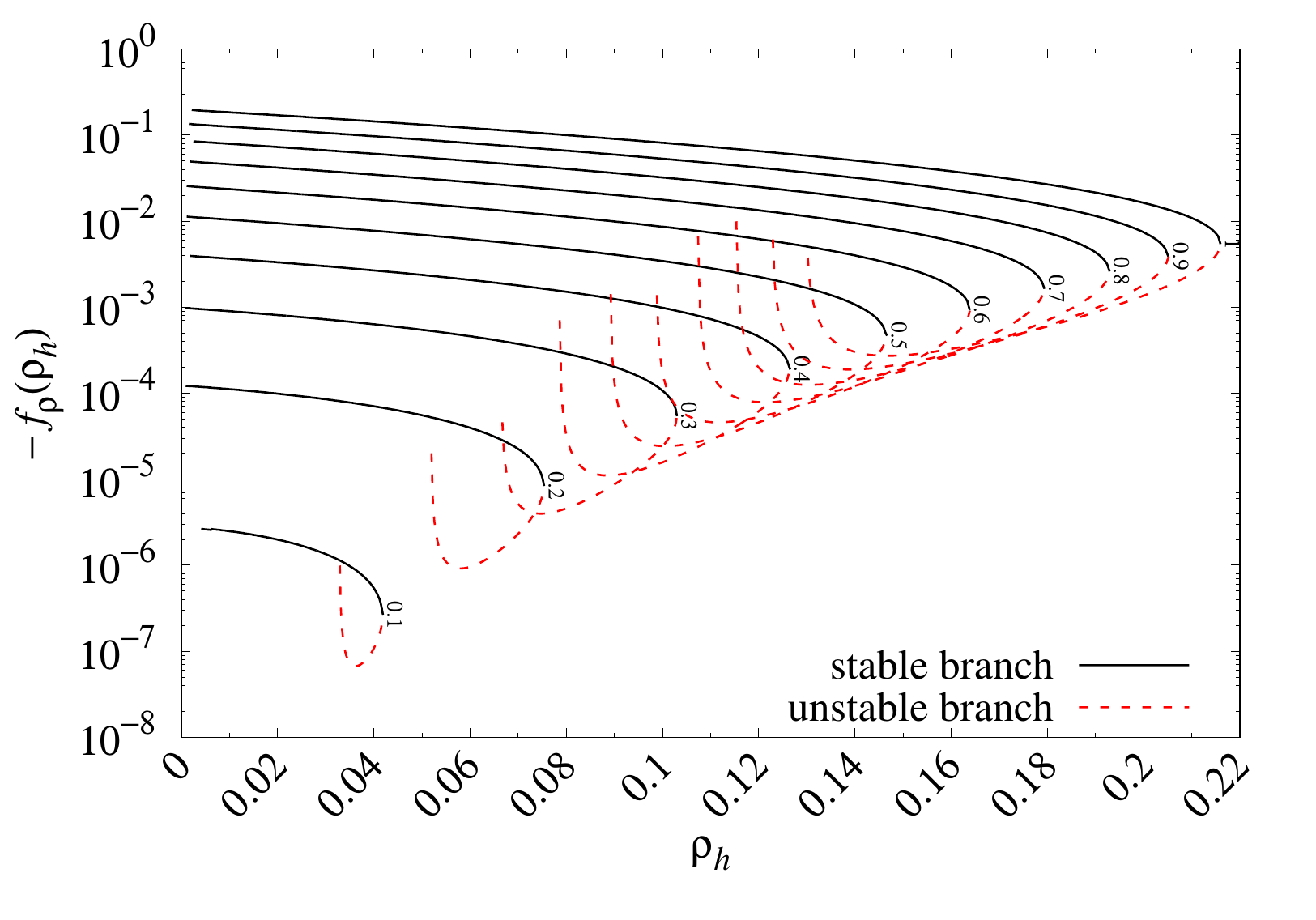}}}
\mbox{\subfloat[]{\includegraphics[width=0.49\linewidth]{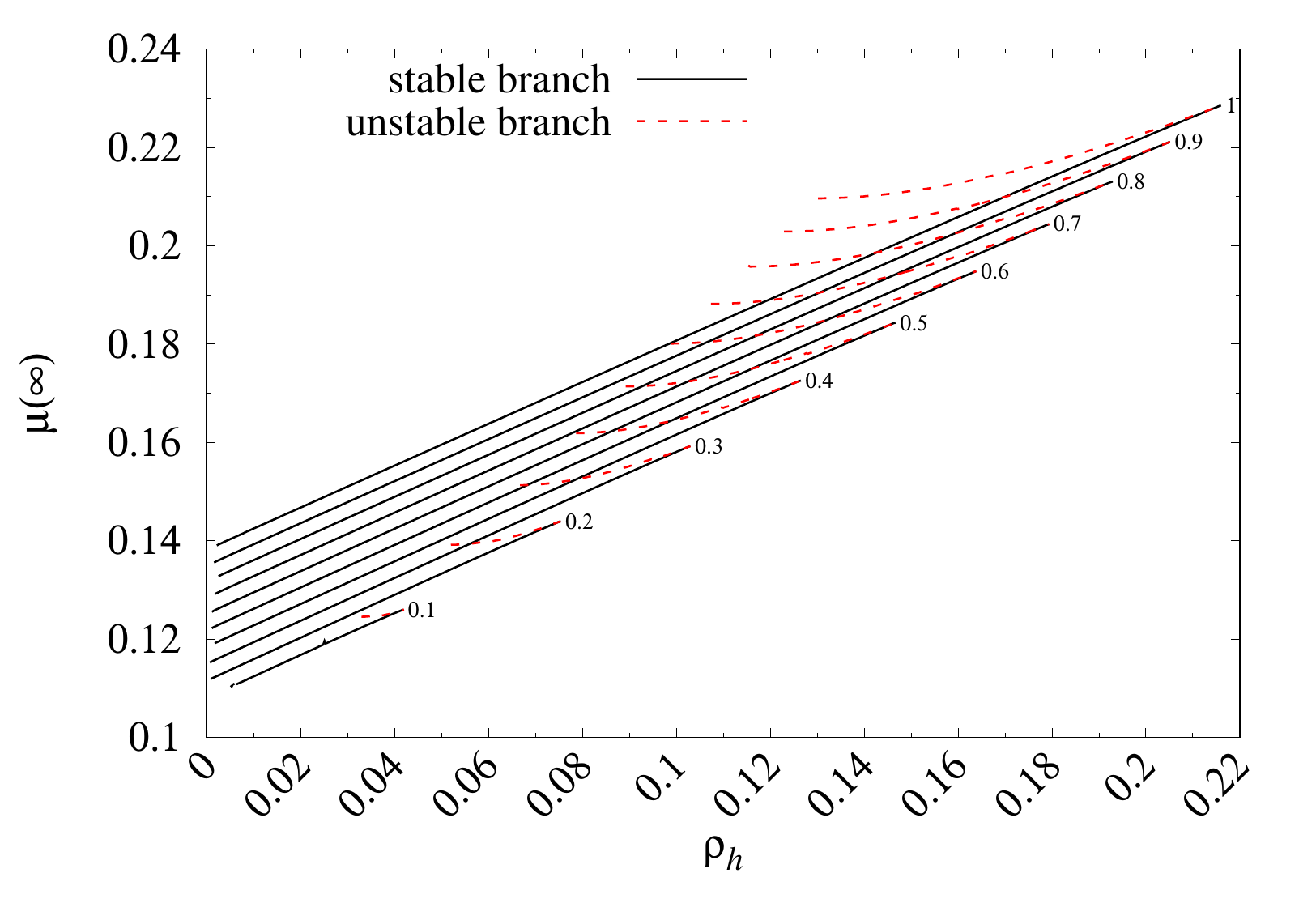}}
\subfloat[]{\includegraphics[width=0.49\linewidth]{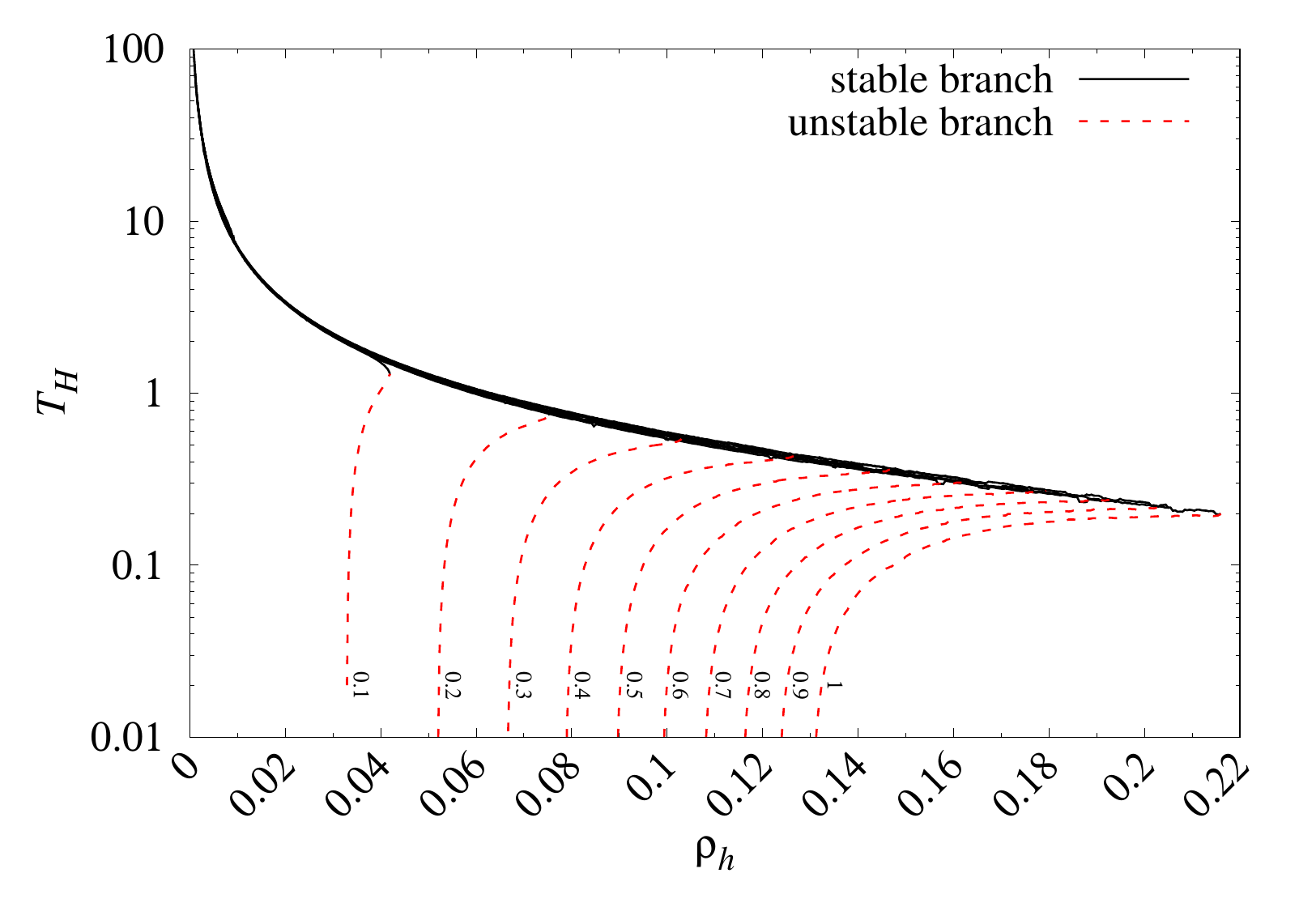}}}
\caption{Stable (black solid lines) and unstable (red dashed lines)
  branches of solutions in the $2+4+10$ model: (a) the value of the
  profile function at the horizon, $f_h$; (b) the derivative of the
  profile function at the horizon, $f_\rho(\rho_h)$; 
  (c) the ADM mass, $\mu(\infty)$;  
  (d) the Hawking temperature $T_H$, all as functions of the size of the
  black hole, i.e.~the horizon radius, $\rho_h$.
  The numbers on the figures indicate the different values of
  $\beta=0.1,0.2,\ldots,1$. }
\label{fig:m2410_rh}
\end{center}
\end{figure}

\begin{figure}[!thp]
\begin{center}
\mbox{\subfloat[]{\includegraphics[width=0.49\linewidth]{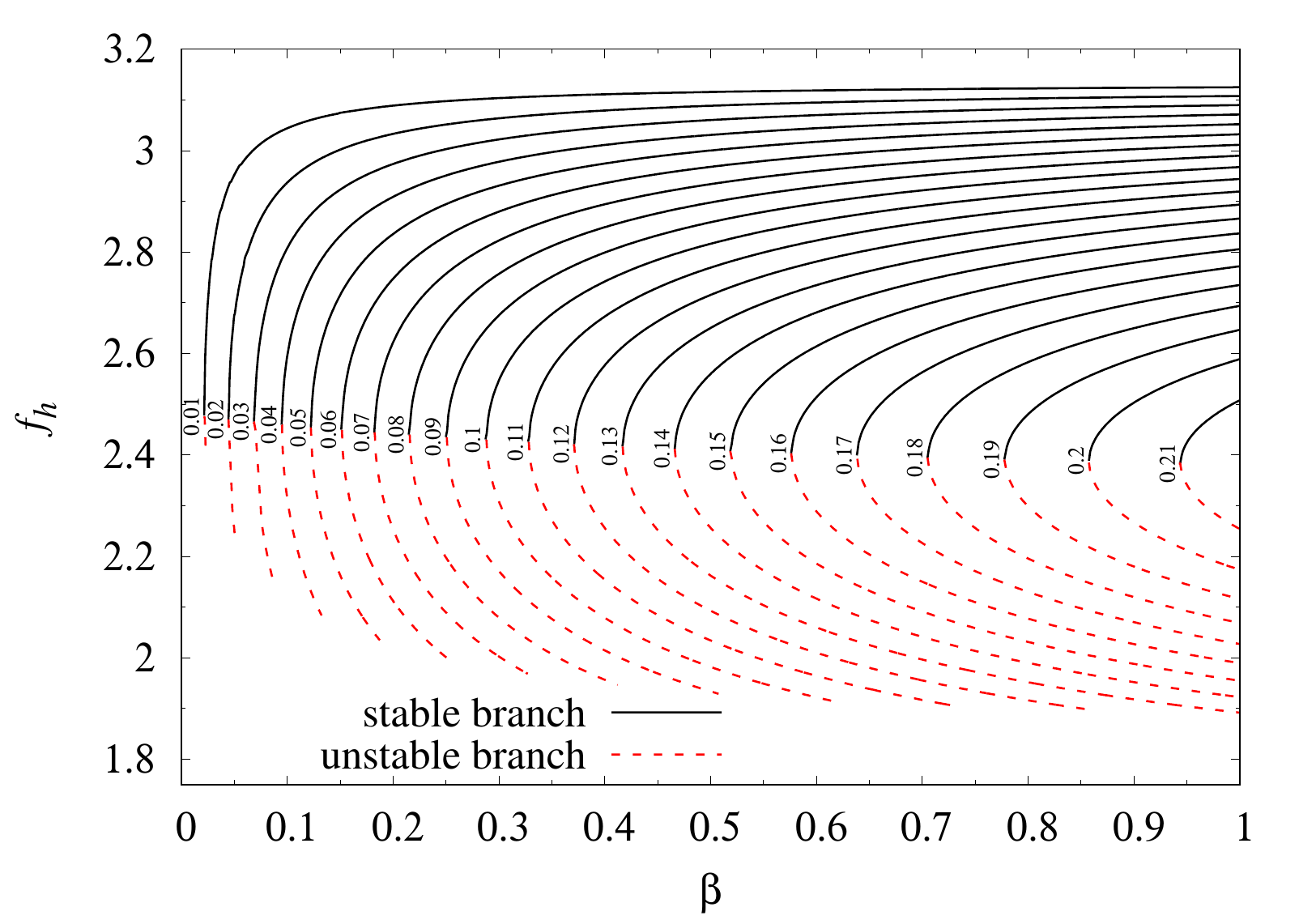}}
\subfloat[]{\includegraphics[width=0.49\linewidth]{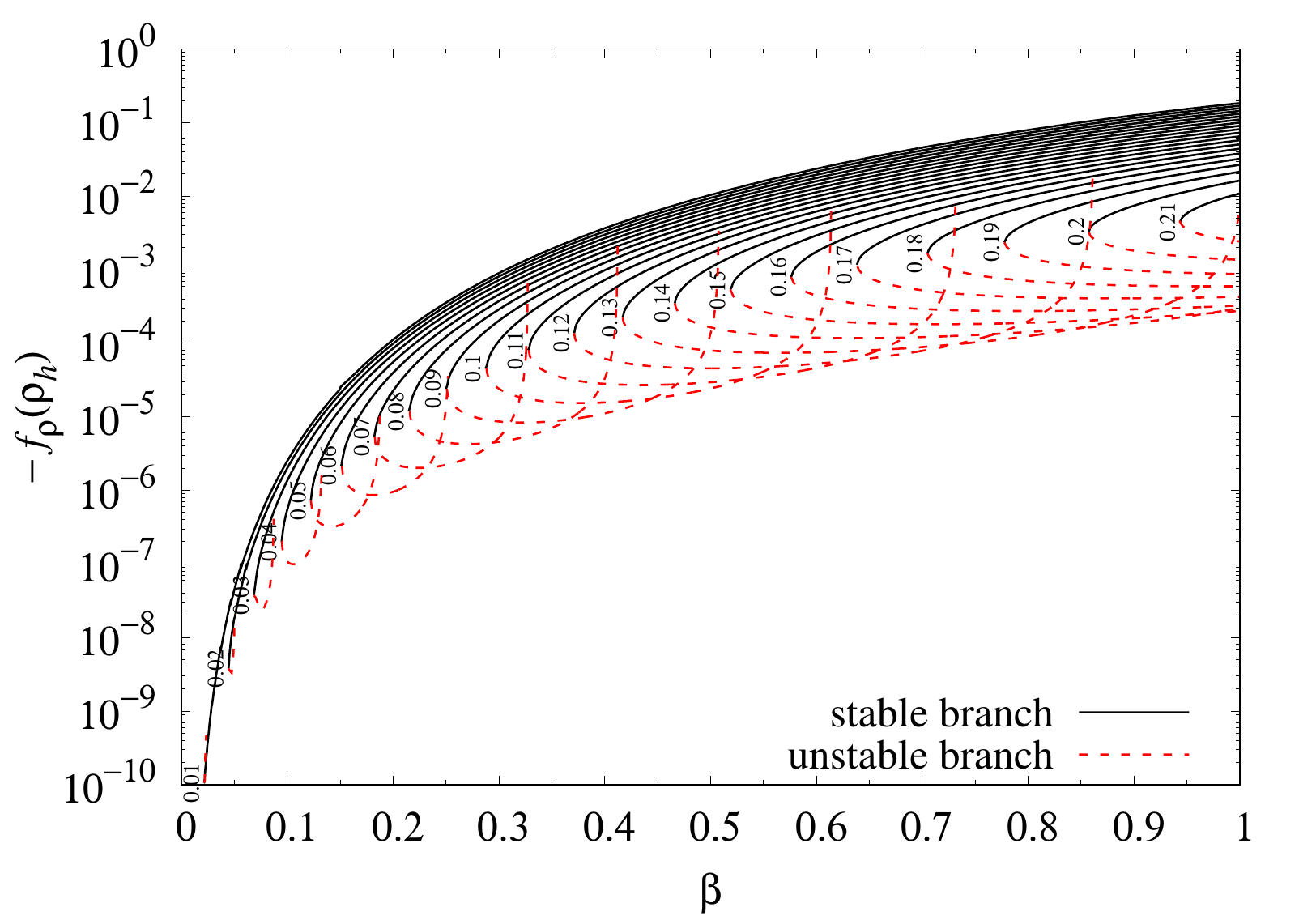}}}
\mbox{\subfloat[]{\includegraphics[width=0.49\linewidth]{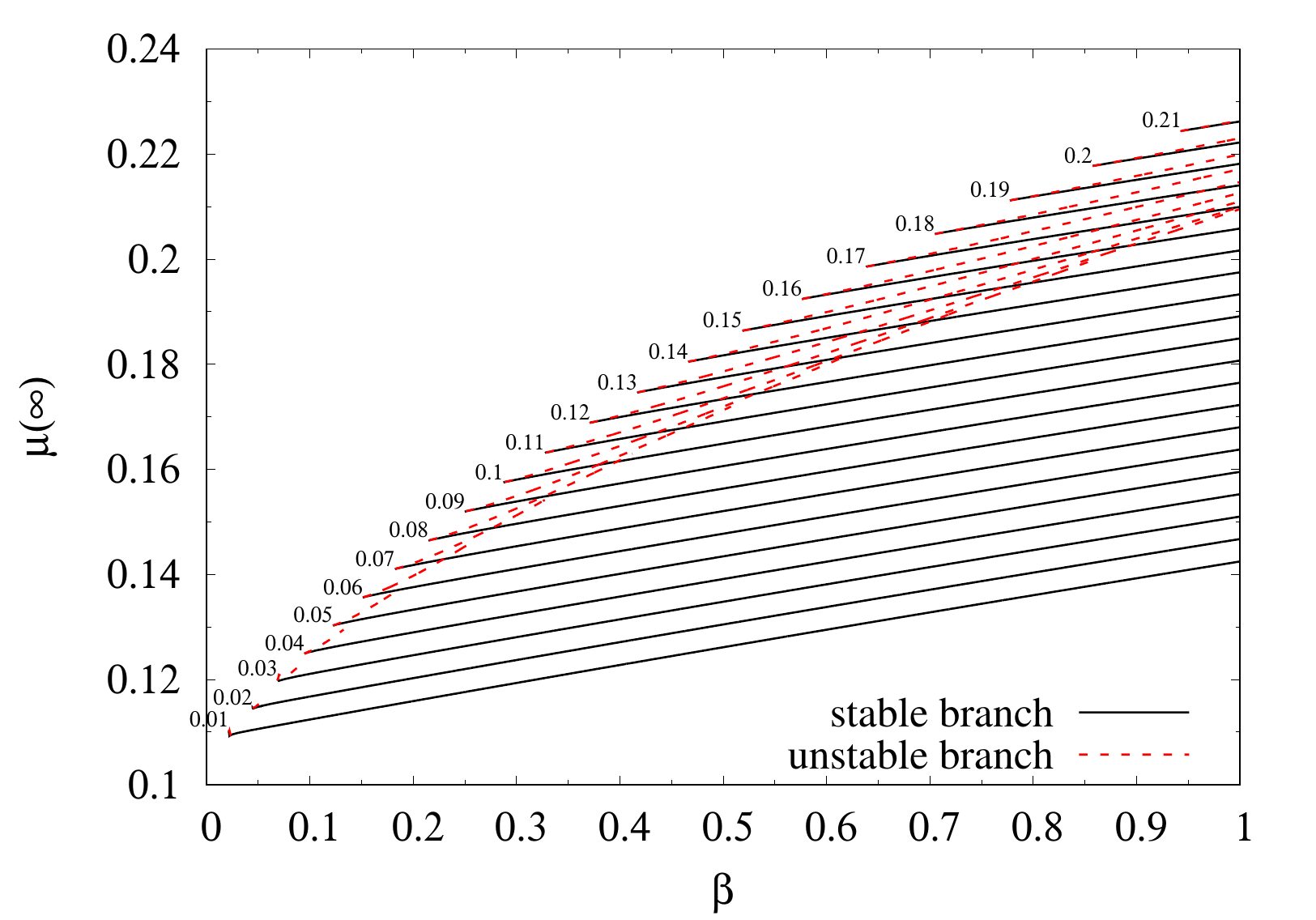}}
\subfloat[]{\includegraphics[width=0.49\linewidth]{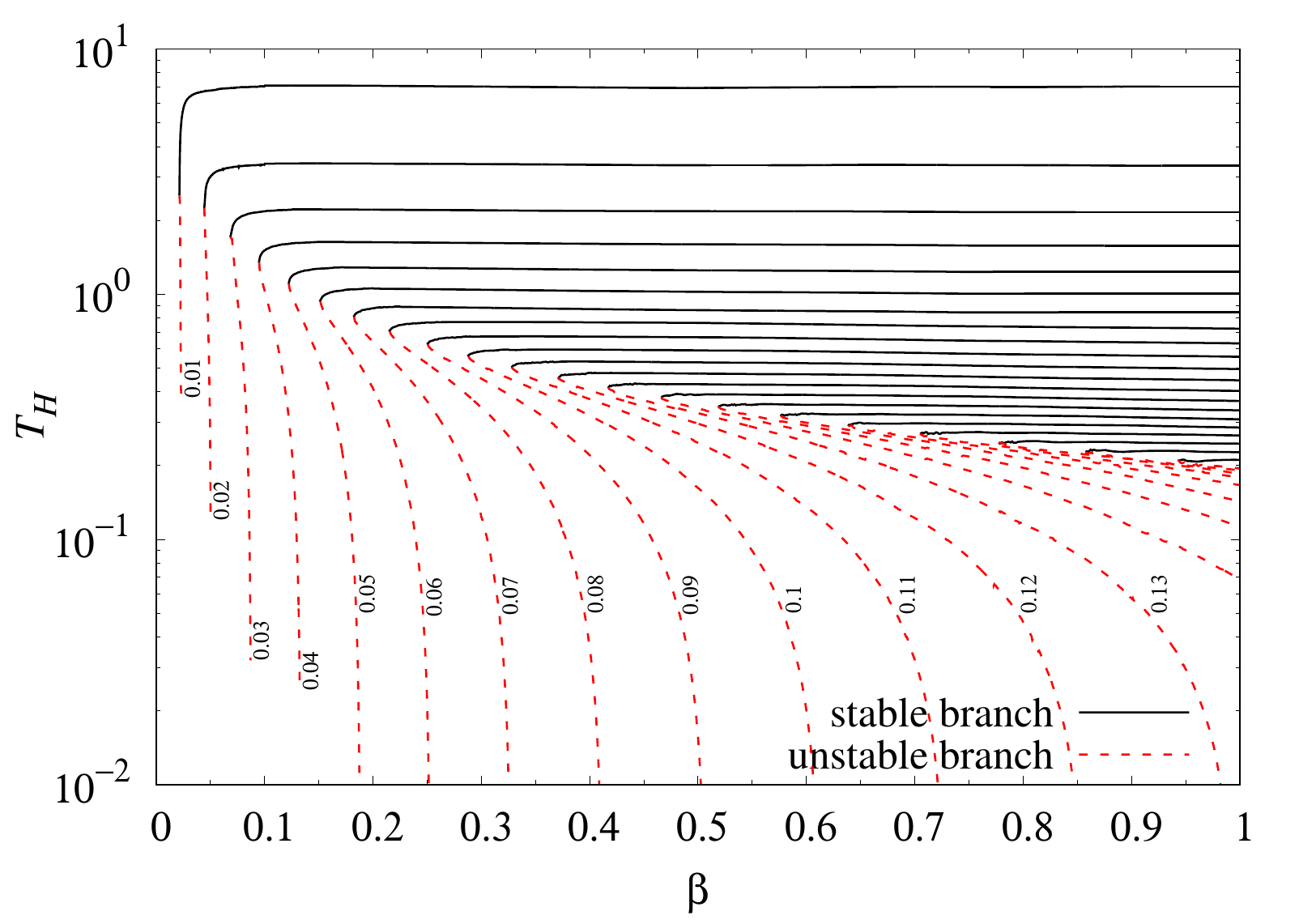}}}
\caption{Stable (black solid lines) and unstable (red dashed lines)
  branches of solutions in the $2+4+10$ model: (a) the value of the
  profile function at the horizon, $f_h$; (b) the derivative of the
  profile function at the horizon, $f_\rho(\rho_h)$; 
  (c) the ADM mass, $\mu(\infty)$;  
  (d) the Hawking temperature $T_H$, all as functions of the
  Skyrme-term coefficient, $\beta$. 
  The numbers on the figures indicate the different values of
  $\rho_h=0.01,0.02,\ldots,0.21$.}
\label{fig:m2410_beta}
\end{center}
\end{figure}

We can see from Figs.~\ref{fig:m2410_rh} and \ref{fig:m2410_beta} that
the $2+4+10$ model is qualitatively similar to the $2+4+6$ model and
the $2+4+8$ model with $\gamma=0$.
Indeed, it does not support BH hair when the Skyrme term is turned
off, i.e.~in the $\beta\to 0$ limit.
As usual for a model that cannot support BH hair (without the Skyrme
term), we see in Fig.~\ref{fig:m2410_beta}(a) that only the smallest
BH sizes remain for small $\beta$ and no branches are left in the
$\beta\to 0$ limit.
From Figs.~\ref{fig:m2410_rh}(b) and \ref{fig:m2410_beta}(b), we can
see that the derivative of the profile function at the horizon,
$f_\rho(\rho_h)$, tends to zero as $\beta$ does. 
We confirm from Figs.~\ref{fig:m2410_rh}(c)
and \ref{fig:m2410_beta}(c) that the unstable branches indeed have
larger ADM mass than their corresponding stable ones.
The Hawking temperature tends asymptotically to zero where the
unstable branches end, see Figs.~\ref{fig:m2410_rh}(d)
and \ref{fig:m2410_beta}(d).
When the temperature tends to zero for the unstable branches, we can
see from Figs.~\ref{fig:m2410_rh}(b) and \ref{fig:m2410_beta}(b) that
the derivative of the profile function at the horizon,
$f_\rho(\rho_h)$, tends to increase sharply until the solution ceases
to exist.

Finally, the Figs.~\ref{fig:m2410_rh} and \ref{fig:m2410_beta} provide
solid evidence for the fact that in the $2+4+10$ model, the BH hair
does not exist in the $\beta\to 0$ limit.

\subsection{The \texorpdfstring{$2+4+12$}{2+4+12} model}

The last model that we will consider in this section, is the $2+4+12$
model, where the higher-order derivative term that is added to the
standard Skyrme model is made of the baryon charge density to the
fourth power -- or equivalently the BPS-Skyrme term squared.
This model also only has one overall coefficient of the highest-order
term, $c_{12}$, which we scaled away in Eq.~\eqref{eq:Lrescaled}.

Completing the Einstein
equations \eqref{eq:EEQ1_242n}-\eqref{eq:EEQ2_242n}, we get
\begin{align}
-C_\rho + \frac{1}{\rho} - \frac{C}{\rho} &=
2\alpha\left[
\rho C f_\rho^2
+\frac{2\sin^2f}{\rho}
+\frac{2\beta\sin^2(f) C f_\rho^2}{\rho}
+\frac{\beta\sin^4f}{\rho^3}
+\frac{\sin^8(f) C^2 f_\rho^4}{\rho^7}
\right],\non
\frac{1}{\alpha}\frac{N_\rho}{N} &=
2\rho f_\rho^2
+\frac{4\beta \sin^2(f)f_\rho^2}{\rho}
+\frac{4\sin^8(f) C f_\rho^4}{\rho^7},
\end{align}
whereas the equation of motion for the profile function is
\begin{align}
C f_{\rho\rho}
+\frac{2C f_\rho}{\rho}
+C_\rho f_\rho
+\frac{N_\rho C f_\rho}{N}
-\frac{\sin 2f}{\rho^2} \non
+\frac{2\beta\sin^2f}{\rho^2}\left(
 C f_{\rho\rho}
 + C_\rho f_\rho
 + \frac{N_\rho C f_\rho}{N}
 - \frac{\sin 2f}{2\rho^2}
 \right)
+\frac{\beta\sin(2f) C f_\rho^2}{\rho^2}
\non
+\frac{2\sin^8(f) C f_\rho^2}{\rho^8}\left(
 3C f_{\rho\rho}
 -\frac{6C f_\rho}{\rho}
 +2C_\rho f_\rho
 +\frac{N_\rho C f_\rho}{N}
 \right)
+\frac{6\sin^6(f)\sin(2f) C^2 f_\rho^4}{\rho^8}
= 0.
\end{align}
Finally, we need the boundary
conditions \eqref{eq:C_rho_at_rh}-\eqref{eq:f_rho_at_rh}, which can be 
written as
\begin{align}
\mu &= \frac{\rho_h}{2}
+\alpha\sin^2(f_h)\left(2
  +\frac{\beta\sin^2f_h}{\rho_h^2}\right)(\rho-\rho_h)
+ \mathcal{O}\left((\rho-\rho_h)^2\right),\label{eq:muprime_12}\\
f &= f_h
+ \rho_h\sin(2f_h)\frac{\rho_h^3 \sin 2f_h + \beta\sin^2f_h}
{\left(\rho_h^2 + 2\beta\sin^2f_h\right)
\Xi_{12}}(\rho-\rho_h)
+ \mathcal{O}\left((\rho-\rho_h)^2\right),\\
\Xi_{12} &\equiv
\rho_h^2 - 4\alpha\sin^2f_h - 2\alpha\beta\sin^4f_h.
\label{eq:Xi12}
\end{align}

\begin{figure}[!thp]
\begin{center}
\mbox{\subfloat[]{\includegraphics[width=0.49\linewidth]{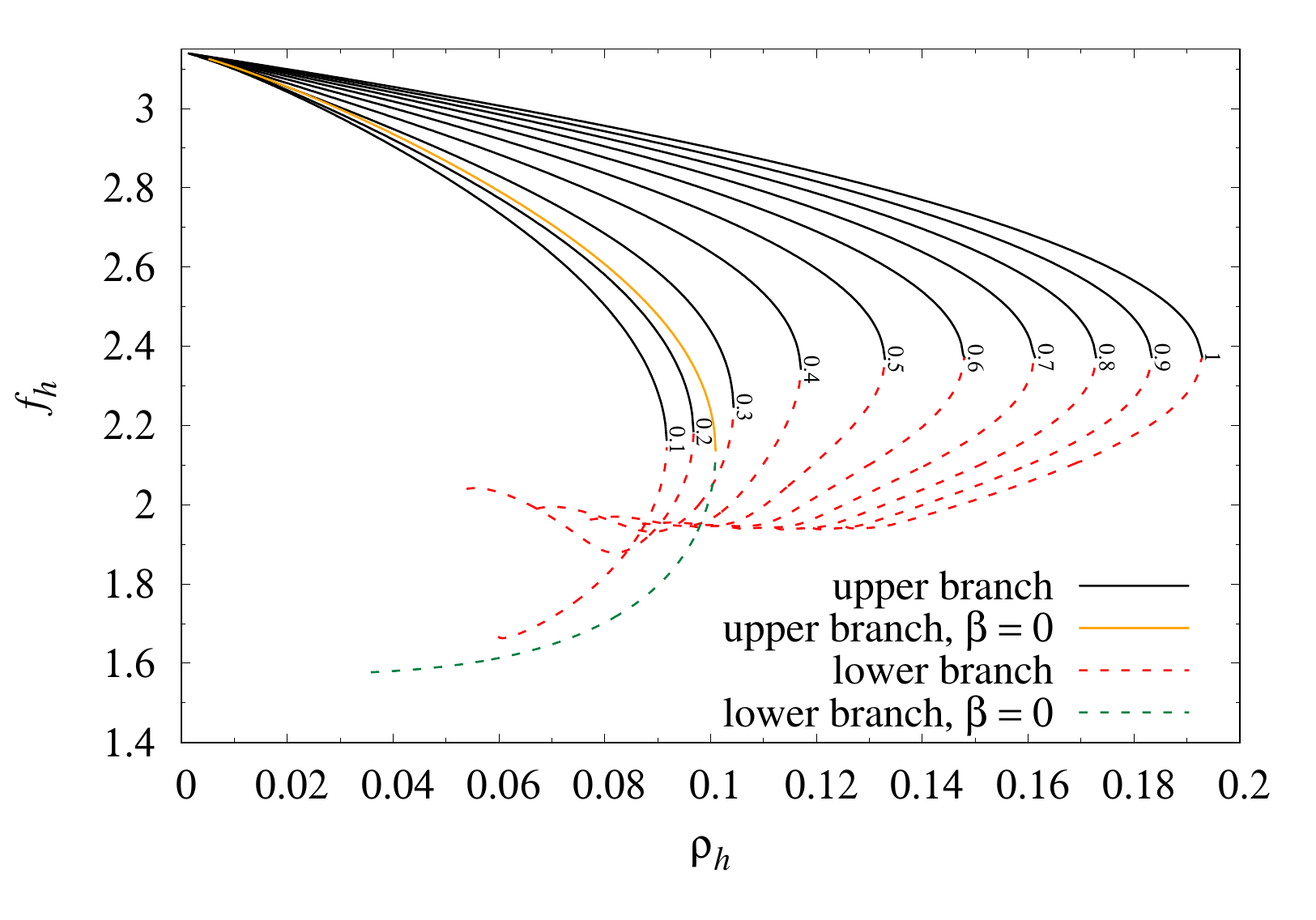}}
\subfloat[]{\includegraphics[width=0.49\linewidth]{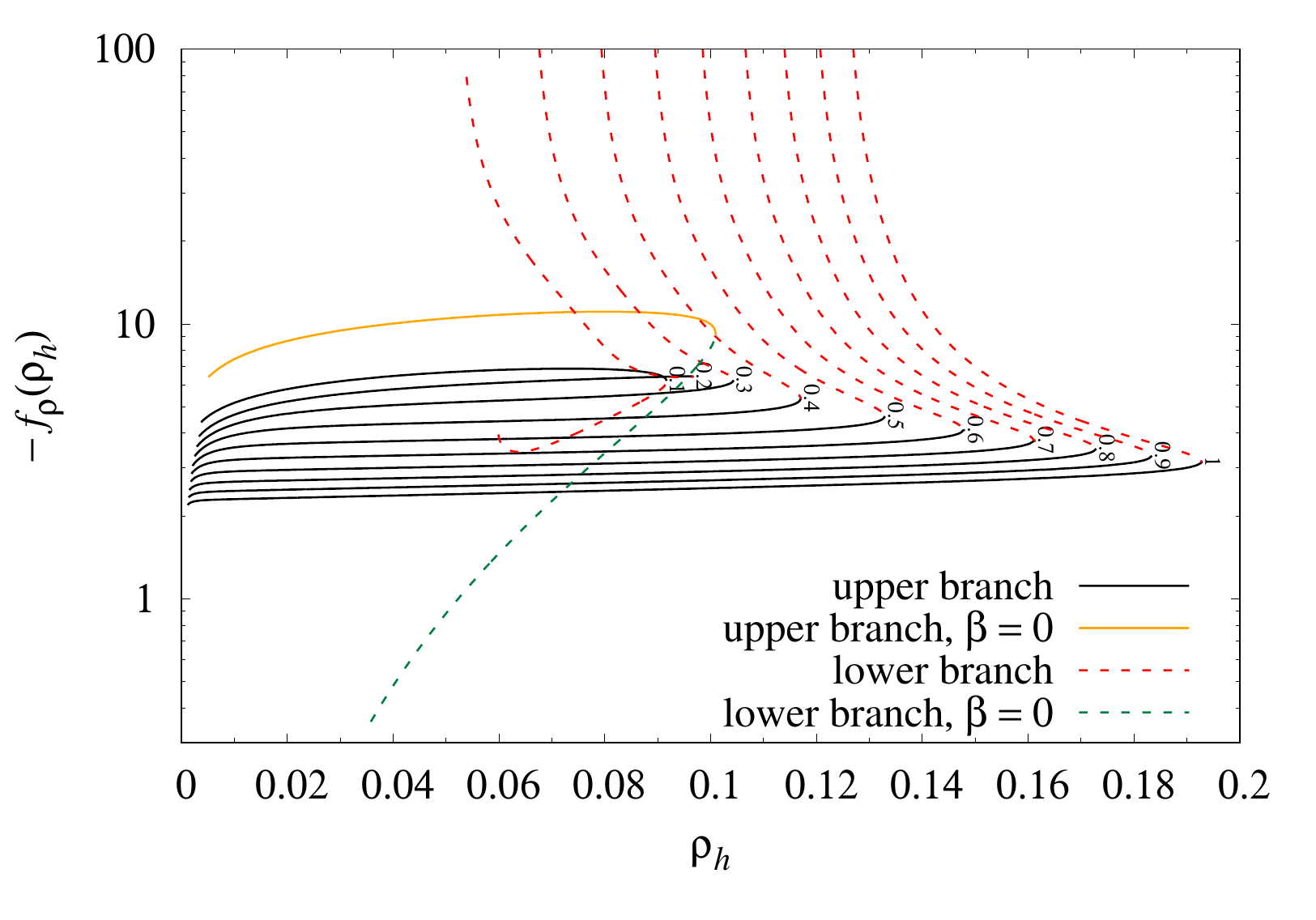}}}
\mbox{\subfloat[]{\includegraphics[width=0.49\linewidth]{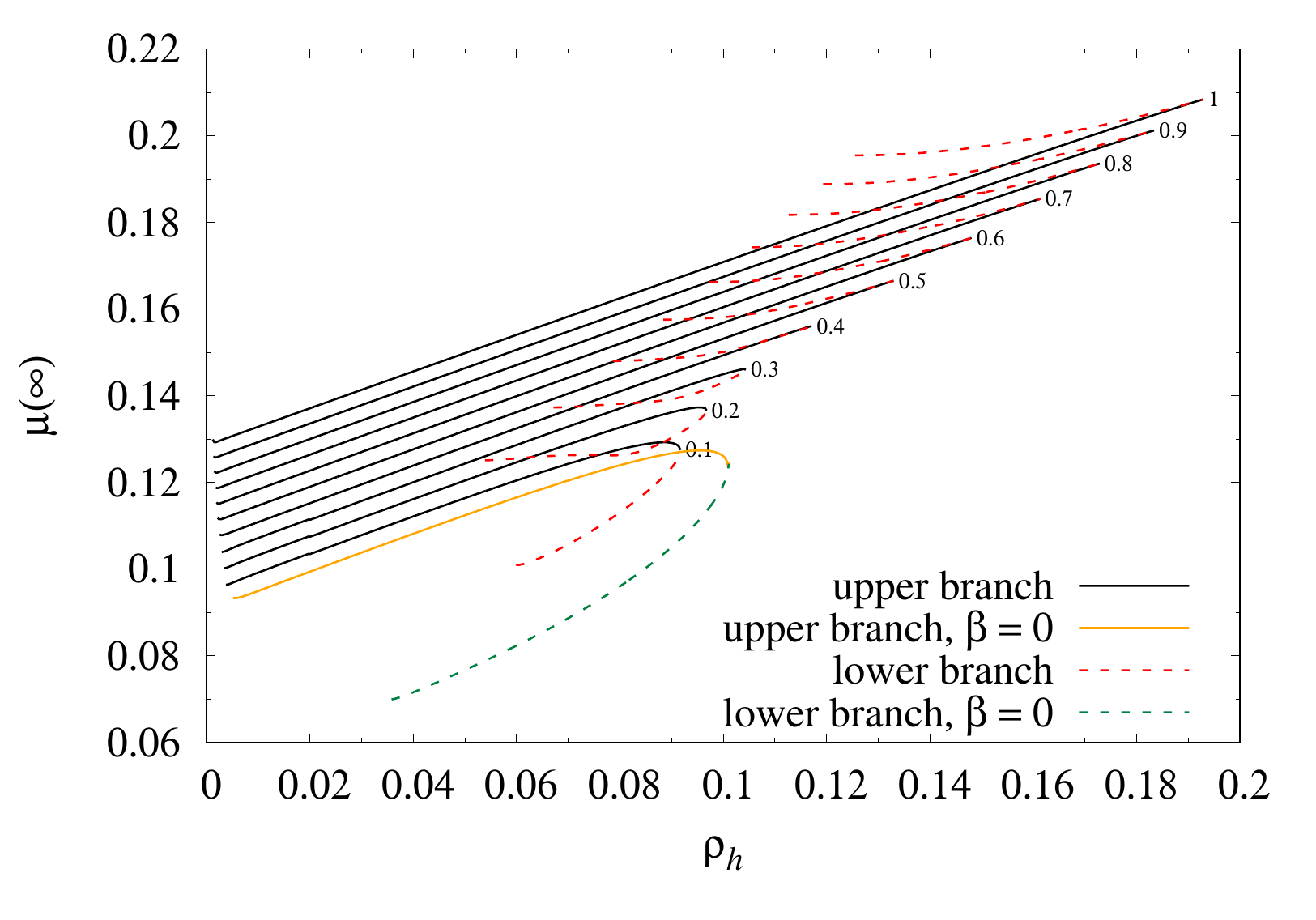}}
\subfloat[]{\includegraphics[width=0.49\linewidth]{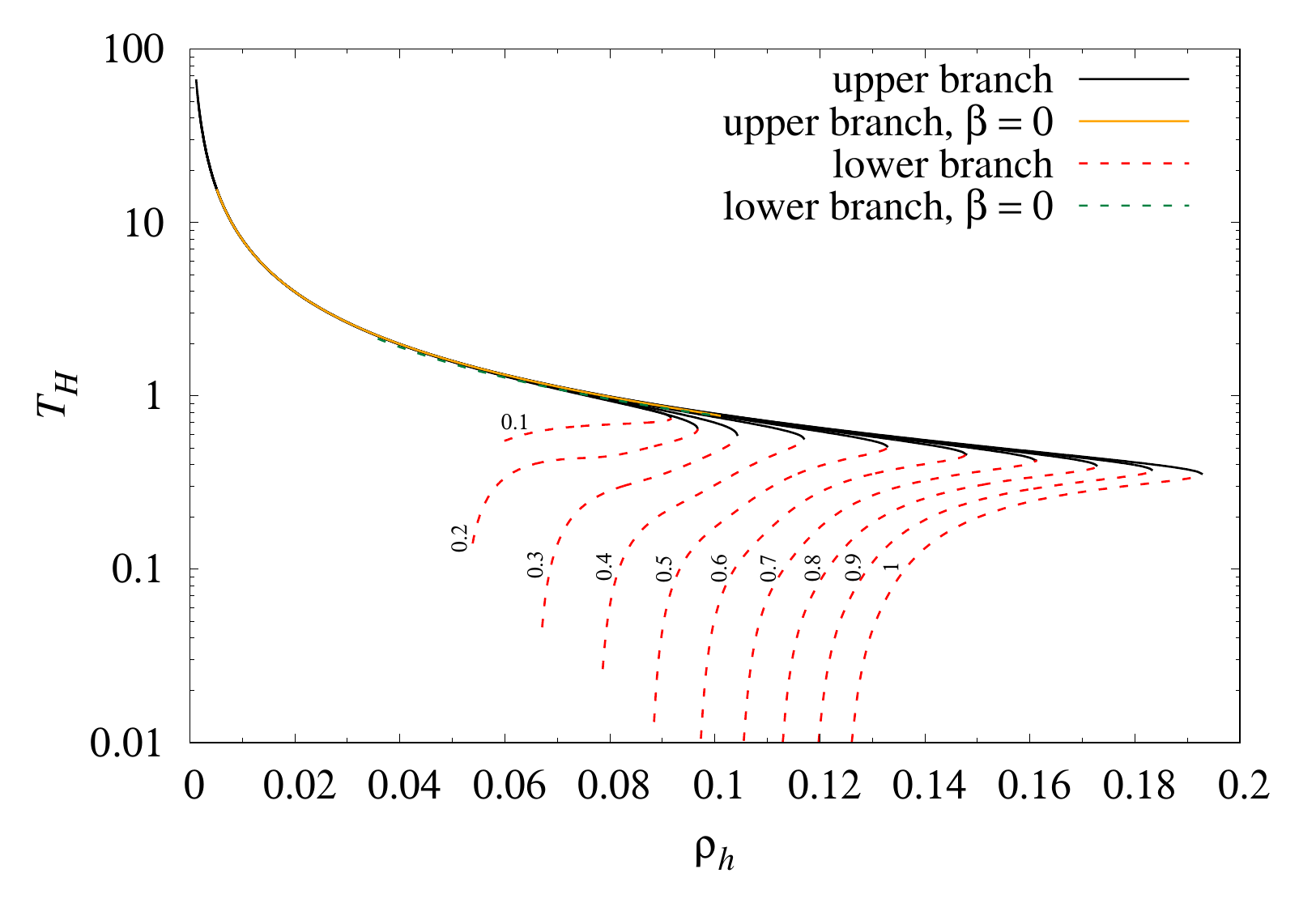}}}
\caption{Upper (black solid lines) and lower (red dashed lines)
  branches of solutions in the $2+4+12$ model: (a) the value of the profile
  function at the horizon, $f_h$; (b) the derivative of the profile
  function at the horizon, $f_\rho(\rho_h)$; (c) the ADM mass,
  $\mu(\infty)$; (d) the Hawking temperature $T_H$, all as functions
  of the size of the black hole, i.e.~the horizon radius, $\rho_h$.
  The numbers on the figures indicate the different values of
  $\beta=0,0.1,0.2,\ldots,1$.
  The upper and lower $\beta=0$ branches are shown in orange and dark
  green colors, respectively. 
}
\label{fig:m2412_rh}
\end{center}
\end{figure}

\begin{figure}[!thp]
\begin{center}
\mbox{\subfloat[]{\includegraphics[width=0.49\linewidth]{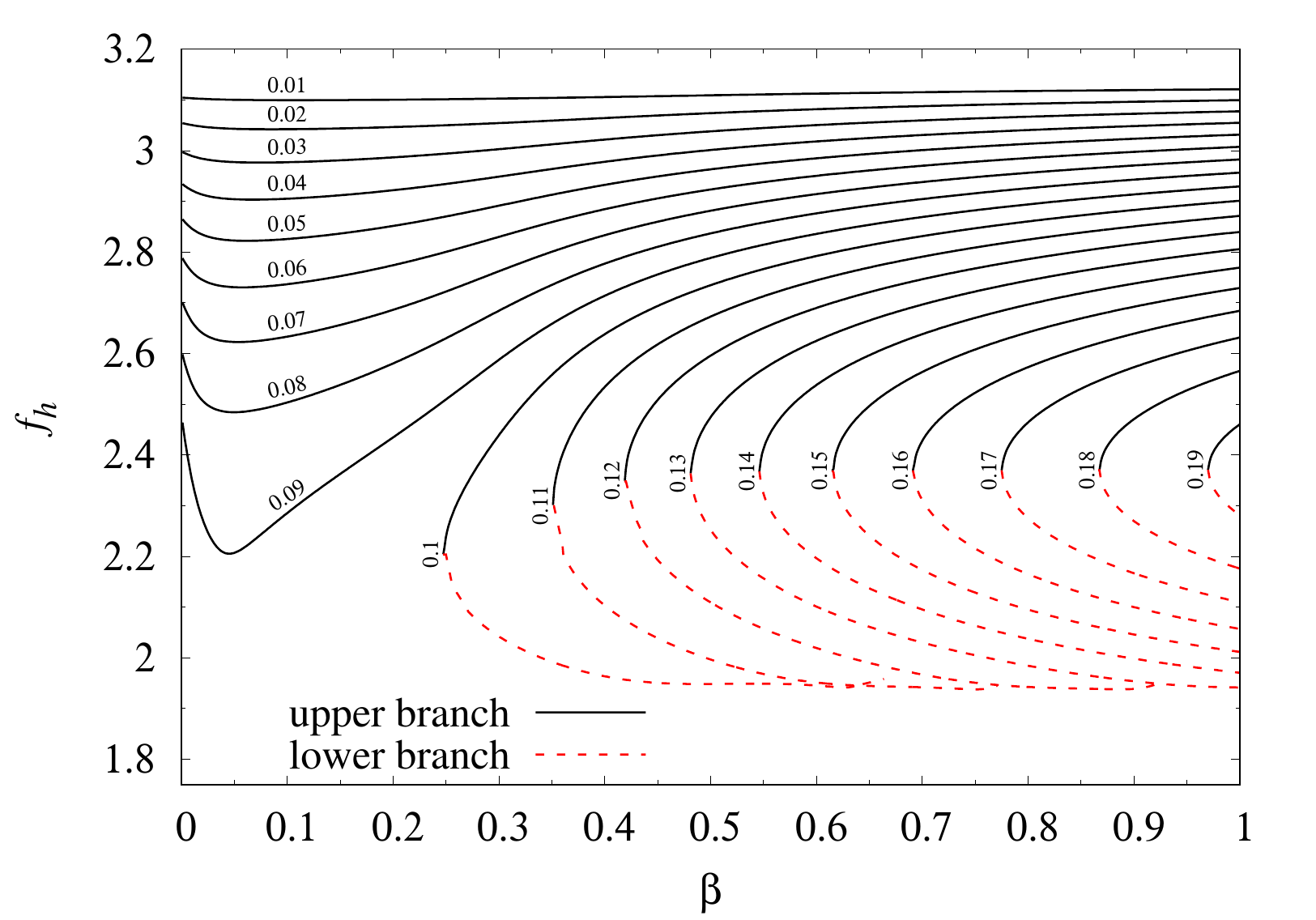}}
\subfloat[]{\includegraphics[width=0.49\linewidth]{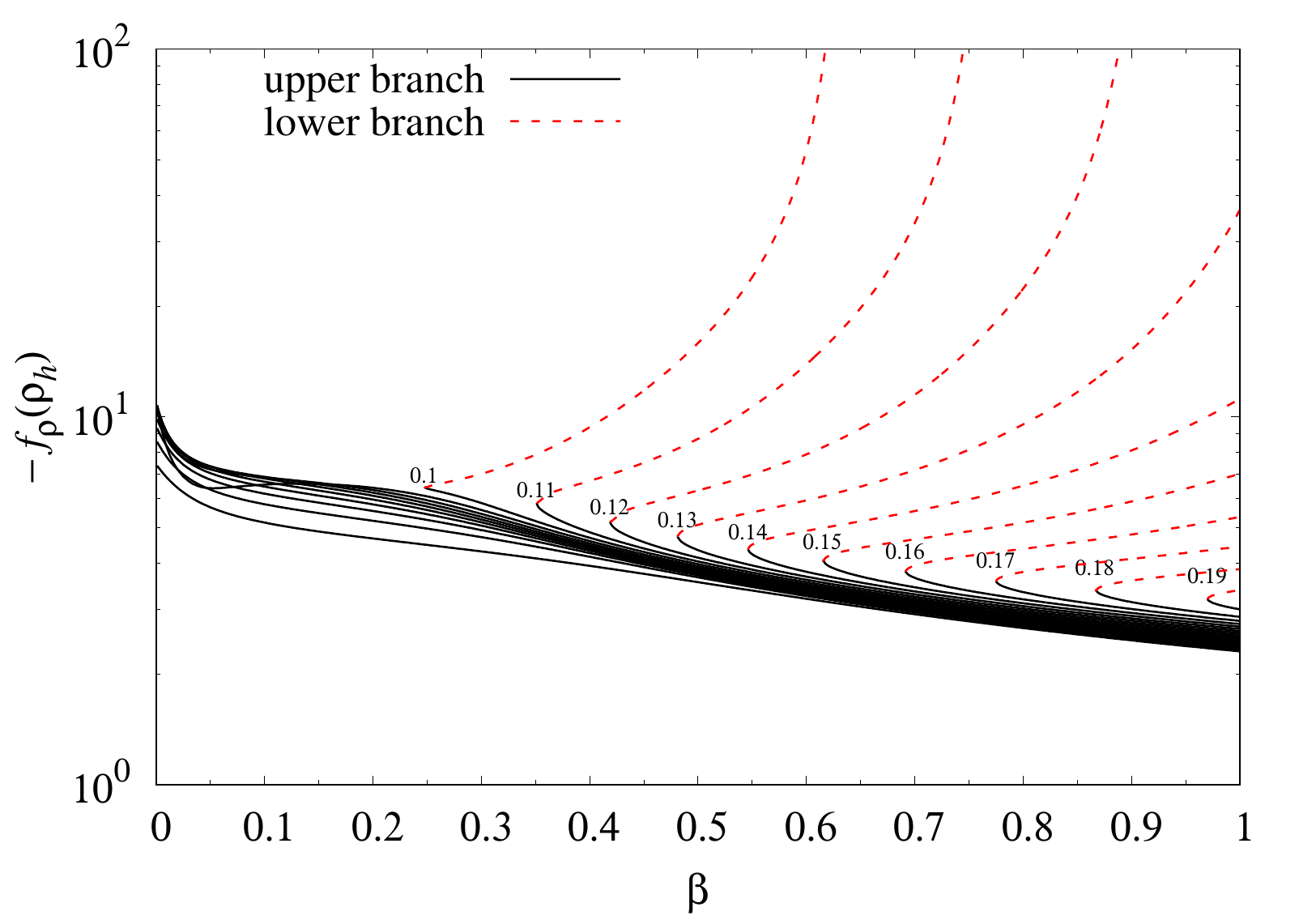}}}
\mbox{\subfloat[]{\includegraphics[width=0.49\linewidth]{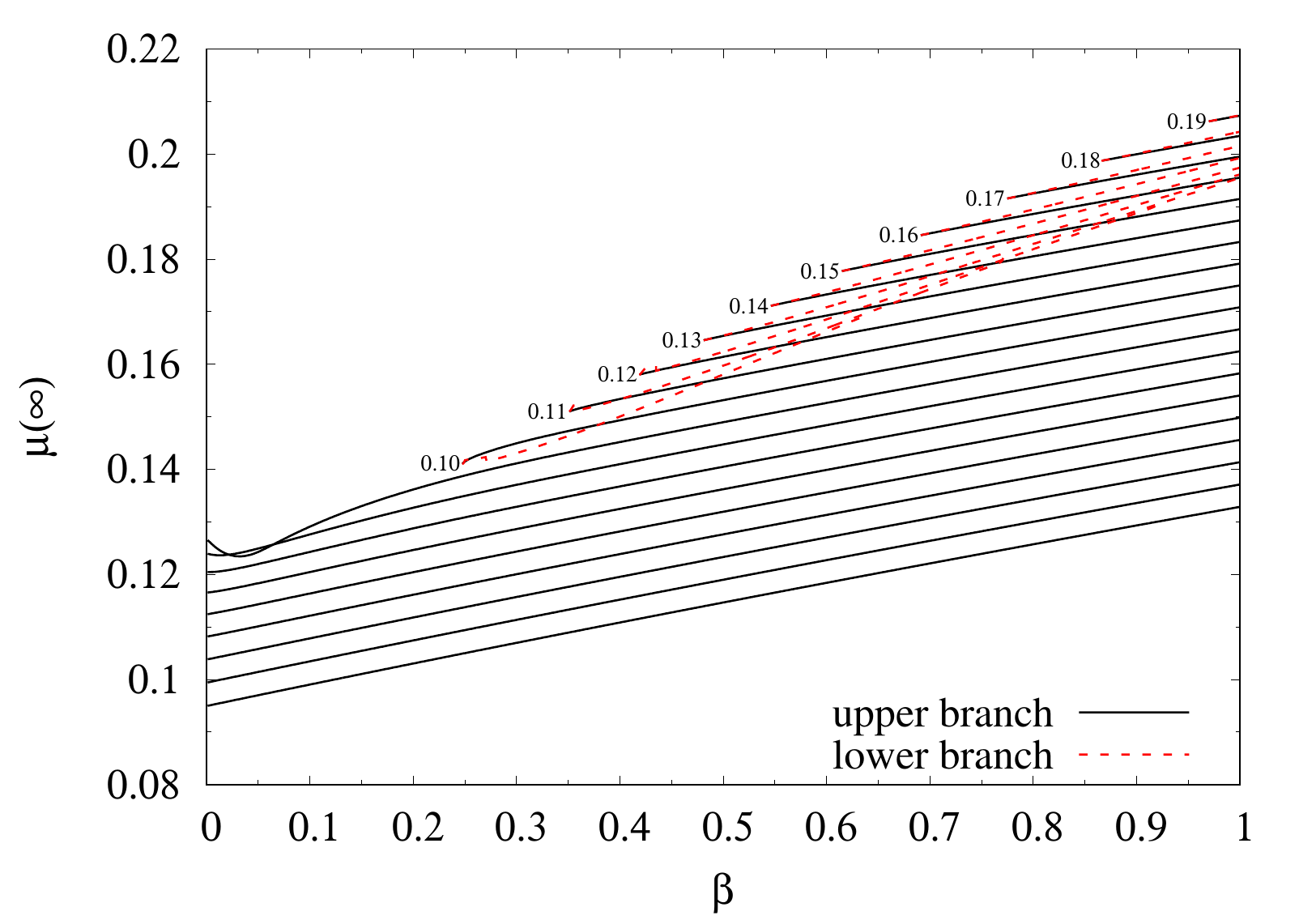}}
\subfloat[]{\includegraphics[width=0.49\linewidth]{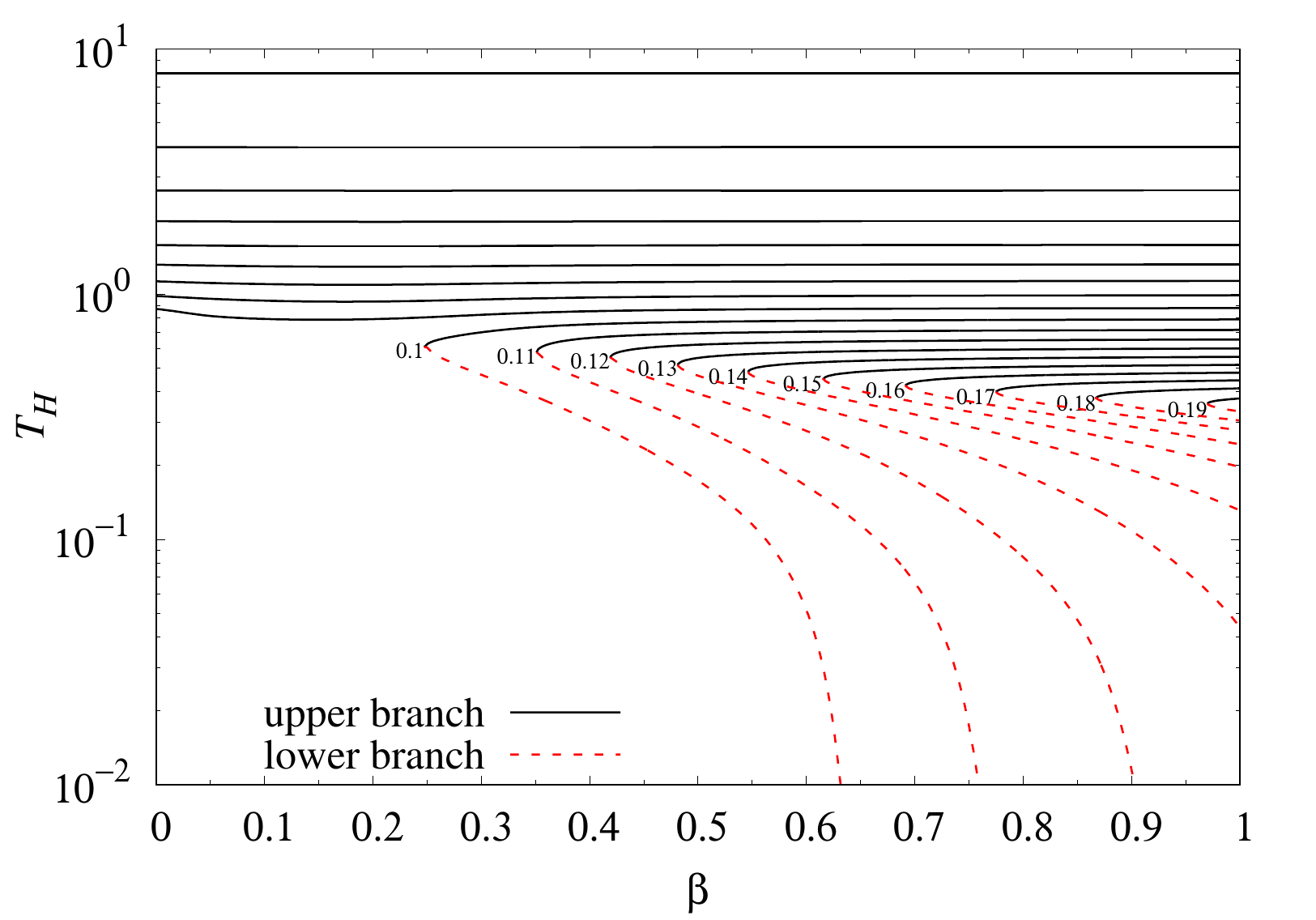}}}
\caption{Upper (black solid lines) and lower (red dashed lines)
  branches of solutions in the $2+4+12$ model: (a) the value of the profile
  function at the horizon, $f_h$; (b) the derivative of the profile
  function at the horizon, $f_\rho(\rho_h)$; (c) the ADM mass,
  $\mu(\infty)$; (d) the Hawking temperature $T_H$, all as functions
  of the Skyrme-term coefficient, $\beta$. 
  The numbers on the figures indicate the different values of
  $\rho_h=0.01,0.02,\ldots,0.19$.
}
\label{fig:m2412_beta}
\end{center}
\end{figure}

\begin{figure}[!thbp]
\begin{center}
\includegraphics[width=0.49\linewidth]{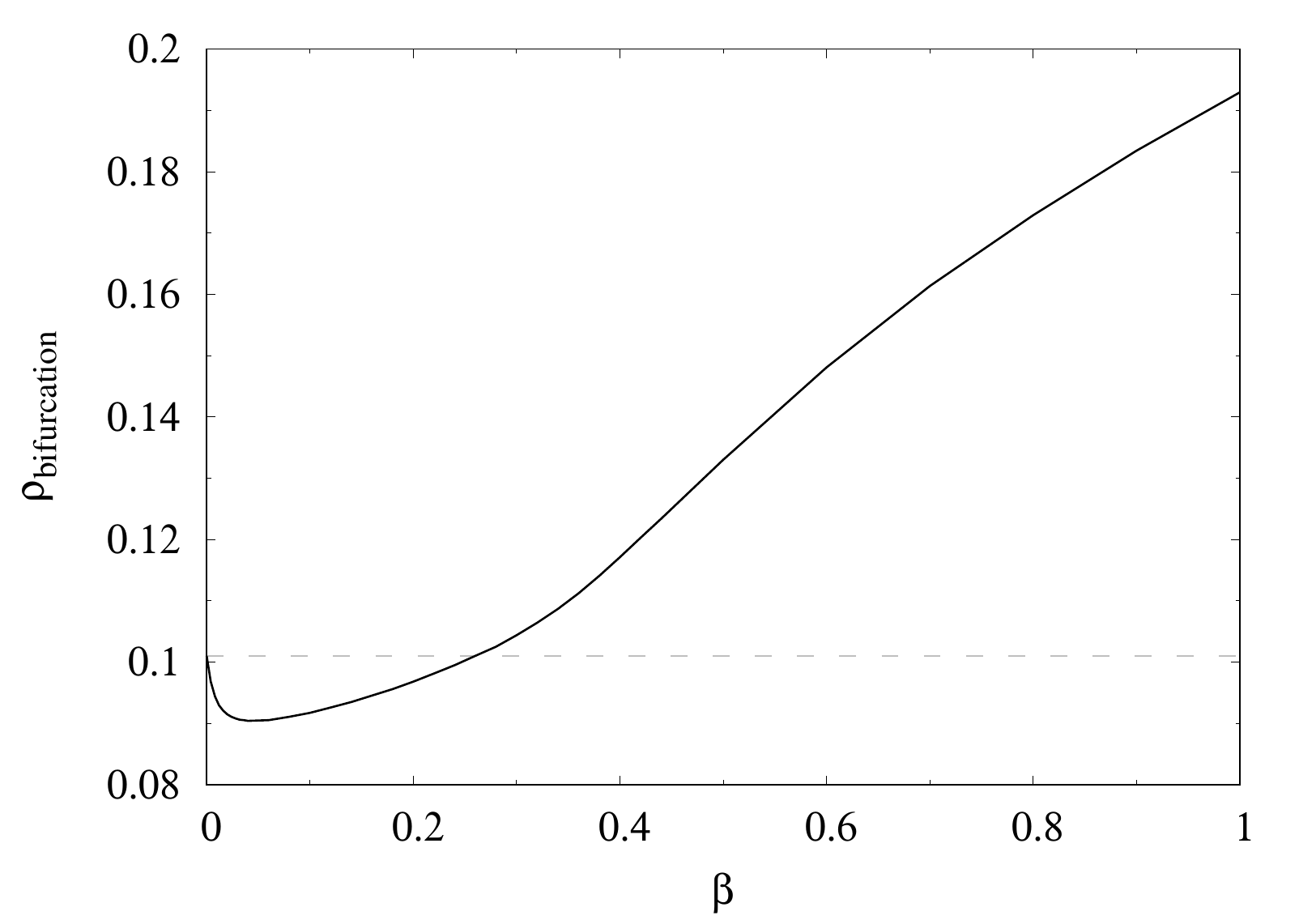}
\caption{The horizon radius at the bifurcation point (maximal BH that
  supports hair) as a function of the Skyrme-term coefficient $\beta$.
  The minimum is around $\rho_h=0.04$.
}
\label{fig:m2412_rh_bifurcation_point}
\end{center}
\end{figure}

\begin{figure}[!thp]
\begin{center}
\mbox{\subfloat[]{\includegraphics[width=0.49\linewidth]{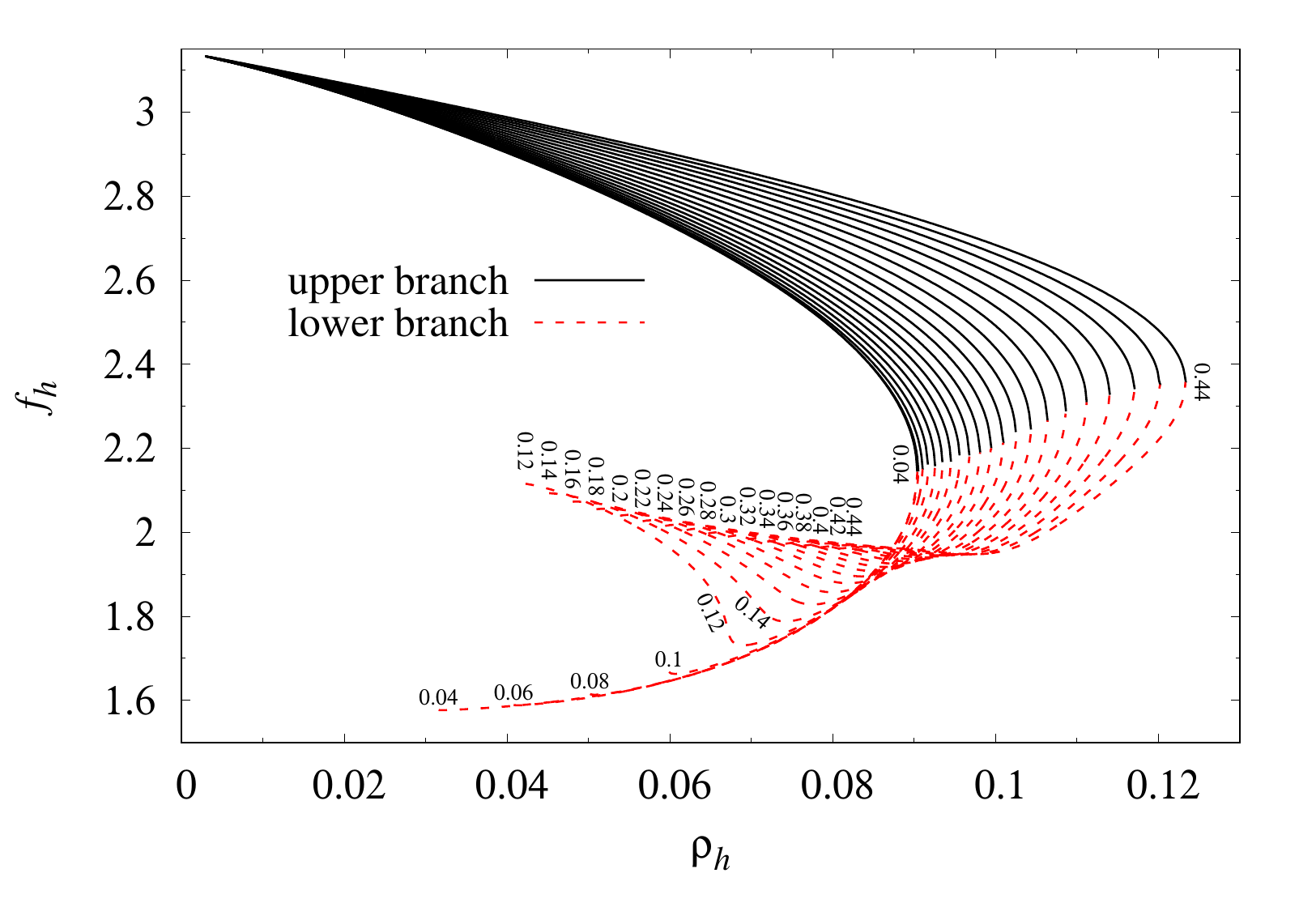}}
\subfloat[]{\includegraphics[width=0.49\linewidth]{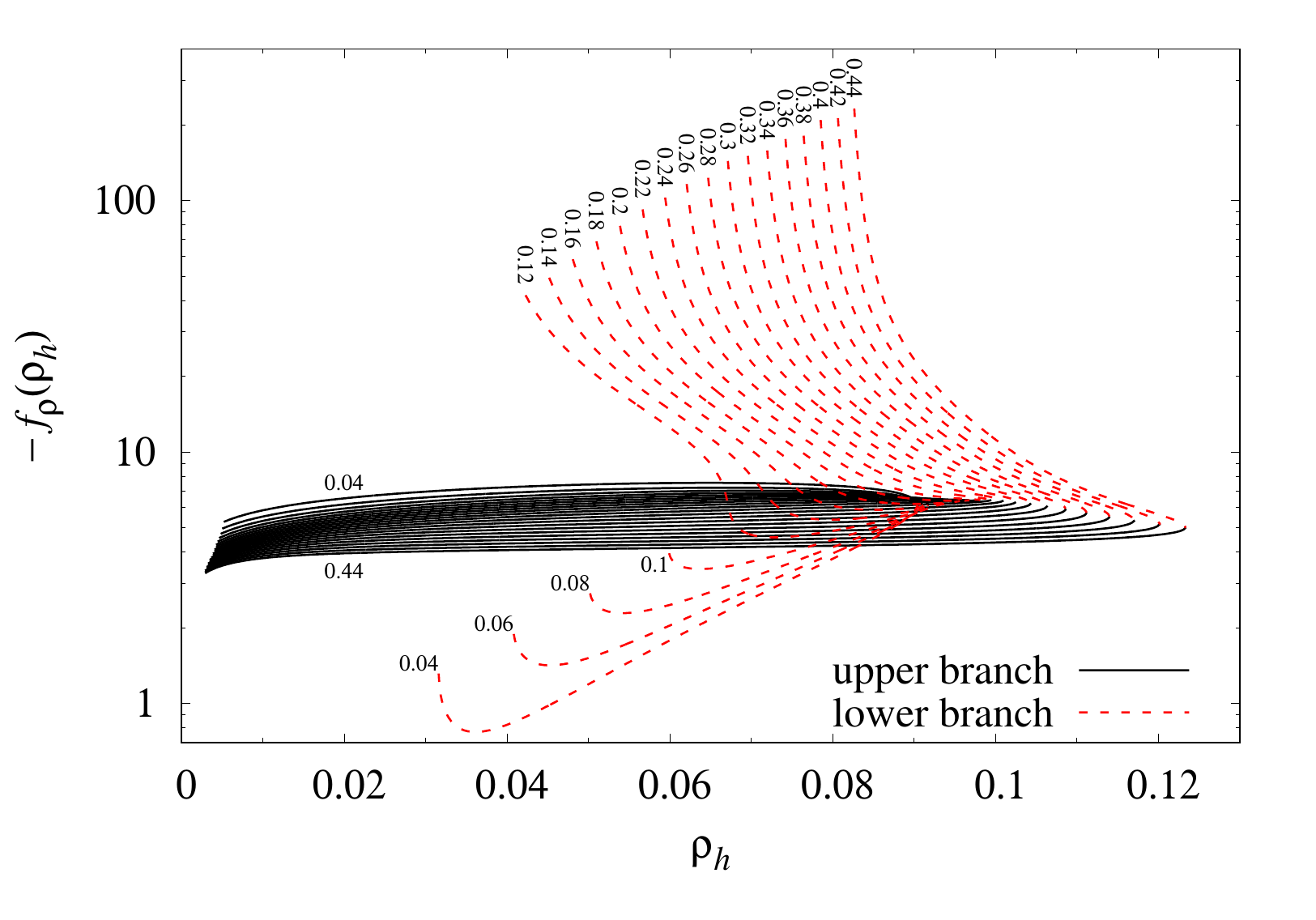}}}
\mbox{\subfloat[]{\includegraphics[width=0.49\linewidth]{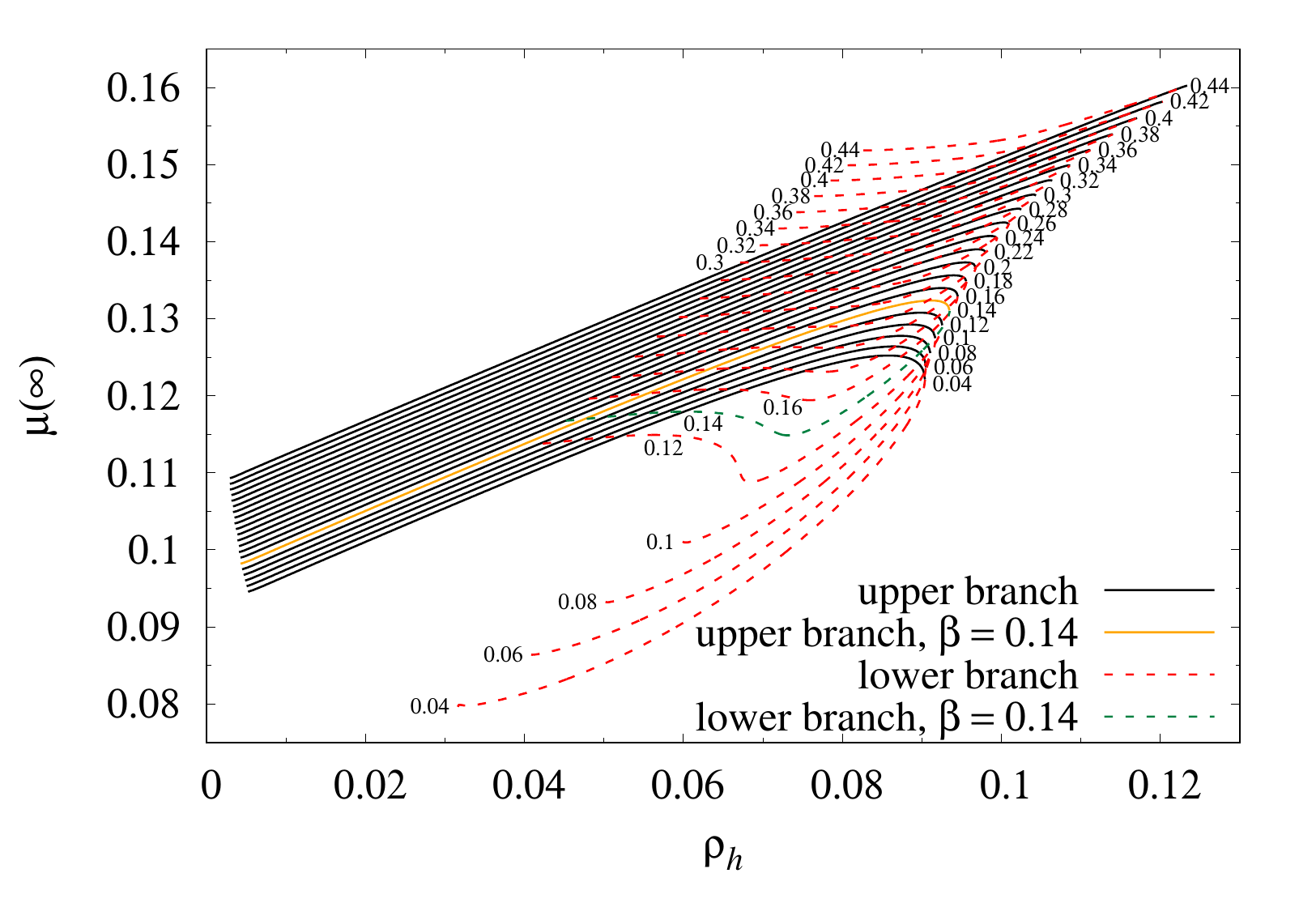}}
\subfloat[]{\includegraphics[width=0.49\linewidth]{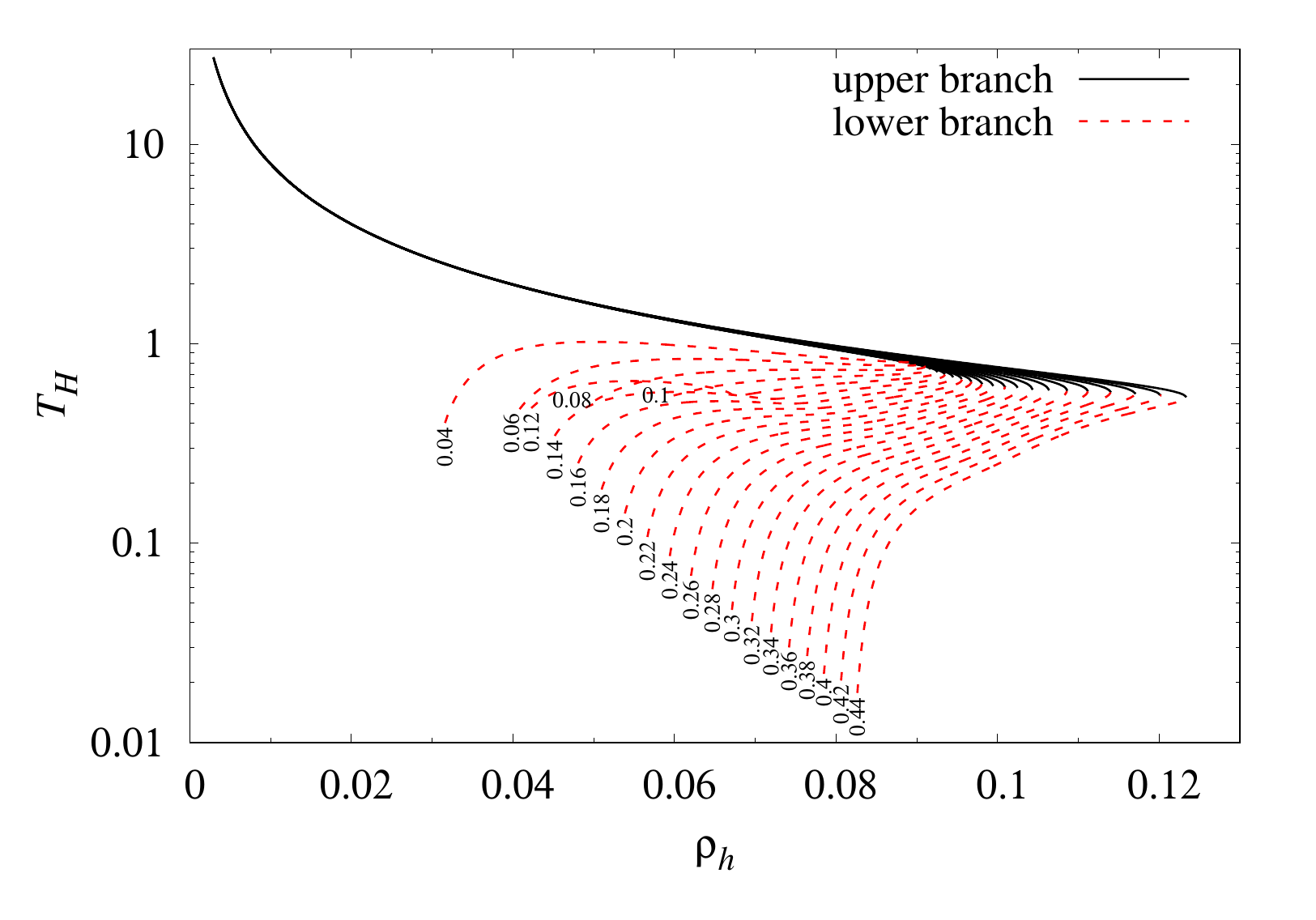}}}
\caption{Details of upper (black solid lines) and lower (red dashed lines)
  branches of solutions in the $2+4+12$ model: (a) the value of the profile
  function at the horizon, $f_h$; (b) the derivative of the profile
  function at the horizon, $f_\rho(\rho_h)$; (c) the ADM mass,
  $\mu(\infty)$; (d) the Hawking temperature $T_H$, all as functions
  of the size of the black hole, i.e.~the horizon radius, $\rho_h$.
  The numbers on the figures indicate the different values of
  $\beta=0.04,0.06,\ldots,0.44$.
}
\label{fig:m2412_rh_detail1}
\end{center}
\end{figure}

\begin{figure}[!thp]
\begin{center}
\mbox{\subfloat[]{\includegraphics[width=0.49\linewidth]{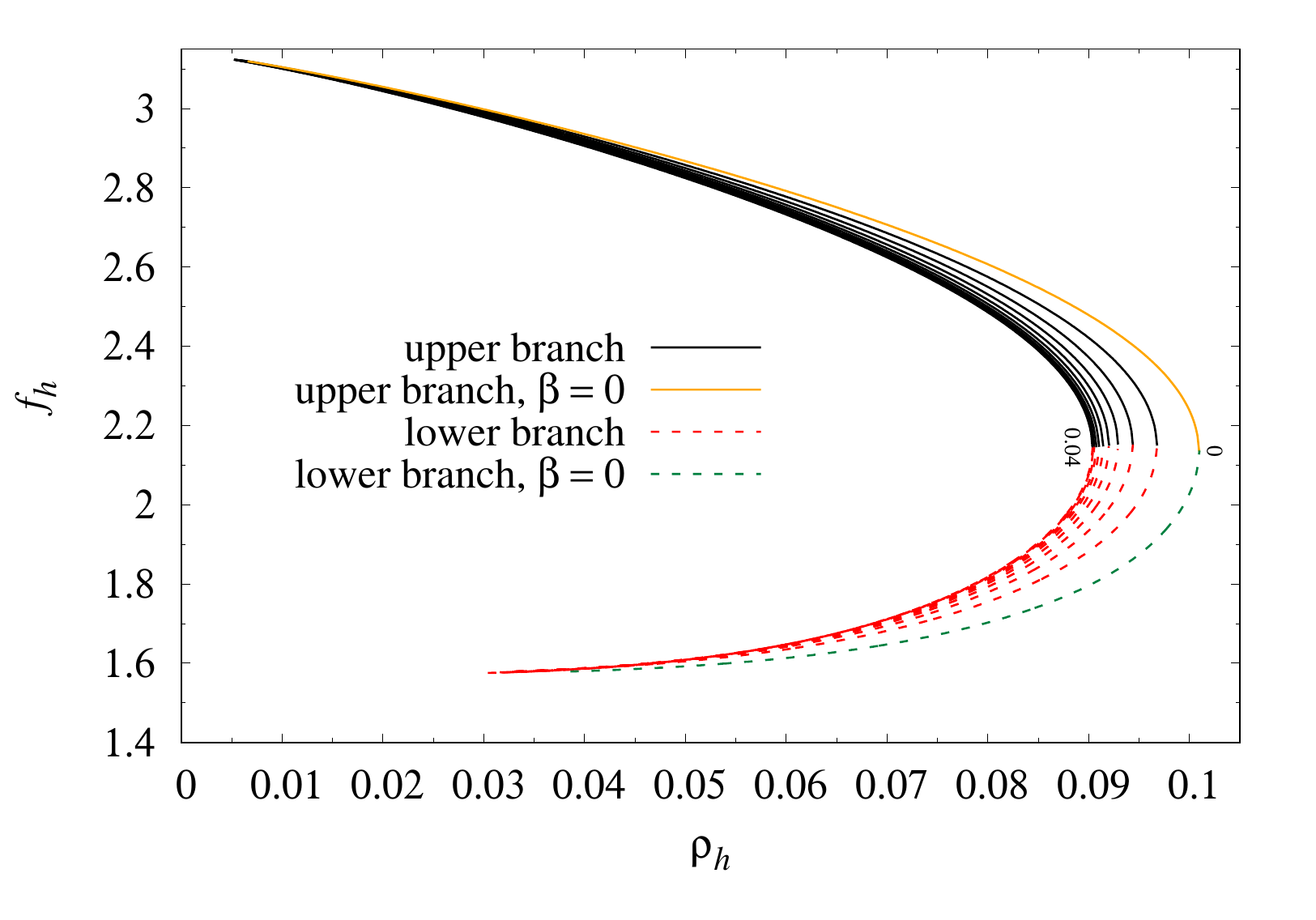}}
\subfloat[]{\includegraphics[width=0.49\linewidth]{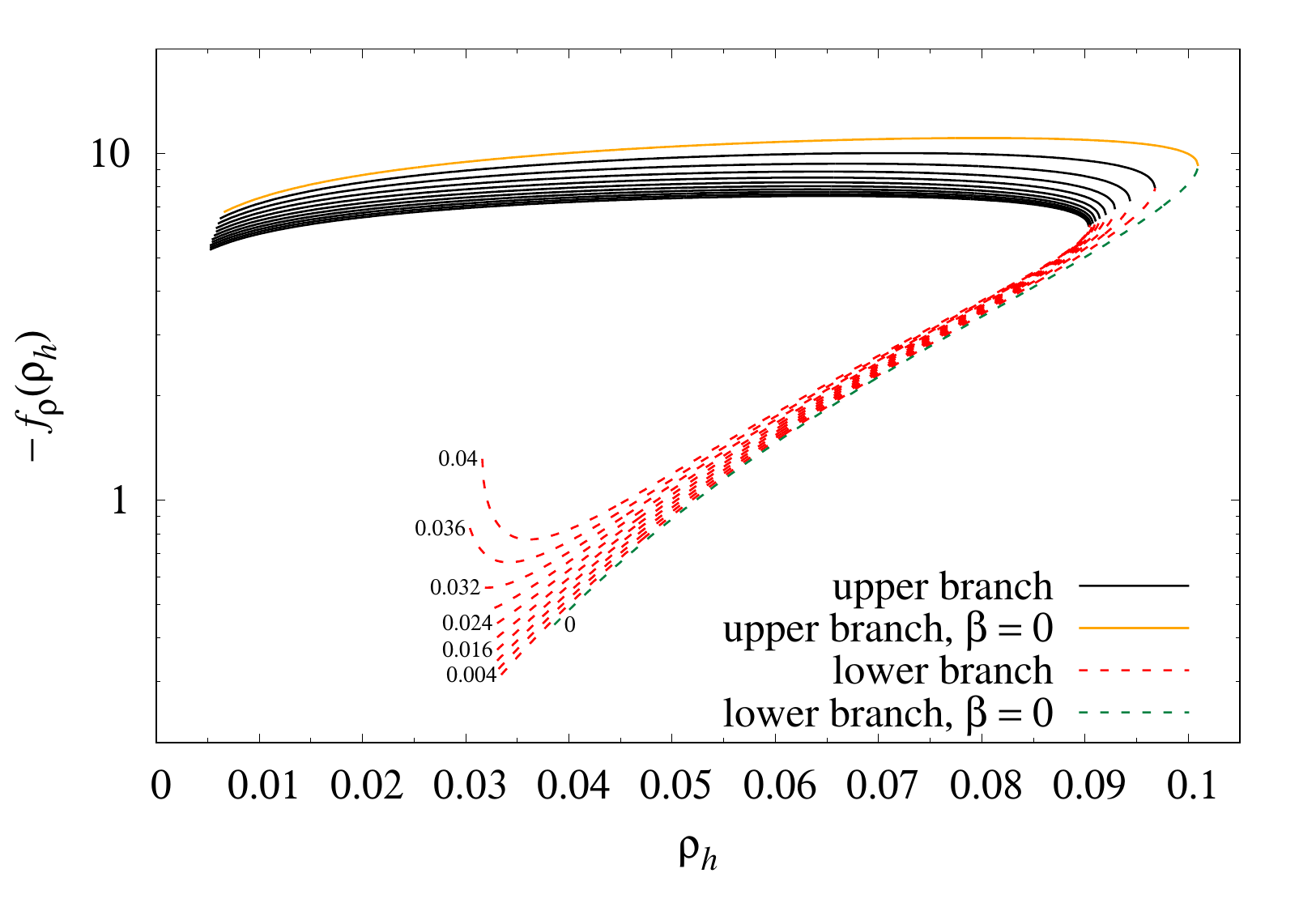}}}
\mbox{\subfloat[]{\includegraphics[width=0.49\linewidth]{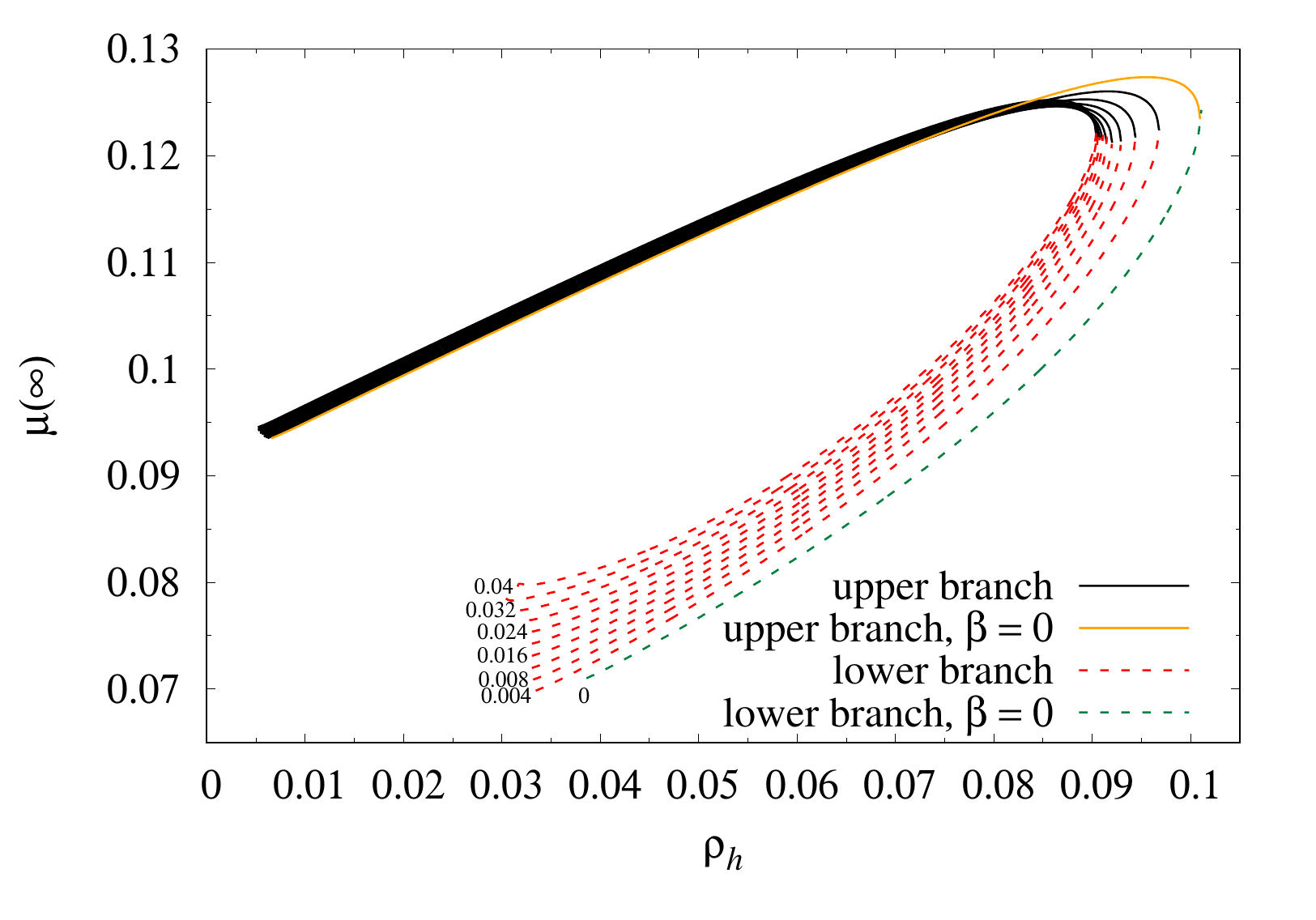}}
\subfloat[]{\includegraphics[width=0.49\linewidth]{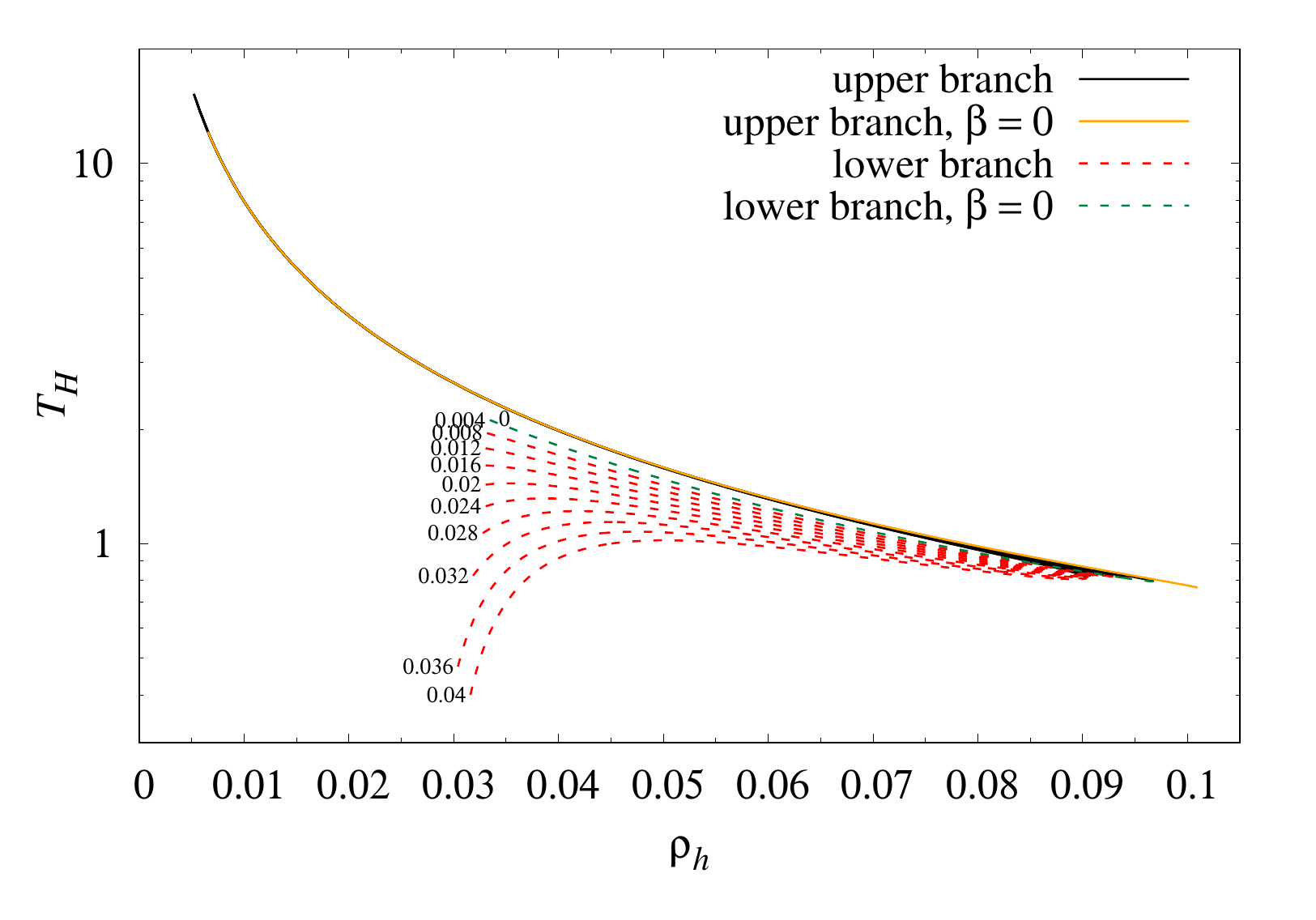}}}
\caption{Details of upper (black solid lines) and lower (red dashed lines)
  branches of solutions in the $2+4+12$ model: (a) the value of the profile
  function at the horizon, $f_h$; (b) the derivative of the profile
  function at the horizon, $f_\rho(\rho_h)$; (c) the ADM mass,
  $\mu(\infty)$; (d) the Hawking temperature $T_H$, all as functions
  of the size of the black hole, i.e.~the horizon radius, $\rho_h$.
  The numbers on the figures indicate the different values of
  $\beta=0,0.004,0.008,\ldots,0.04$.
  The upper and lower $\beta=0$ branches are shown in orange and dark
  green colors, respectively. 
}
\label{fig:m2412_rh_detail2}
\end{center}
\end{figure}

We are now ready to present the results of the numerical calculations
for the $2+4+12$ model; they are shown in Figs.~\ref{fig:m2412_rh}
and \ref{fig:m2412_beta}. This model is unique among the models we
have considered in this paper. Indeed it has existing branches of
solutions in the $\beta\to 0$ limit, but the unforeseen twist is that 
the lower branches start to go below the upper branches in the
ADM mass when $\beta$ becomes smaller than about $0.3$, see
Fig.~\ref{fig:m2412_rh}(c). 
This is evidence for a kind of phase transition, where the lower
branch becomes the stable one.
We note that the general tendency is that as $\beta$ becomes smaller,
the ADM mass decreases.
Before the phase transition takes place ($\beta>\beta^{\rm crit}$),
the lower branches have higher ADM mass for the same value of horizon
radius as their corresponding upper branches.
There is a range in $\beta$ where the ADM mass of the lower branch
intersects that of the upper branch.
Then finally, as mentioned just above, in the $\beta\to 0$ limit, the
lower branch has a lower ADM mass than the corresponding upper branch,
for all values of $\rho_h$ that the branch exists.

As common for many of the models with higher higher-order derivative
terms than fourth order, the unstable branches terminate on the way
back from the bifurcation point at a \emph{finite} horizon radius
($\rho_h>0$), see e.g.~Fig.~\ref{fig:m2412_rh}(a). 
This is related to the line in the $(f_h,\rho_h)$ diagram where $\Xi$
vanishes, see e.g.~Eq.~\eqref{eq:Xi12}. At that point, the first
derivative of the profile function at the horizon, $f_\rho(\rho_h)$,
would tend to minus infinity, which in turn would prevent a smooth
soliton solution \cite{Gudnason:2016kuu}.
The implication here, of the unstable branch ending at a finite
horizon radius, is that the upper branch (in $f_h$) will be the stable
solution for small enough $\rho_h$ and then become unstable at a
horizon radius where the lower branch starts, see
Fig.~\ref{fig:m2412_rh}(a).
This means that if a BH possesses Skyrme hair in this model and has a
size $\rho_h<\rho_h^{\rm crit}$, the upper branch will dictate the
value of the profile function.
If we now throw in some stuff, say a piano, the BH grows and at the
point where $\rho_h>\rho_h^{\rm crit}$, the Skyrme hair will become
unstable and decay to the stable solution, i.e.~the lower branch. 
In order to check our claims, we will perform a stability analysis of
this model in the next section.

In Fig.~\ref{fig:m2412_rh}(b) we can see another peculiar feature
happening after our claimed phase transition. Namely, (minus) the
first derivative of the profile function at the horizon 
$-f_\rho(\rho_h)$, for the lower branch with $\beta=0$ tends towards
zero, whereas before the transition $\beta>\beta^{\rm crit}$, they
tend to become very large.

The Hawking temperature for $\beta>\beta^{\rm crit}$ has the usual
feature for a stable model, e.g.~the $2+4+8$ model with $\gamma=1$ or
$\gamma=\frac{1}{3}$; that  is, the temperature for the upper branches
all rise as the horizon radius tends to zero and decreases for larger
BHs until the bifurcation point of the particular branch is reached,
see Fig.~\ref{fig:m2412_rh}(d). 
The lower branches for $\beta>\beta^{\rm crit}$ tend backwards in the
$(T_H,\rho_h)$ diagram (see Fig.~\ref{fig:m2412_rh}(d)) until they
suddenly drop and terminate. 
What happens when the phase transition, $\beta<\beta^{\rm crit}$ is
reached, is that the lower branches no longer tend to vanishing
temperature and in the $\beta\to 0$ limit, the Hawking temperature is
at the same level as the upper branches.

We will now plot all the standard quantities as functions of the
Skyrme-term coefficient, $\beta$, instead of the horizon radius,
$\rho_h$, see Fig.~\ref{fig:m2412_beta}.
The main point to notice is that for $\rho_h\lesssim 0.1$, the upper
branches do not have a bifurcation point at any finite $\beta$ and
thus have a well-defined $\beta\to 0$ limit.
The last branch with a bifurcation point in
Fig.~\ref{fig:m2412_beta}(a) is the $\rho_h=0.1$ branch and said point
is around $\beta=0.25$.
This indicates that $\beta^{\rm crit}$ is probably between $0.2$ and
$0.25$. 

All the lower branches depicted in Fig.~\ref{fig:m2412_beta} are
connected to the upper branches with $\beta>\beta^{\rm crit}$ and
therefore do not exhibit the peculiar behavior seen in
Fig.~\ref{fig:m2412_rh}.
The derivative of the profile function at the horizon goes up
slightly, but remains of order one in the $\beta\to 0$ limit, see
Fig.~\ref{fig:m2412_beta}(b) and the Hawking temperature remains
basically constant in the limit, see Fig.~\ref{fig:m2412_beta}(d).
This applies to the upper branches.
Since the lower branches possess somewhat more complicated behavior
around and after the phase transition, they are not depicted in
Fig.~\ref{fig:m2412_beta}.

We can see from Fig.~\ref{fig:m2412_rh}(a) that the bifurcation point
moves to smaller BH radii as $\beta$ decreases from $1$ to about
$0.04$ and then it turns around and continues to increase as $\beta$
goes to zero.
The bifurcation point, which corresponds to the largest BH size
possessing hair, is plotted in
Fig.~\ref{fig:m2412_rh_bifurcation_point}.

We will now make a more detailed plot of $\beta$ branches as functions
of $\rho_h$ and in order to avoid clutter, we will make the first
series of solutions with $\beta\in[0.04,0.44]$, see
Fig.~\ref{fig:m2412_rh_detail1}, and the second series of solutions
with $\beta\in[0,0.04]$, see Fig.~\ref{fig:m2412_rh_detail2}.
In Fig.~\ref{fig:m2412_rh_detail1} the bifurcation point is decreasing
in $\rho_h$, as $\beta$ decreases, and in
Fig.~\ref{fig:m2412_rh_detail2} the bifurcation point is increasing in 
$\rho_h$. 

The physics in Fig.~\ref{fig:m2412_rh_detail1} is very versatile;
three different situations arise.
The classification is based on the ADM masses, see
Fig.~\ref{fig:m2412_rh_detail1}(c). 
For $\beta\geq 0.44$ the ADM mass of the lower branch is always above
that of the upper branch. This makes lower branches of solutions
unstable and everything is as usual.
In a range of $\beta\in(0.44,0.12)$ (approximately), the lower branch
has a lower ADM mass than that of the upper branch for a finite range
in BH sizes $\rho_h\in[\rho_h^\star,\rho^{\rm birfurcation}]$.
At $\rho_h=\rho_h^\star$ the ADM masses of the lower and upper
branches cross and the lower branch becomes the unstable branch for
$\rho_h<\rho_h^\star$. 
Finally, for $\beta\in[0,0.12]$ the lower branch possesses a lower ADM
mass than that of the upper branch for its entire domain of
existence.
The last branch, i.e.~the smallest value of $\beta$ for which the
ADM mass of the lower branch crosses that of the upper branch is
$\beta=0.14$.
For $\beta=0.12$, the branch terminates just about where the ADM mass
of the lower branch would cross over that of the upper branch.
For $\beta<0.12$, the lower branches behave drastically
different than those with $\beta\geq 0.12$.
In the case of the ADM mass, instead of bending down, going up and
then continuing back with a constant ADM mass as $\rho_h$ decreases
(for $\beta\geq 0.12$), it just goes downwards in an arc-like shape
until the branch terminates, see Fig.~\ref{fig:m2412_rh_detail1}(c).

This same transition point -- $\beta\geq 0.12$ -- also has an impact
on $f_h$, see Fig.~\ref{fig:m2412_rh_detail1}(a).
Indeed, the lower branches keep a relatively high value ($\sim 2$) of
the profile function at the horizon.
For $\beta<0.12$, approximately, the lower branches experience a
paradigm shift and they continue downwards in an arc-like shape until
they terminate.
The derivative of the profile function at the horizon,
$-f_\rho(\rho_h)$, also experiences the same paradigm shift around
$\beta\sim 0.12$, see Fig.~\ref{fig:m2412_rh_detail1}(b).
For $\beta\geq 0.12$, the derivatives go upwards until the lower
branches terminate whereas for $\beta<0.12$, the derivatives go below
that of the upper branches and only shortly before they terminate,
they start to go upwards.
Finally, this ``paradigm shift'' can also be seen in the Hawking
temperature, see Fig.~\ref{fig:m2412_rh_detail1}(d).
For $\beta\geq 0.12$ the end point of the lower branches moves
downwards in $\rho_h$ in a monotonic fashion as $\beta$ is decreased,
whereas for $\beta<0.12$ the behavior is different. Instead, the
temperature is raised a bit and the branches move in a higher level,
see Fig.~\ref{fig:m2412_rh_detail1}(d).

Fig.~\ref{fig:m2412_rh_detail1} is the most interesting one as it
contains the ``paradigm shifting'' behavior and possible a phase
transition.
Fig.~\ref{fig:m2412_rh_detail2}, however, is the remaining details for
$\beta\in[0,0.04]$ which -- as we have mentioned above -- is separated
from Fig.~\ref{fig:m2412_rh_detail1} in order to avoid clutter
(overlapping curves) due to the fact that the bifurcation point turns
around and starts moving up again, see
Fig.~\ref{fig:m2412_rh_bifurcation_point}.

Indeed the behavior of the curves in Fig.~\ref{fig:m2412_rh_detail2}
is practically monotonic and the most interesting fact is that the
$\beta\to 0$ limit exists.
The catch in this model is that the lower branch possesses a lower ADM
mass than its corresponding upper branch.
In order to verify the stability and be able to claim which of them is
the stable branch of solutions, we will perform a linear stability
analysis in the next section.

\section{Linear stability of the \texorpdfstring{$2+4+12$}{2+4+12} model}\label{sec:linstability}

In all models, except the $2+4+12$ model, the unstable branches had
lower values of the profile function at the horizon, $f_h$, and a
higher ADM mass.
Hence, we have not made a detailed stability analysis for those cases.
The $2+4+12$ model is different. 
In this model, the there is a critical value of $\beta$ for which the
upper (in $f_h$) branches have a higher ADM mass and hence are
unstable (or at best metastable).
For $\beta<\beta^{\rm crit}$ this is the case and the lower (in $f_h$)
branches start to have a lower ADM mass for a finite range in $\rho_h$
than their corresponding upper ones. 
In the limit of $\beta\to 0$, the lower branch retains a lower ADM
mass than the upper branch throughout its entire domain of existence.

Our claim is that there is a phase transition and the upper branch
becomes unstable (or metastable) for a range in BH sizes (horizon
radii). 
In order to back up this claim, we will here make a linear stability
analysis of the model in question.

Starting with the Lagrangian of the $2+4+12$ model and turning on time
dependence of the profile function $f(\rho)\to f(\rho,\tau)$, we can
write the matter part of the action as
\begin{equation}
S^{\rm matter} = 4\pi \int_{\rho_h}^\infty d\rho\; e^\delta
\left[
u\left(\frac{1}{e^{2\delta}C} f_\tau^2 - C f_\rho^2\right)
-w\left(\frac{1}{e^{2\delta}C} f_\tau^2 - C f_\rho^2\right)^2
-v
\right],
\end{equation}
where we have defined the functions
\begin{align}
u &\equiv \rho^2 + 2\beta\sin^2f,\\
w &\equiv \frac{\sin^8f}{\rho^6},\\
v &\equiv \sin^2(f)\left(2 + \frac{\beta\sin^2f}{\rho^2}\right),
\end{align}
and the dimensionless time coordinate
$\tau \equiv a t$, as well as
\beq
\delta \equiv \log N,
\eeq
which will be convenient shortly.

Turning on time dependence of the metric functions
$\delta(\rho)\to \delta(\rho,\tau)$ and $C(\rho)\to C(\rho,\tau)$ does
not alter the metric part of the usual two combinations of the
Einstein equations 
\begin{align}
-\rho C_\rho + 1 - C
&= 2\alpha\bigg[
u C\left(f_\rho^2 + \frac{f_\tau^2}{e^{2\delta}C^2}\right)
+v
+w\left(C^2 f_\rho^4
  - \frac{3f_\tau^4}{e^{4\delta}C^2}
  + \frac{2f_\rho^2 f_\tau^2}{e^{2\delta}}\right)
\bigg],\label{eq:m2412_EEQ1}\\
\frac{1}{\alpha}\delta_\rho &=
\frac{2u}{\rho}\left(f_\rho^2 + \frac{f_\tau^2}{e^{2\delta}C^2}\right)
+\frac{4w}{\rho}\left(C f_\rho^4 - \frac{f_\tau^4}{e^{4\delta}C^3}\right),
\label{eq:m2412_EEQ2}
\end{align}
but it gives and additional equation from the $\tau\rho$ component of the
Einstein equations
\beq
-\frac{C_\tau}{C}
= 2\alpha\rho T_{\tau\rho}
= 4\alpha\left[u + 2w C\left(f_\rho^2 - \frac{f_\tau^2}{e^{2\delta}C^2}\right)
\right] \frac{f_\rho f_\tau}{\rho}.
\label{eq:m2412_EEQtr}
\eeq
Finally, we also need the full time-dependent equation of motion for
the profile function
\begin{align}
\p_\rho\left(e^\delta u C f_\rho\right)
-2\p_\rho\left(e^{-\delta} w f_\rho f_\tau^2\right)
+2\p_\rho\left(e^\delta w C^2 f_\rho^3\right)
-\p_\tau\left(\frac{u}{e^\delta C} f_\tau\right)
+2\p_\tau\left(\frac{w}{e^{3\delta} C^2} f_\tau^3\right) \non
\mathop-2\p_\tau\left(e^{-\delta} w f_\rho^2 f_\tau\right)
+\frac{u_f}{2e^\delta C} f_\tau^2
-\frac{1}{2}e^\delta u_f C f_\rho^2
-\frac{w_f}{2e^{3\delta} C^2} f_\tau^4
-\frac{1}{2} e^\delta w_f C^2 f_\rho^4
+e^{-\delta} w_f f_\rho^2 f_\tau^2 \non
\mathop-\frac{1}{2}e^\delta v_f = 0,
\label{eq:m2412_full_eom}
\end{align}
where $u_f\equiv\frac{\p u}{\p f}$, $u_\rho\equiv\frac{\p u}{\p\rho}$
is the \emph{partial} derivative of $u$ (the total derivative of $u$
is $\frac{d u}{d\rho}=u_\rho + u_f f_\rho$), and similarly for the
other functions. 

Armed with the full time-dependent system of equations, it is now easy
to write down the linearized perturbations.
Hence, let us define
\begin{align}
f(\rho,\tau) &\equiv \bar{f}(\rho) + f'(\rho,\tau), \\
\delta(\rho,\tau) &\equiv \bar{\delta}(\rho) + \delta'(\rho,\tau), \\
C(\rho,\tau) &\equiv \bar{C}(\rho) + C'(\rho,\tau),
\end{align}
where $\bar{f}$, $\bar{N}$ and $\bar{C}$ are the solutions of the
soliton background while $f'$, $N'$ and $C'$ are
fluctuation fields about the background soliton. 
The linearization greatly simplifies especially the high powers of the
time derivatives since
\beq
f_\tau = f'_\tau,
\eeq
and hence any power larger than one will eliminate the term.

Following Refs.~\cite{Heusler:1991xx,Heusler:1992av,Shiiki:2005pb}, we
start by determining the perturbation of the radial metric function,
$C'$, by integrating
Eq.~\eqref{eq:m2412_EEQtr},
\beq
-\rho C' =
4\alpha\left(\bar{u} + 2\bar{w}\bar{C}\bar{f}_\rho^2\right)
\bar{C}\bar{f}_\rho f' + q(\rho),
\label{eq:m2412_Cprime}
\eeq
where $\bar{u}$ means that $u(f)$ is evaluated with the background
field: $u(\bar{f})$, and $q(\rho)$ is an integration constant.
We will now prove that the integration constant vanishes.
In order to do so, we will first combine the two Einstein
equations \eqref{eq:m2412_EEQ1} and \eqref{eq:m2412_EEQ2} to get
\beq
-\rho C_\rho + 1 - C - \rho C \delta_\rho
= 2\alpha\left[v
  - w\left(C f_\rho^2 - \frac{f_\tau^2}{e^{2\delta}C}\right)^2\right],
\eeq
whose linearization reads
\beq
-\p_\rho\left(\rho e^{\bar{\delta}} C'\right)
=\rho\bar{C}e^{\bar{\delta}}\delta'_\rho
+2\alpha e^{\bar{\delta}}\left[
  \bar{v}_f f'
  -\bar{w}_f\bar{C}^2\bar{f}_\rho^4 f'
  -2\bar{w}\bar{C}\bar{f}_\rho^4 C'
  -4\bar{w}\bar{C}^2\bar{f}_\rho^3 f'_\rho
\right].
\label{eq:m2412_linearized_comb_EEQ}
\eeq
Next, we will linearize Eq.~\eqref{eq:m2412_EEQ2}:
\beq
\frac{\rho}{2\alpha}\delta'_\rho =
\bar{u}_f \bar{f}_\rho^2 f'
+2\bar{u}\bar{f}_\rho f'_\rho
+2\bar{w}_f \bar{C} \bar{f}_\rho^4 f'
+2\bar{w} C' \bar{f}_\rho^4
+8\bar{w} \bar{C} \bar{f}_\rho^3 f'_\rho,
\eeq
and insert this as well as the equation of
motion \eqref{eq:m2412_full_eom} evaluated on the background soliton
(i.e.~$\p_\tau=0$) into Eq.~\eqref{eq:m2412_linearized_comb_EEQ},
yielding
\beq
-\p_\rho\left(\rho e^{\bar{\delta}} C'\right)
= 4\alpha\p_\rho\left(e^{\bar{\delta}}\bar{u}\bar{C}\bar{f}_\rho f'\right)
+8\alpha\p_\rho\left(e^{\bar{\delta}}\bar{w}\bar{C}^2\bar{f}_\rho^3 f'\right),
\eeq
which when integrated yields
\beq
-\rho C' = 4\alpha\left(
\bar{u} + 2\bar{w}\bar{C}\bar{f}_\rho^2\right)
\bar{C}\bar{f}_\rho f' + e^{-\bar{\delta}}\tilde{q}(\tau).
\eeq
Comparing the above equation with Eq.~\eqref{eq:m2412_Cprime}, we can
conclude that the integration constant cannot be time dependent.
Using now that the appropriate boundary conditions for the
fluctuations are that all fluctuations vanish at spatial infinity, we
can finally conclude that $q=\tilde{q}=0$.

Finally, we need the linearized equation of motion for the
perturbation of the profile function, $f'$.
After some massage, we can write the full perturbation in terms of
that of the profile function as
\begin{align}
\p_\rho\left[e^{\bar{\delta}}\left(\bar{u}
  +6\bar{w}\bar{C}\bar{f}_\rho^2\right) \bar{C} f'_\rho\right]
- \bar{U} f'
= \left(\bar{u} + 2\bar{w}\bar{C}\bar{f}_\rho^2\right)
  \frac{1}{e^{\bar{\delta}}C} f'_{\tau\tau},
\label{eq:m2412_lin_perturb_f}
\end{align}
where the potential is
\begin{align}
\bar{U} &\equiv
-\p_\rho\left(e^{\bar{\delta}}\bar{u}_f\bar{C}\bar{f}_\rho\right)
-2\p_\rho\left(e^{\bar{\delta}}\bar{w}_f\bar{C}^2\bar{f}_\rho^3\right)
+\frac{1}{2}e^{\bar{\delta}}\bar{v}_{ff}
+\frac{1}{2}e^{\bar{\delta}}\bar{u}_{ff}\bar{C}\bar{f}_\rho^2
+\frac{1}{2}e^{\bar{\delta}}\bar{w}_{ff}\bar{C}^2\bar{f}_\rho^4 \non
&\phantom{\equiv\ }
+4\alpha e^{-\bar{\delta}}\p_\rho\left(\frac{e^{2\bar{\delta}}\bar{u}^2\bar{C}\bar{f}_\rho^2}{\rho}\right)
+24\alpha e^{-\bar{\delta}}\p_\rho\left(\frac{e^{2\bar{\delta}}\bar{u}\bar{w}\bar{C}^2\bar{f}_\rho^4}{\rho}\right)
+32\alpha e^{-\bar{\delta}}\p_\rho\left(\frac{e^{2\bar{\delta}}\bar{w}^2\bar{C}^3\bar{f}_\rho^6}{\rho}\right) \non
&\phantom{\equiv\ }
-\frac{4\alpha e^{\bar{\delta}}\bar{u}\bar{u}_f\bar{C}\bar{f}_\rho^3}{\rho}
-\frac{8\alpha e^{\bar{\delta}}\bar{w}\bar{u}_f\bar{C}^2\bar{f}_\rho^5}{\rho}
-\frac{8\alpha e^{\bar{\delta}}\bar{u}\bar{w}_f\bar{C}^2\bar{f}_\rho^5}{\rho}
-\frac{16\alpha e^{\bar{\delta}}\bar{w}\bar{w}_f\bar{C}^3\bar{f}_\rho^7}{\rho}\non
&\phantom{\equiv\ }
-\frac{16\alpha^2 e^{\bar{\delta}}}{\rho^2}\left(
\frac{1}{2}\bar{u}^3\bar{C}\bar{f}_\rho^4
+3\bar{u}^2\bar{w}\bar{C}^2\bar{f}_\rho^6
+6\bar{u}\bar{w}^2\bar{C}^3\bar{f}_\rho^8
+4\bar{w}^3\bar{C}^4\bar{f}_\rho^{10}
\right),
\label{eq:m2412_lin_perturb_f_pot}
\end{align}
where we have used the equation of motion for the background field,
$\bar{f}$, to eliminate $\bar{v}_f$. 

We will now set 
\beq
f'(\rho,\tau) = \xi(\rho) e^{i\omega\tau},
\eeq
for which Eq.~\eqref{eq:m2412_lin_perturb_f} is a regular
Sturm-Liouville problem with a nontrivial (non-constant) weight or
density function
\beq
\p_\rho\left[e^{\bar{\delta}}\left(\bar{u}
  +6\bar{w}\bar{C}\bar{f}_\rho^2\right) \bar{C} \xi_\rho\right]
- \bar{U} \xi
= -\left(\bar{u} + 2\bar{w}\bar{C}\bar{f}_\rho^2\right)
  \frac{1}{e^{\bar{\delta}}C} \omega^2 \xi.
\label{eq:m2412_SL}
\eeq
The right-hand side of the above equation is the weight function of
the Sturm-Liouville problem and $\omega^2\in\mathbb{R}$ is the
eigenvalue and by Sturm-Liouville theory it has to be real.
Physically, if $\omega^2<0$ then $\omega$ is imaginary and the
perturbation mode signals an instability at the linear level.
Full nonlinear stability requires the absence of linear instabilities,
although that is generally not sufficient for claiming nonlinear
stability.
In this paper, we will content ourselves with the above linear
stability analysis.

In the case of the standard Skyrme model, which corresponds to the
case of $w=0$, the stability is considerably simpler and the
Sturm-Liouville problem can be transformed into a free eigenvalue
problem with a complicated potential by setting
$\xi=\zeta/\sqrt{\bar{u}}$ and using tortoise coordinates defined by
$dx=d\rho/(e^\delta C)$ \cite{Shiiki:2005pb}. 
The latter transformation is able to simplify the problem
\beq
\p_x\left(p(x) \xi_x\right) - q(x)\xi = -\omega^2 r(x)\xi,
\eeq
considerably because the weight function, $r(x)$, and the kernel
function, $p(x)$, are equal to each other.

In our case with $w\neq 0$, the kernel and the weight functions are
different and we have not been able to find a suitable transformation
to simplify the problem. 
We will therefore solve the Sturm-Liouville
problem \eqref{eq:m2412_SL} directly using a numerical finite
difference method.

The reason why the kernel and weight functions are different in our
higher-order model as compared to the standard Skyrme model is due to
the linearization of a higher-order derivative operator on a static
background.
That is, the linearization of $f_\rho^4$ yields
$4\bar{f}_\rho^3f'_\rho$, whereas the linearization of $f_\tau^4$
vanishes because of the static background.
The higher-order term still gives a contribution to the time
derivative of the fluctuation, but it comes from a cross term,
$f_\rho^2f_\tau^2$ and hence does not give the same factor as the
radial derivative of the fluctuation does.

\begin{figure}[!htp]
\begin{center}
\includegraphics[width=0.5\linewidth]{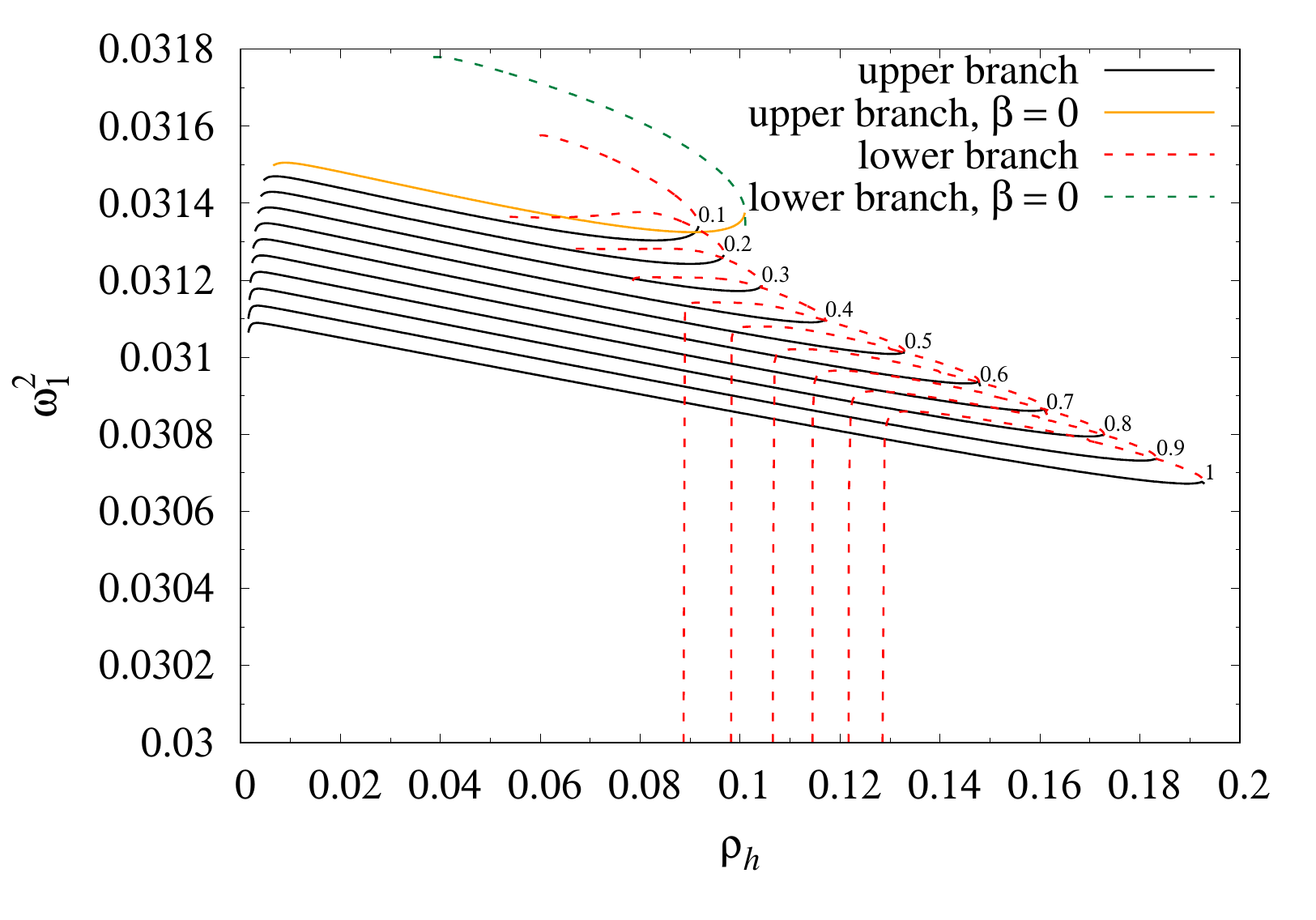}
\caption{The lowest eigenvalue, $\omega_1^2$, of the linear
  perturbation of the profile function, $f'$, in the $2+4+12$ model. 
  The numbers on the figures indicate the different values of
  $\beta=0,0.1,\ldots,1$.
  The upper and lower $\beta=0$ branches are shown in orange and dark
  green colors, respectively. }
\label{fig:m2412_rh_omegasq}
\end{center}
\end{figure}

We are now ready to present the numerical results for the lowest
eigenvalue, $\omega_1^2$, of the perturbation of the profile function,
$f'$, shown in Fig.~\ref{fig:m2412_rh_omegasq}.
If we start with the upper branches, they start off at a small horizon
radius, $\rho_h$, with a positive lowest eigenvalue ($\omega_1^2>0$),
which decreases as the horizon radius is increased.
We would expect the eigenvalue to go towards zero and at the
bifurcation point meet a negative eigenvalue coming from below zero.
This does not happen in this model.
If we start with the small values of $\beta\in[0,0.4]$, the behavior
is as follows.
Instead the lowest eigenvalue of the upper branch goes upwards and
meets that of the lower branch at the bifurcation point. The lower
branch hence does not have a negative mode and the lowest eigenvalue
of the fluctuation for the lower branch is higher than that of the
upper branch.

Now, when $\beta$ is increased to above about $0.5$, the lower branch
has a hybrid behavior. Instead of having a negative lowest eigenvalue
of the fluctuation, it emanates from bifurcation point over that of
the upper branch -- stays there for a finite range in horizon radius
-- and then turns negative shortly before the branch terminates at a
smaller horizon radius, see Fig.~\ref{fig:m2412_rh_omegasq}.
There are only quite few data points with a negative eigenvalue of the
fluctuation, but enough to conclude that the lower branch develops an
unstable mode for large enough $\beta\gtrsim 0.5$.

The fact that the lower branch (in $f_h$) does not have an unstable
mode over its entire domain of existence is indeed a surprise.
Since the analysis carried out here is limited to a \emph{linear}
stability analysis, this does not rule out a nonlinear instability.
Physically, the absence of an unstable mode in the fluctuation
spectrum could be interpreted as the lower branch being a \emph{local}
minimum or vice versa.
In order to determine which local minimum is the global minimum of
solution space, we would still turn to the ADM mass.
Our explanation at this point is simply that the highly nonlinear
nature of our model has created a situation where the two branches
both become local minima.
Except for very particular points in parameter space, only one of them
is a global minimum.

\begin{figure}[!thp]
\begin{center}
\includegraphics[width=0.5\linewidth]{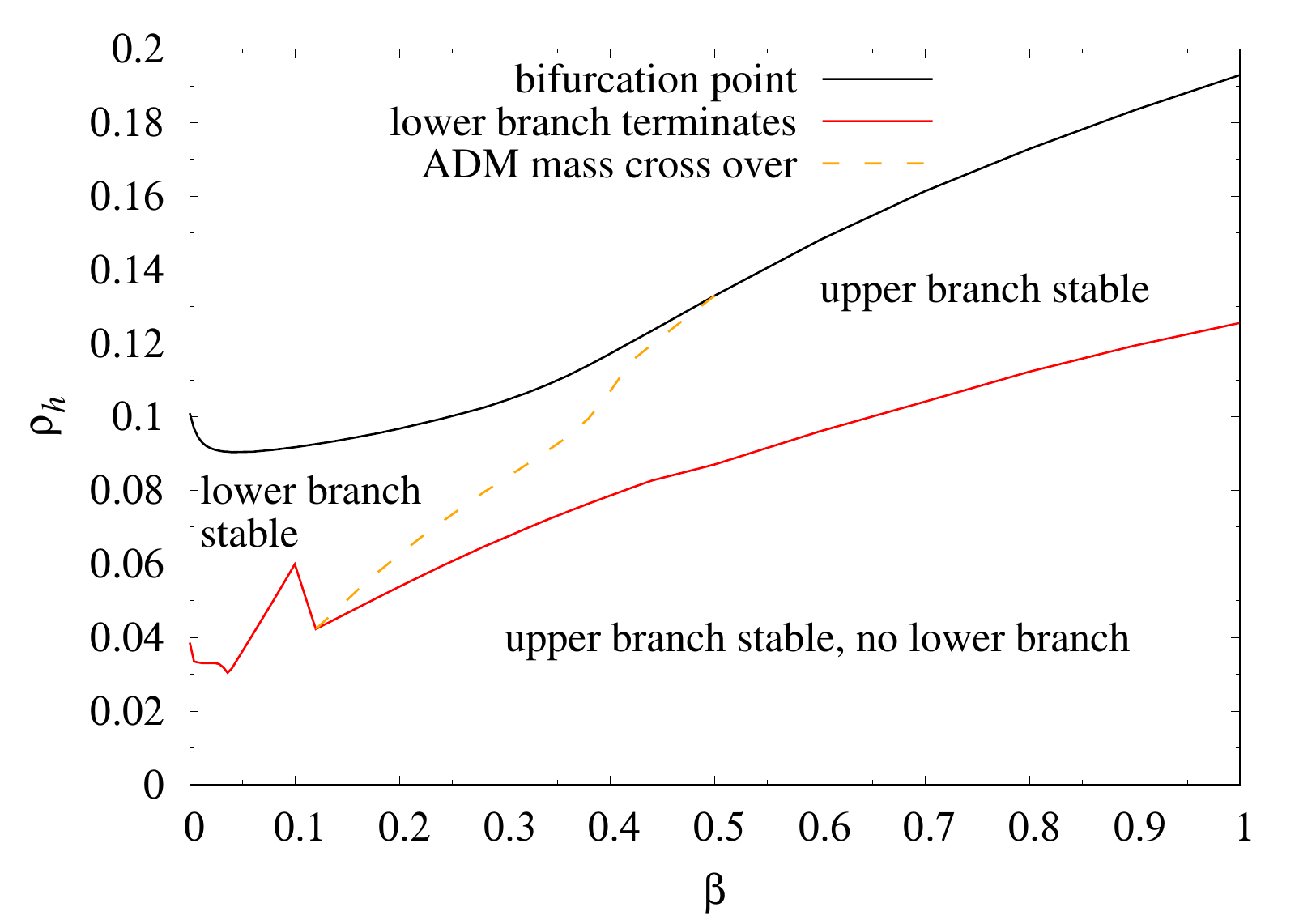}
\caption{Phase diagram for the $2+4+12$ model based on the lowest ADM
  mass.
}
\label{fig:m2412_phasediagram}
\end{center}
\end{figure}

With the above discussion in mind, we will now make a phase diagram of
the $2+4+12$ model, see Fig.~\ref{fig:m2412_phasediagram}.
The expected features are the radius of the bifurcation point (black
upper curve) and the point where the lower branch terminates (red
lower curve). The crossover from where the lower branch has a lower
ADM mass happens at $\beta\lesssim 0.5$ and the curve decreases in
$\beta$ as the horizon radius, $\rho_h$, gets smaller (orange dashed
curve).

Since it is puzzling that the lower branches do not possess the
expected unstable modes, we will consider the limit of turning off the
twelfth-order term (but keeping the Skyrme term) in the $2+4+12$ model
in the next section.

\section{Taking the Skyrme model limit of
the \texorpdfstring{$2+4+12$}{2+4+12} model}\label{sec:Skyrme_model_limit}

In this section we will take the limit of the $2+4+12$ model becoming
the standard Skyrme model; i.e.~turning off the twelfth-order
derivative term.
In order to do so, we need to rescale the Lagrangian such that the
free parameter is the coefficient of the twelfth-order term and not
the Skyrme term.

The model is defined as
\begin{align}
\Lag &= \Lag_2 + \Lag_4 + \Lag_{12} + \frac{R}{16\pi G}\non
&= -c_2\cC_1 - \frac{c_4}{2}\cC_2 - \frac{c_{12}}{9}\cC_3^2
  + \frac{R}{16\pi G},
\end{align}
where the $\cC$s are defined in Eqs.~\eqref{eq:C1}-\eqref{eq:C3}.
Similarly to Sec.~\ref{sec:m242n}, we will now switch to dimensionless
units, $\rho\equiv a r$, for which we get
\beq
\Lag = a^3 M_0\left(
-\frac{c_2}{a M_0}\cC_1
-\frac{a c_4}{2M_0}\cC_2
-\frac{a^9 c_{12}}{9M_0}\cC_3^2
+\frac{R}{4\alpha}
\right),
\eeq
where the effective (dimensionless) gravitational coupling is still given by
Eq.~\eqref{eq:alpha_def} and $M_0$ is the mass scale of the soliton. 
Instead of letting $c_4$ be the free parameter, we will fix the
coefficient of $\Lag_4$ and let $c_{12}$ be the free parameter.
This is done by
\beq
a M_0 = c_2, \qquad
\frac{M_0}{a} = c_4,
\eeq
which gives
\beq
a = \sqrt{\frac{c_2}{c_4}}, \qquad
M_0 = \sqrt{c_2c_4},
\eeq
which both have mass dimension 1 as they must.

The rescaled Lagrangian is now written as
\beq
\Lag = a^3M_0\left(
-\cC_1
-\frac{1}{2}\cC_2
-\frac{\eta}{9}\cC_3^2
+\frac{R}{4\alpha}
\right),
\eeq
where the new parameter, $\eta$ is defined as
\beq
\eta \equiv c_{12}\frac{c_2^4}{c_4^5}.
\eeq
In the limit of $\eta\to 0$, this model becomes the Skyrme model.

The soliton mass then reads
\beq
M = 4\pi M_0\int_{\rho_h}^\infty d\rho \; \rho^2 N\left(
\cC_1
+\frac{1}{2}\cC_2
+\frac{\eta}{9}\cC_3^2
\right).
\eeq
The dimensionless Einstein equations read
\begin{align}
-C_\rho + \frac{1}{\rho} - \frac{C}{\rho} &=
2\alpha\left[
\rho C f_\rho^2
+\frac{2\sin^2f}{\rho}
+\frac{2\sin^2(f) C f_\rho^2}{\rho}
+\frac{\sin^4f}{\rho^3}
+\frac{\eta\sin^8(f) C^2 f_\rho^4}{\rho^7}
\right],\non
\frac{1}{\alpha}\frac{N_\rho}{N} &=
2\rho f_\rho^2
+\frac{4\sin^2(f)f_\rho^2}{\rho}
+\frac{4\eta\sin^8(f) C f_\rho^4}{\rho^7},
\end{align}
and the equation of motion becomes
\begin{align}
C f_{\rho\rho}
+\frac{2C f_\rho}{\rho}
+C_\rho f_\rho
+\frac{N_\rho C f_\rho}{N}
-\frac{\sin 2f}{\rho^2} \non
+\frac{2\sin^2f}{\rho^2}\left(
 C f_{\rho\rho}
 + C_\rho f_\rho
 + \frac{N_\rho C f_\rho}{N}
 - \frac{\sin 2f}{2\rho^2}
 \right)
+\frac{\sin(2f) C f_\rho^2}{\rho^2}
\non
+\frac{2\eta\sin^8(f) C f_\rho^2}{\rho^8}\left(
 3C f_{\rho\rho}
 -\frac{6C f_\rho}{\rho}
 +2C_\rho f_\rho
 +\frac{N_\rho C f_\rho}{N}
 \right)
+\frac{6\eta\sin^6(f)\sin(2f) C^2 f_\rho^4}{\rho^8}
= 0.
\end{align}
Finally, the boundary conditions are simply given by
Eqs.~\eqref{eq:muprime_12}-\eqref{eq:Xi12} with $\beta=1$ (they do not
depend on $\eta$).

\begin{figure}[!thp]
\begin{center}
\mbox{\subfloat[]{\includegraphics[width=0.49\linewidth]{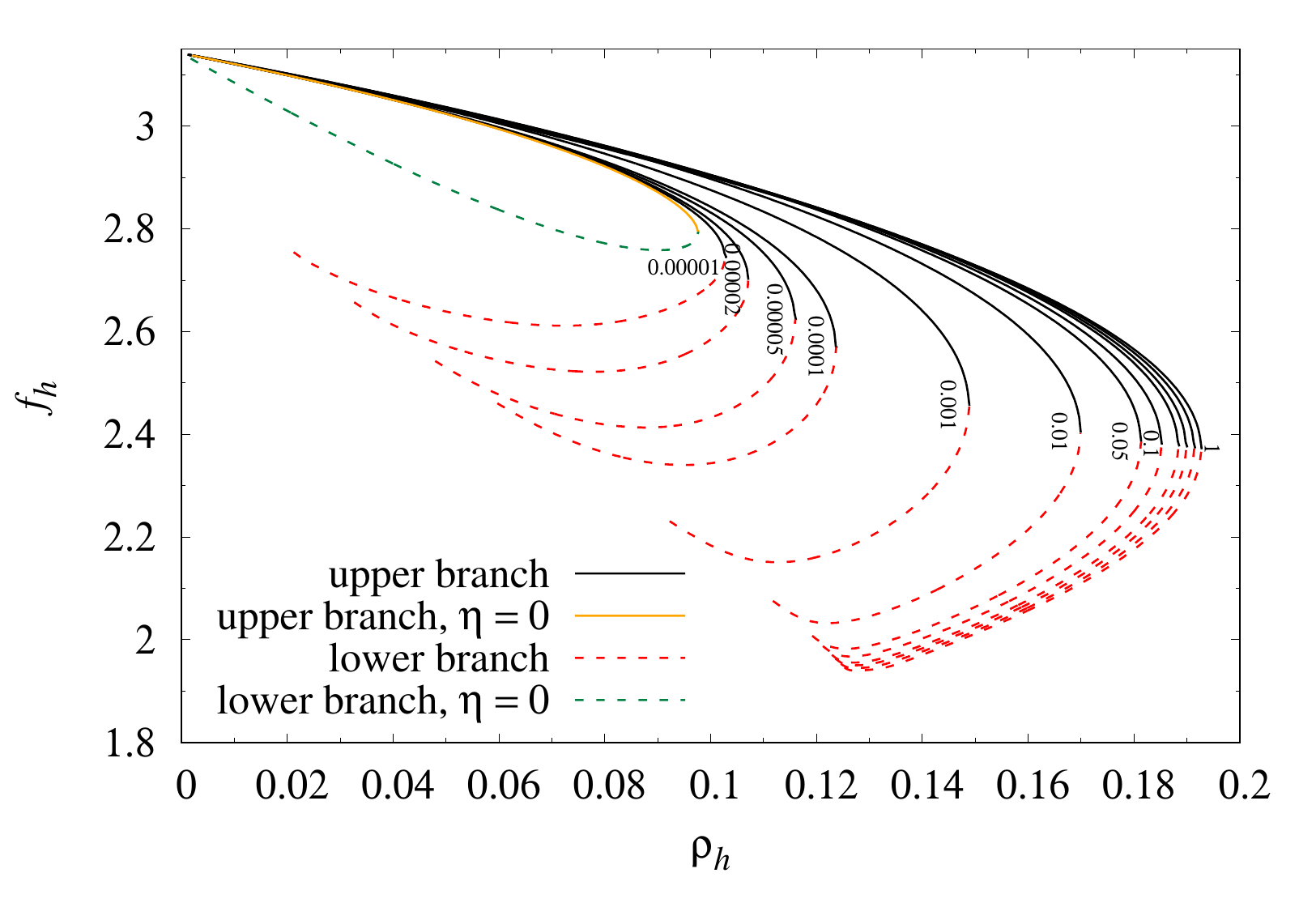}}
\subfloat[]{\includegraphics[width=0.49\linewidth]{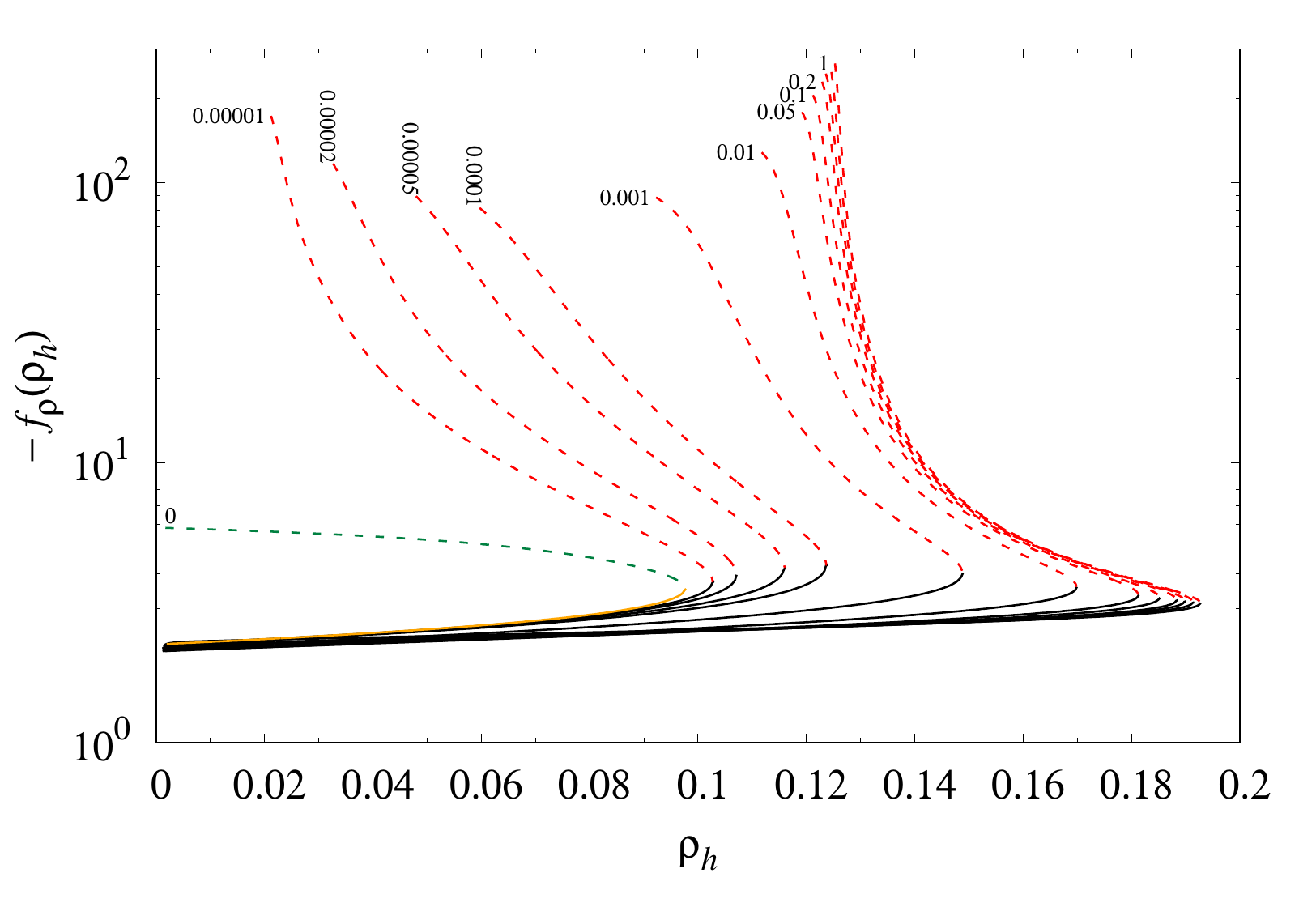}}}
\mbox{\subfloat[]{\includegraphics[width=0.49\linewidth]{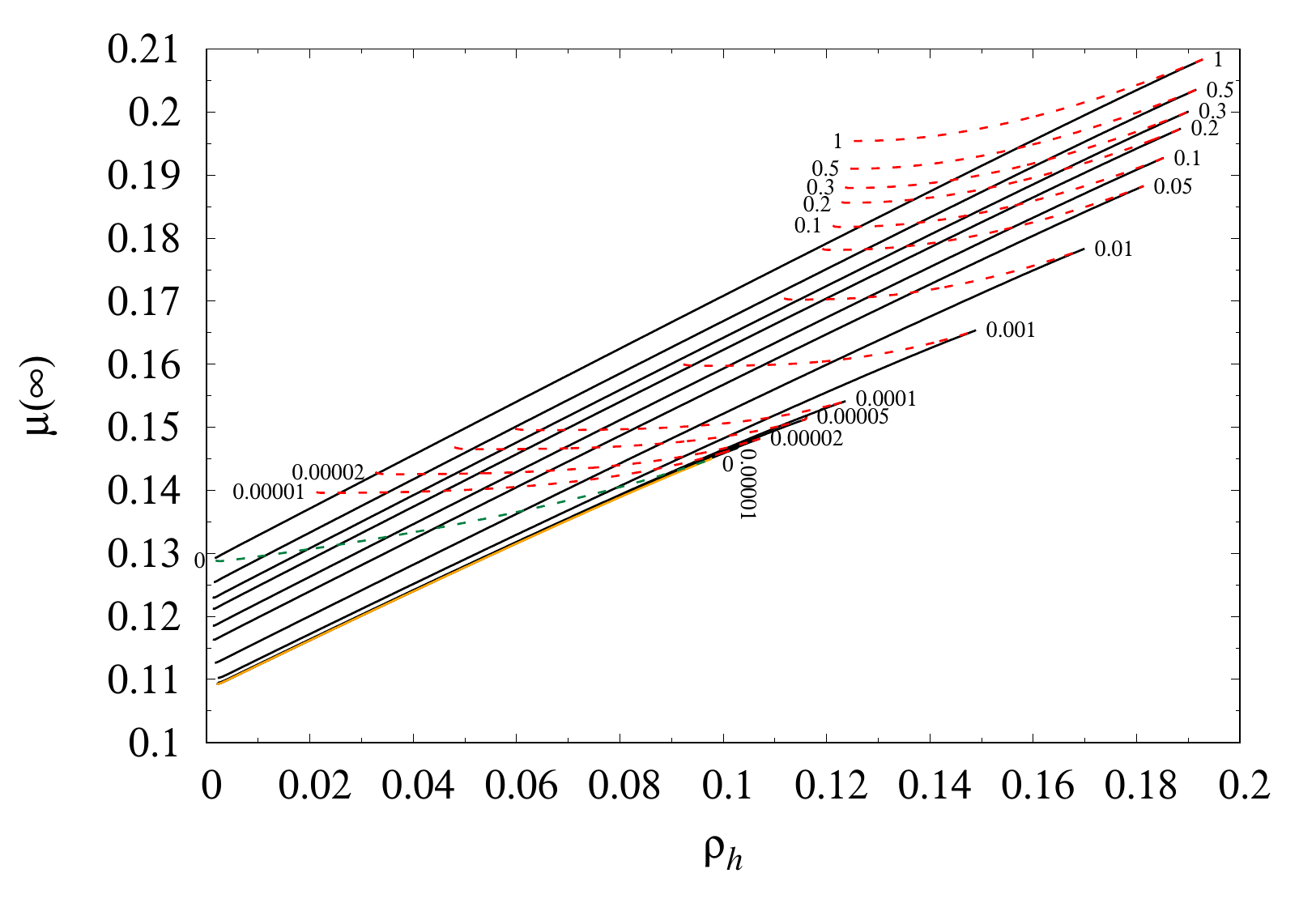}}
\subfloat[]{\includegraphics[width=0.49\linewidth]{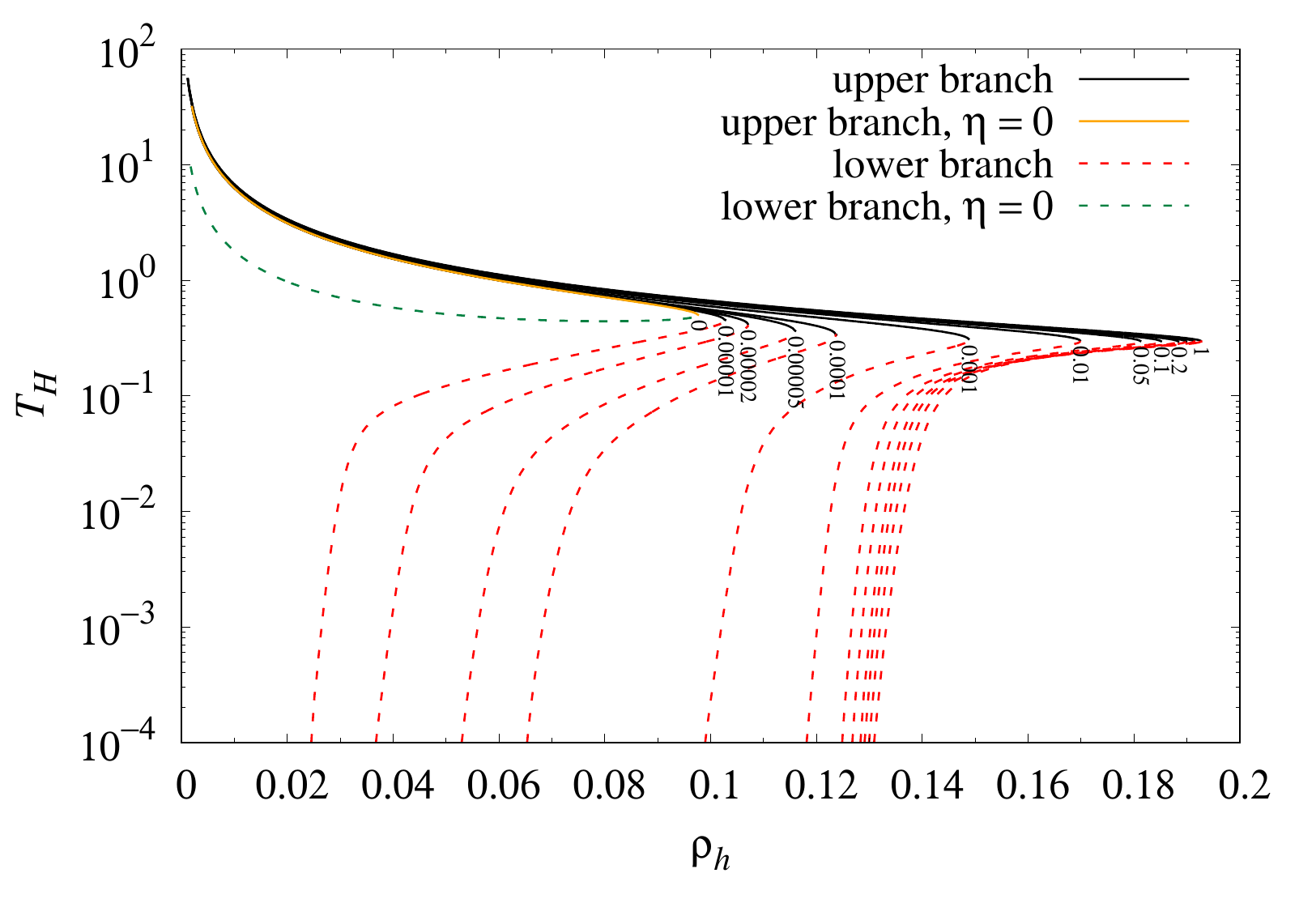}}}
\caption{Upper (black solid lines) and lower (red dashed lines)
  branches of solutions in the $2+4+12$ model, where instead of
  decreasing $\beta$, we will send $\eta$ to zero: (a) the value of
  the profile function at the horizon, $f_h$; (b) the derivative of
  the profile function at the horizon, $f_\rho(\rho_h)$; (c) the ADM
  mass, $\mu(\infty)$; (d) the Hawking temperature $T_H$, all as
  functions of the size of the black hole, i.e.~the horizon radius,
  $\rho_h$. 
  The numbers on the figures indicate the different values of
  $\eta=0$, $10^{-5}$, $2\times 10^{-5}$, $5\times 10^{-5}$,
  $10^{-4}$, $0.001$, $0.01$, $0.05$, $0.1$, $0.2$, $0.3$,
  $0.5$, $1$.
  The upper and lower $\eta=0$ branches are shown in orange and dark
  green colors, respectively. 
}
\label{fig:m2412_c12_rh}
\end{center}
\end{figure}

\begin{figure}[!thp]
\begin{center}
\mbox{\subfloat[]{\includegraphics[width=0.49\linewidth]{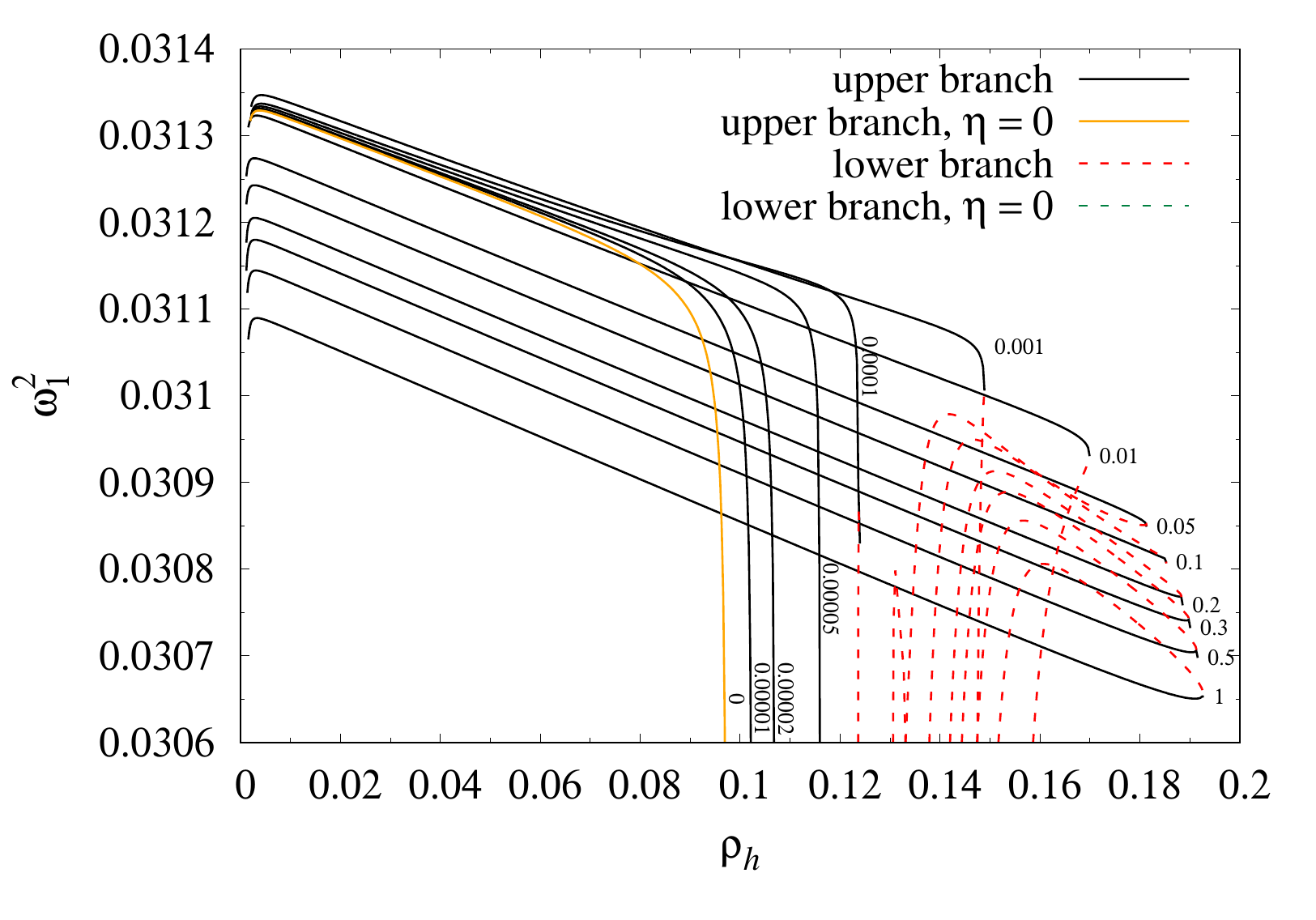}}
\subfloat[]{\includegraphics[width=0.49\linewidth]{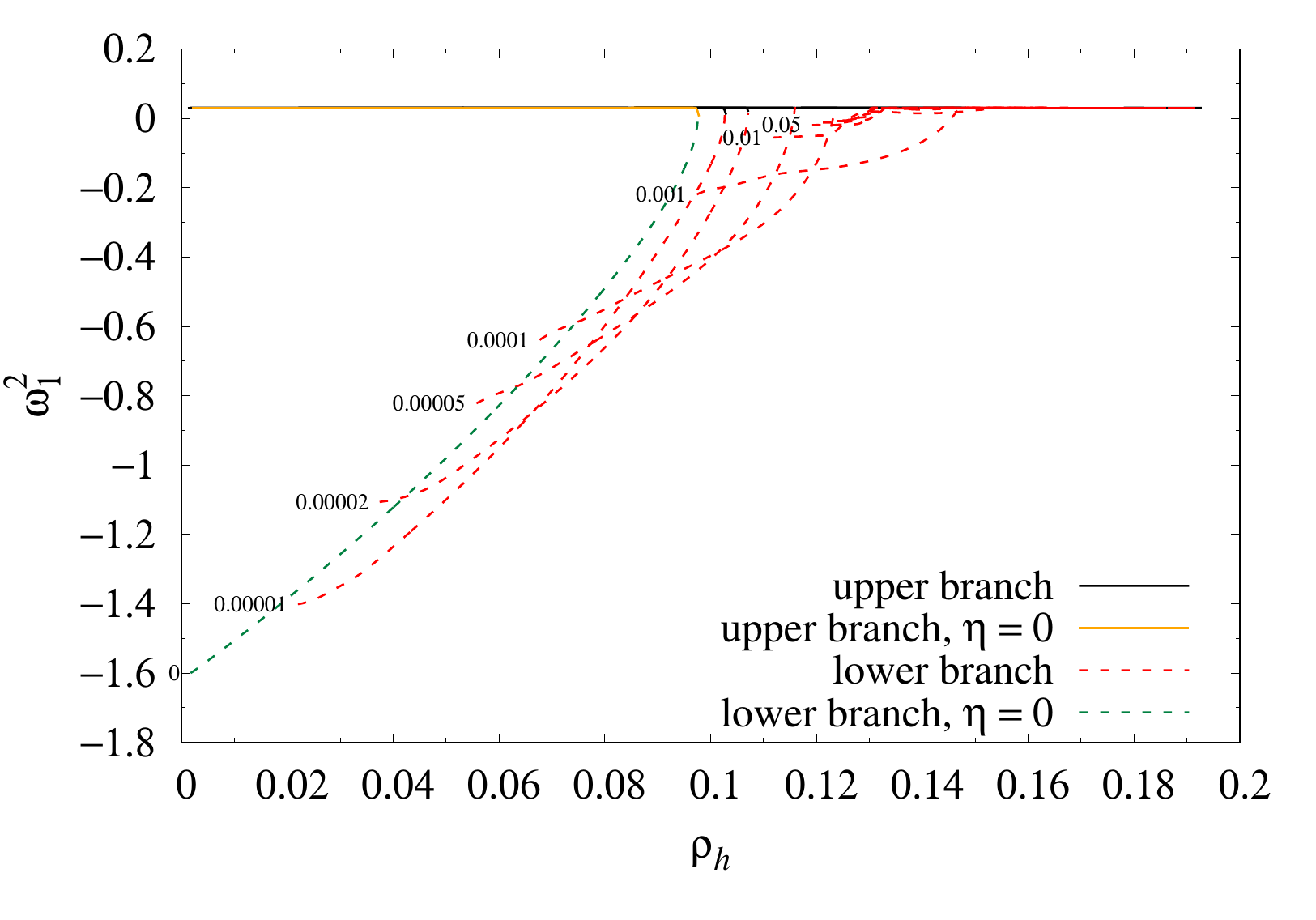}}}
\caption{The lowest eigenvalue, $\omega_1^2$, of the linear
  perturbation of the profile function, $f'$, in the $2+4+12$ model
  with $\eta\to 0$.
  The numbers on the figures indicate the different values of
  $\eta=0$, $10^{-5}$, $2\times 10^{-5}$, $5\times 10^{-5}$,
  $10^{-4}$, $0.001$, $0.01$, $0.05$, $0.1$, $0.2$, $0.3$,
  $0.5$, $1$.
  The upper and lower $\eta=0$ branches are shown in orange and dark
  green colors, respectively. 
}
\label{fig:m2412_c12_rh_omegasq}
\end{center}
\end{figure}

The numerical results for this model, i.e.~the $2+4+12$ model with
variable $\eta$ instead of $\beta$ (variable twelfth-order term
coefficient instead of Skyrme-term coefficient), are shown in 
Fig.~\ref{fig:m2412_c12_rh}.
The standard quantities are displayed in the usual four panels.
We are interested in what happens in the $\eta\to 0$ limit, for which
the model becomes the standard Skyrme model (coupled to gravity).
First we note that varying $\eta$ in the range $[0.1,1]$ has little
effect on both the upper and lower branches, see
Fig.~\ref{fig:m2412_c12_rh}(a).
The approach from the branch with $\eta\sim 1$ to the Skyrme model is
almost logarithmic in $\eta$.
The $\eta=0$ (i.e.~the standard Skyrme model limit) branches of
solutions are distinct from the nonzero $\eta$ ones by the fact that
the lower branch of the $\eta=0$ returns smoothly to $f\sim\pi$ in the
limit of vanishing horizon radius, $\rho_h\to 0$, see the dark green
dashed curve in Fig.~\ref{fig:m2412_c12_rh}(a). 
Once a tiny nonzero $\eta$ is turned on, the lower branch does not
return all the way when $\rho_h\to 0$, but terminates at a finite
horizon radius.
The derivative of the profile function at the horizon also behaves
differently for the lower $\eta=0$ branch as compared to the nonzero
$\eta$ branches, see Fig.~\ref{fig:m2412_c12_rh}(b).
The lower $\eta=0$ branch remains almost constant in the $\rho_h\to 0$ 
limit, whereas the lower nonzero $\eta$ branches raise up (in negative 
values) about two orders of magnitude before they terminate at a finite
horizon radius.
A final distinction between the lower $\eta=0$ branch and the nonzero
ones is seen in Fig.~\ref{fig:m2412_c12_rh}(d), where the Hawking
temperature goes smoothly back from the bifurcation point in the
$\rho_h\to 0$ limit. 
The lower nonzero $\eta$ branches on the other hand, drop very
drastically and suddenly in temperature where they terminate at a
finite horizon radius. 

The ADM masses for all the lower branches in this model are larger
than all those of the upper branches, which is the expected (from the
standard Skyrme model scenario) behavior, see
Fig.~\ref{fig:m2412_c12_rh}(c).
So far, except for the distinct behavior of the $\eta=0$ branch,
everything seems in line with the results for the $2+4+6$ model
studied in Refs.~\cite{Gudnason:2016kuu,Adam:2016vzf}.
The expectation from the standard Skyrme model is that the lower
branches have larger ADM masses than the upper branches, and they also
possess a negative lowest eigenvalue in the linear fluctuation
spectrum -- signaling an instability at the linear level. 

Now we will consider the lowest eigenvalue for this model, see
Fig.~\ref{fig:m2412_c12_rh_omegasq}.
Although all the ADM masses for the lower branches are higher, they do
not all have a negative eigenvalue in their linear fluctuation
spectrum over the entire range of horizon radii.
If we start with the $\eta=0$ branch, everything is as expected. The
upper branch has a positive eigenvalue which drops suddenly near the
bifurcation point (see Fig.~\ref{fig:m2412_c12_rh_omegasq}(a)) and the
lower branch has one single negative mode all the way in $\rho_h$ and
it only increases near the bifurcation point to meet with that of the
upper branch. 
If we turn on a very tiny $\eta$ in the range $[10^{-5},10^{-4}]$, the
trend continues as just described.
Then for $\eta$ around $10^{-3}$ a transition occurs and the
eigenvalue for the upper branch no longer drops to zero near the
bifurcation point. Instead the lower branch now possesses only
(linearly) stable modes for a finite range in $\rho_h$ from a finite
value larger than where the branch terminates, up to the bifurcation
point.
For $\eta=10^{-3}$, the upper branch still has the largest eigenvalue
compared to the lower branch, but that quickly changes and for
$\eta\gtrsim 0.01$, there is a finite range in $\rho_h$ where the
eigenvalue of the lower branch is larger than that of the upper
branch. After this range, the eigenvalue of the lower branch drops
suddenly to negative values where it remains until the branch
terminates at a finite horizon radius, see
Fig.~\ref{fig:m2412_c12_rh_omegasq}(b).

\section{The \texorpdfstring{$2+2n$}{2+2n} model}\label{sec:m22n}

In this section, we will study the dependence of the existing models
with  $\beta=0$ on the effective gravitational
coupling \eqref{eq:alpha_def}.
These models are thus new Skyrme-type models that possess BH hair
without having the Skyrme term component in the Lagrangian density. 
The $2+2n$ model is the $\beta\to 0$ limit of the $2+4+2n$ model of
Sec.~\ref{sec:m242n}.
Only two models survive in the limit, namely the $2+4+8$ model with 
$\gamma\neq 0$ and the $2+4+12$ model.
The $\gamma\neq 0$ condition in the $2+4+8$ model corresponds to
having a nonvanishing $c_{8|4,4}$ term in the Lagrangian density.
Here we will consider the $2+8$ model with $\gamma=1$ and
$\gamma=\frac13$ as well as the $2+12$ model.
The $\gamma=\frac13$ case of the $2+8$ model  corresponds to the
higher-derivative term being the Skyrme term squared, whereas the
$\gamma=1$ case is the three curvatures of the Skyrme term
individually squared without the corresponding cross terms. 
After rescaling, and in the case of the $2+8$ model, fixing $\gamma$,
the effective gravitational coupling is the only parameter of the
model.
As the $\beta=0$ branches for fixed $\alpha=0.01$ have been studied
already in Sec.~\ref{sec:m242n}, here we will consider only the
standard quantities as functions of $\alpha$, for a few different BH
sizes (different horizon radii). 
The equations of motion and the boundary conditions for the $2+2n$
models with $n=4$ and $n=6$ are simply given in Sec.~\ref{sec:m242n}
with $\beta=0$. Thus we will not repeat them here.

\begin{figure}[!thp]
\begin{center}
\mbox{\subfloat[]{\includegraphics[width=0.49\linewidth]{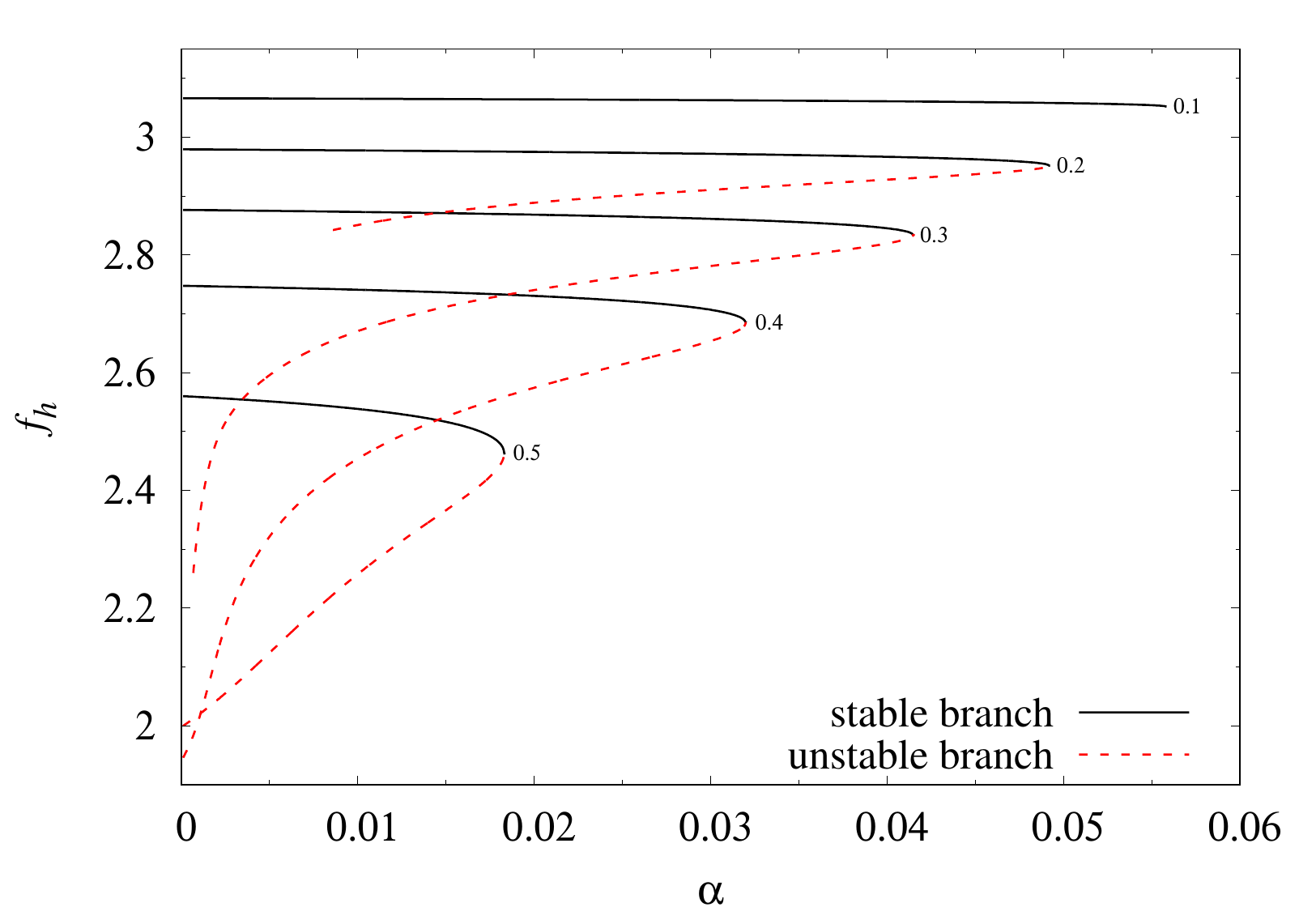}}
\subfloat[]{\includegraphics[width=0.49\linewidth]{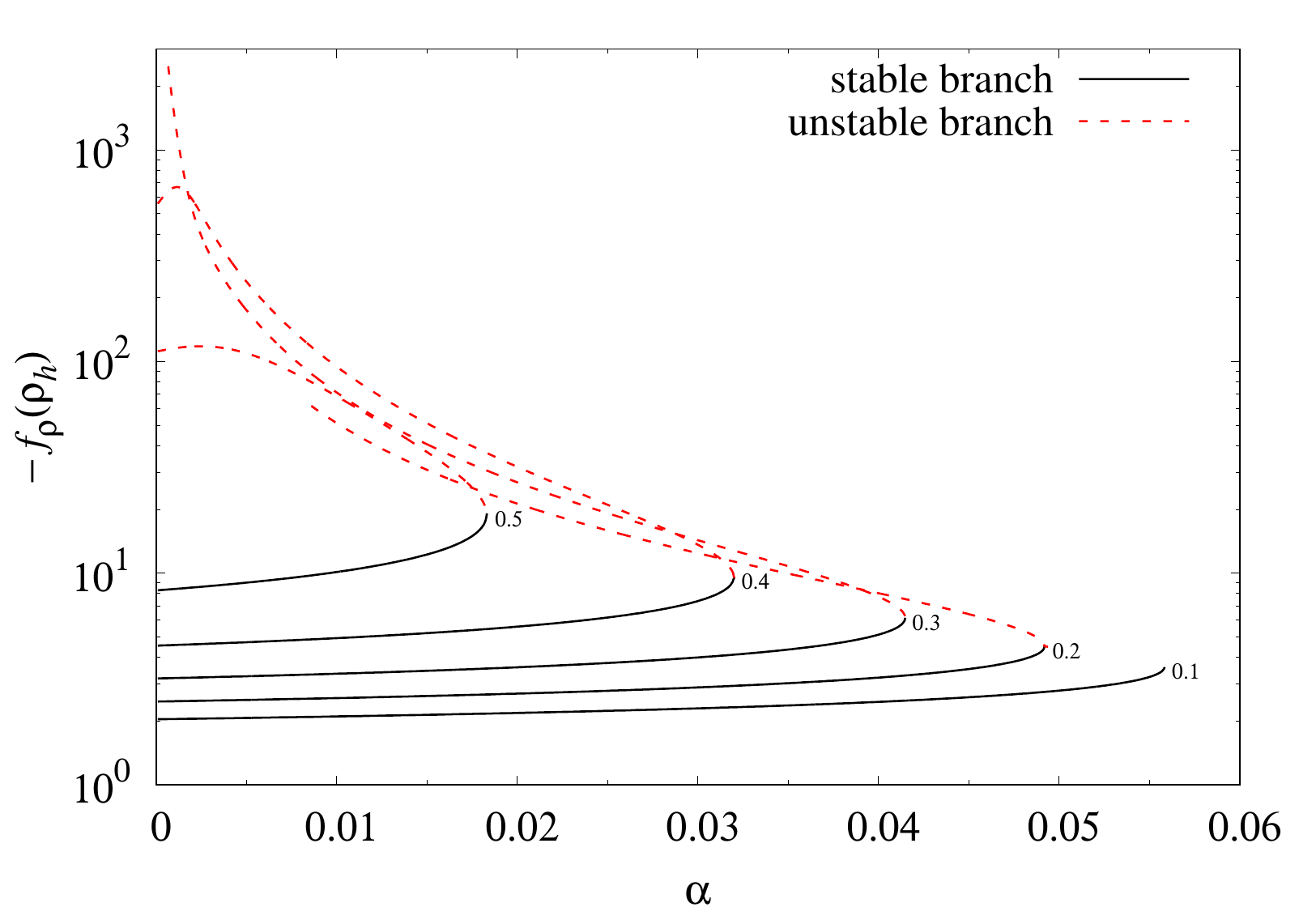}}}
\mbox{\subfloat[]{\includegraphics[width=0.49\linewidth]{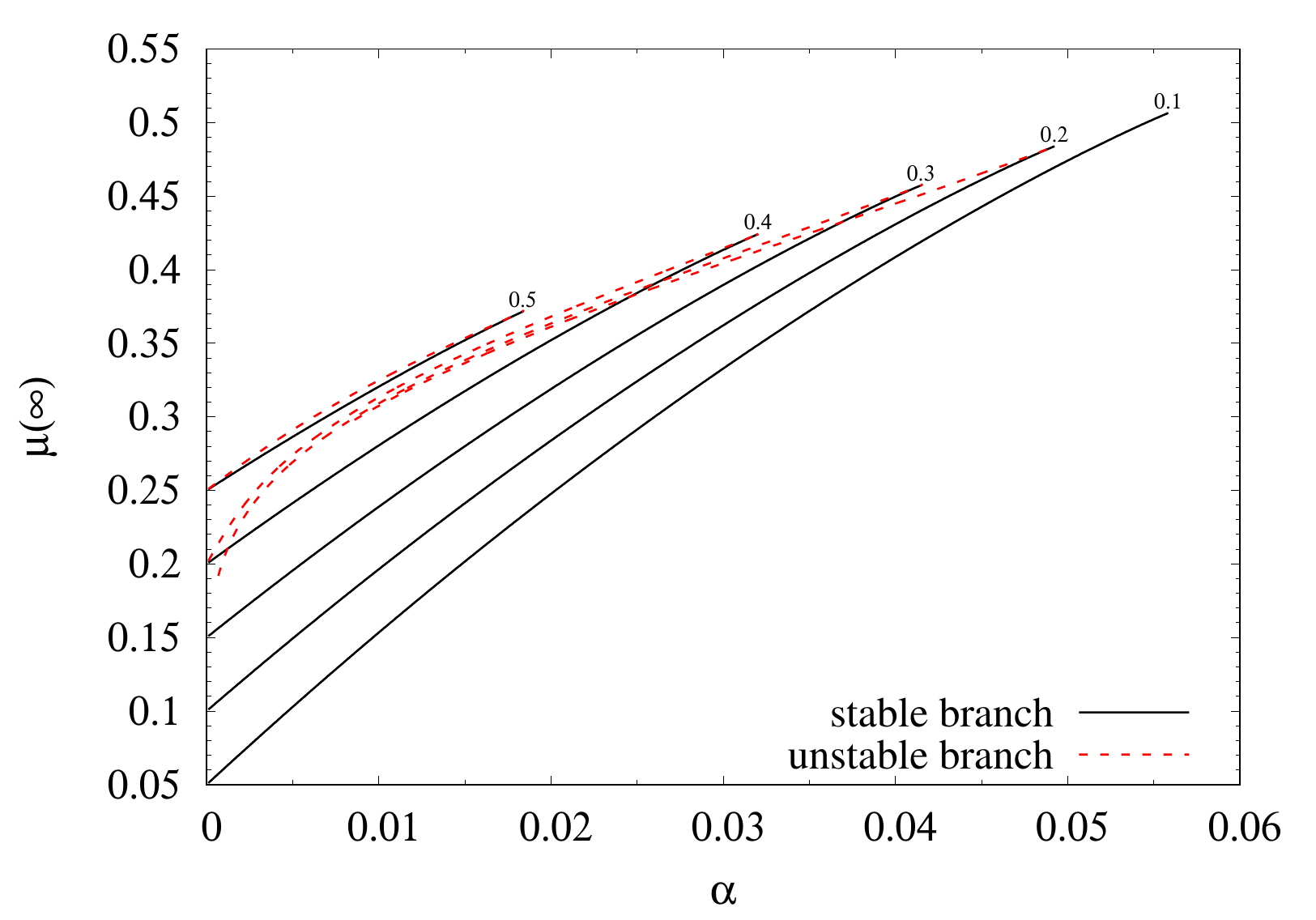}}
\subfloat[]{\includegraphics[width=0.49\linewidth]{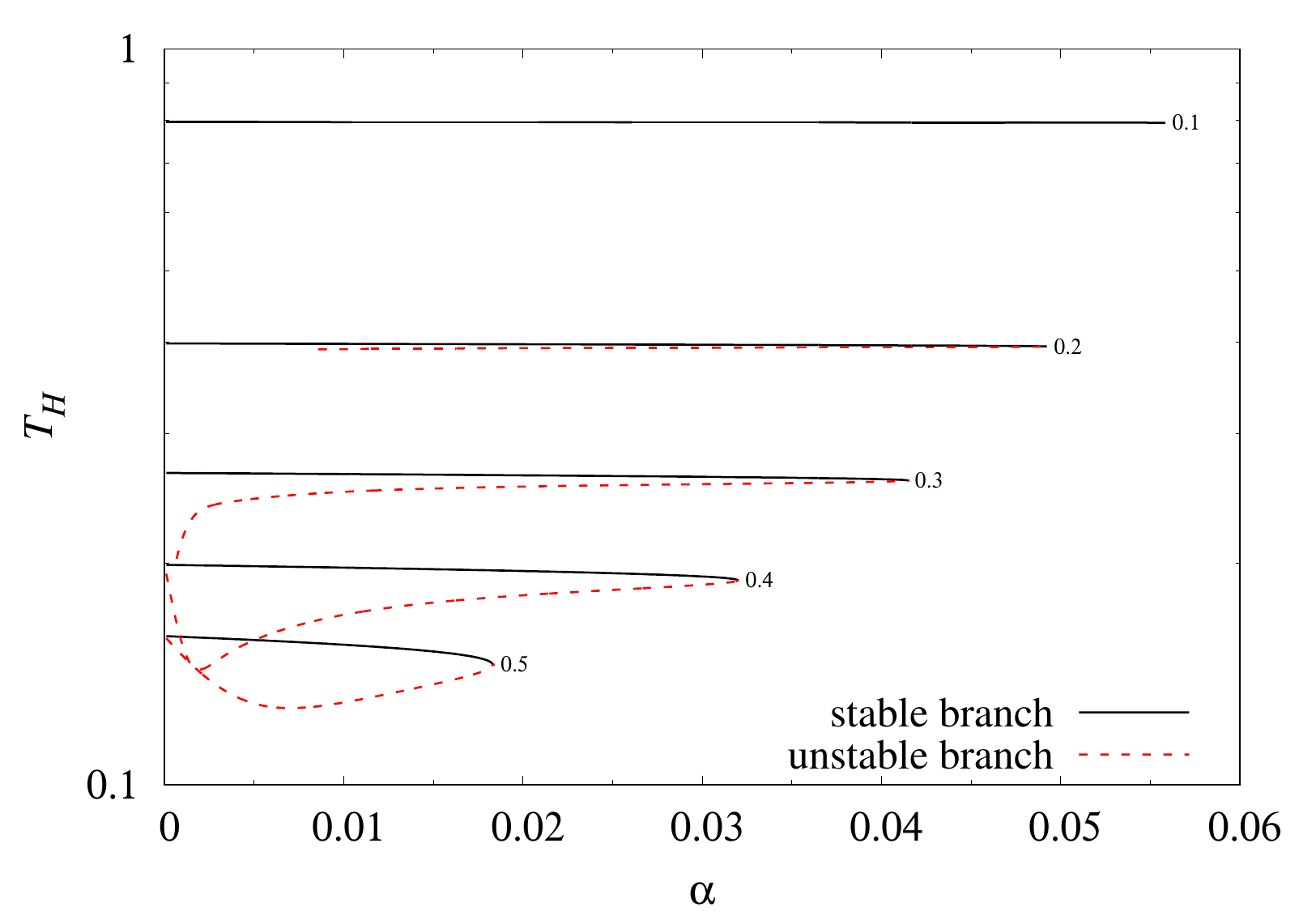}}}
\caption{Stable (black solid lines) and unstable (red dashed lines)
  branches of solutions in the $2+8$ model with $\gamma=1$: (a) the
  value of the profile function at the horizon, $f_h$; (b) the
  derivative of the profile function at the horizon, $f_\rho(\rho_h)$;
  (c) the ADM mass, $\mu(\infty)$; (d) the Hawking temperature $T_H$,
  all as functions of the gravitational coupling, $\alpha$. 
  The numbers on the figures indicate the different values of
  $\rho_h=0.1,0.2,\ldots,0.5$.
}
\label{fig:m28_alpha_gamma1}
\end{center}
\end{figure}

\begin{figure}[!thp]
\begin{center}
\mbox{\subfloat[]{\includegraphics[width=0.49\linewidth]{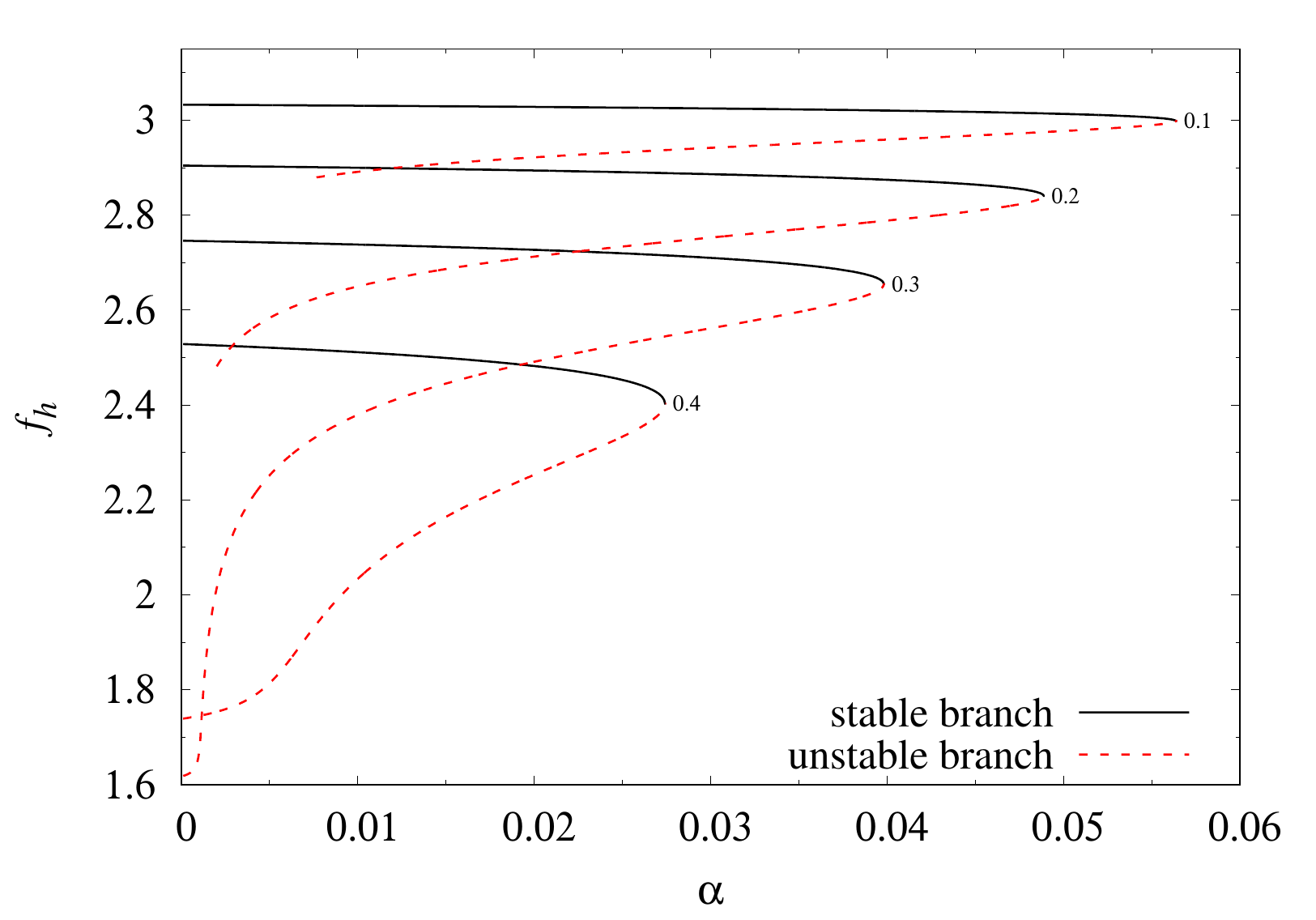}}
\subfloat[]{\includegraphics[width=0.49\linewidth]{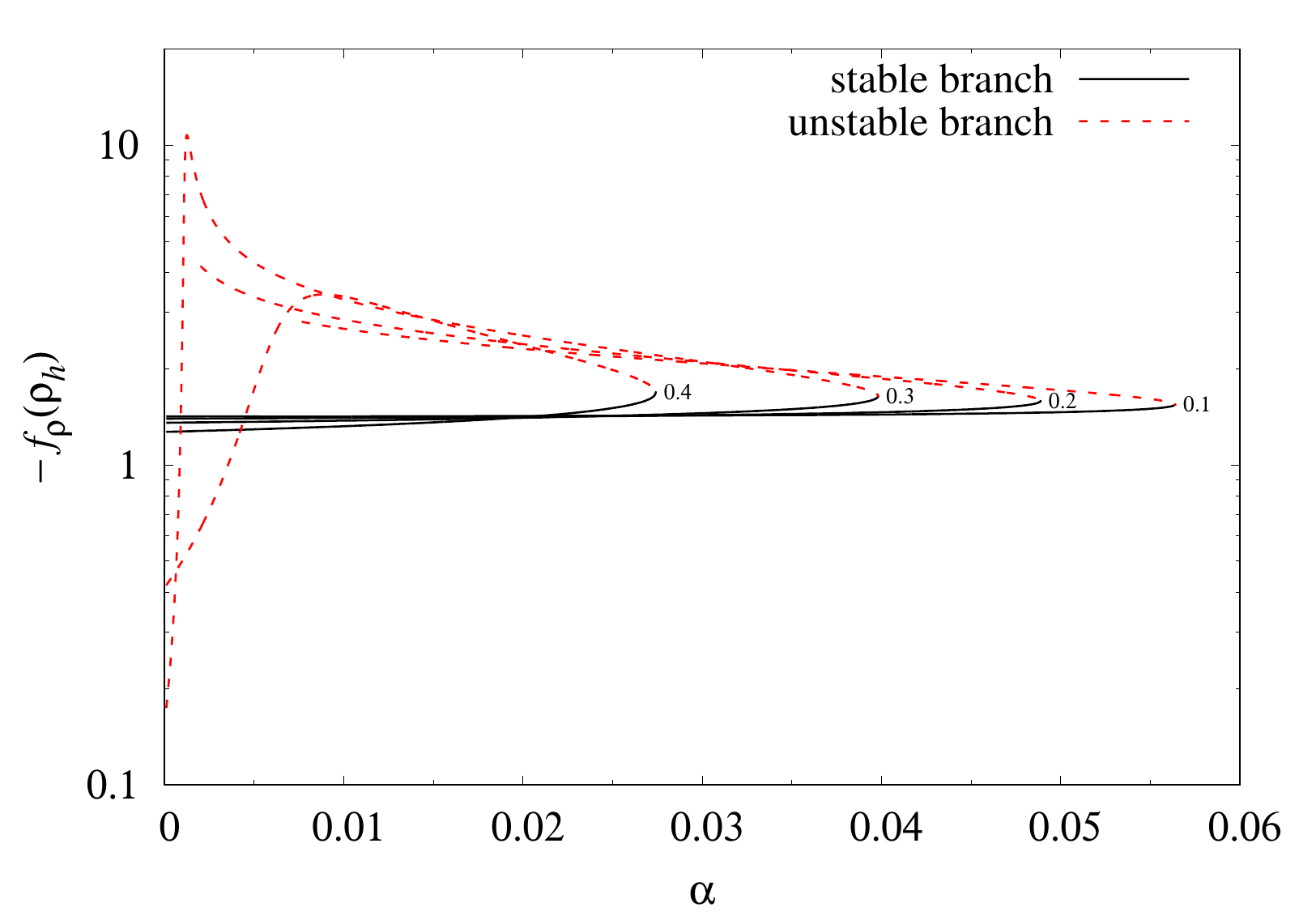}}}
\mbox{\subfloat[]{\includegraphics[width=0.49\linewidth]{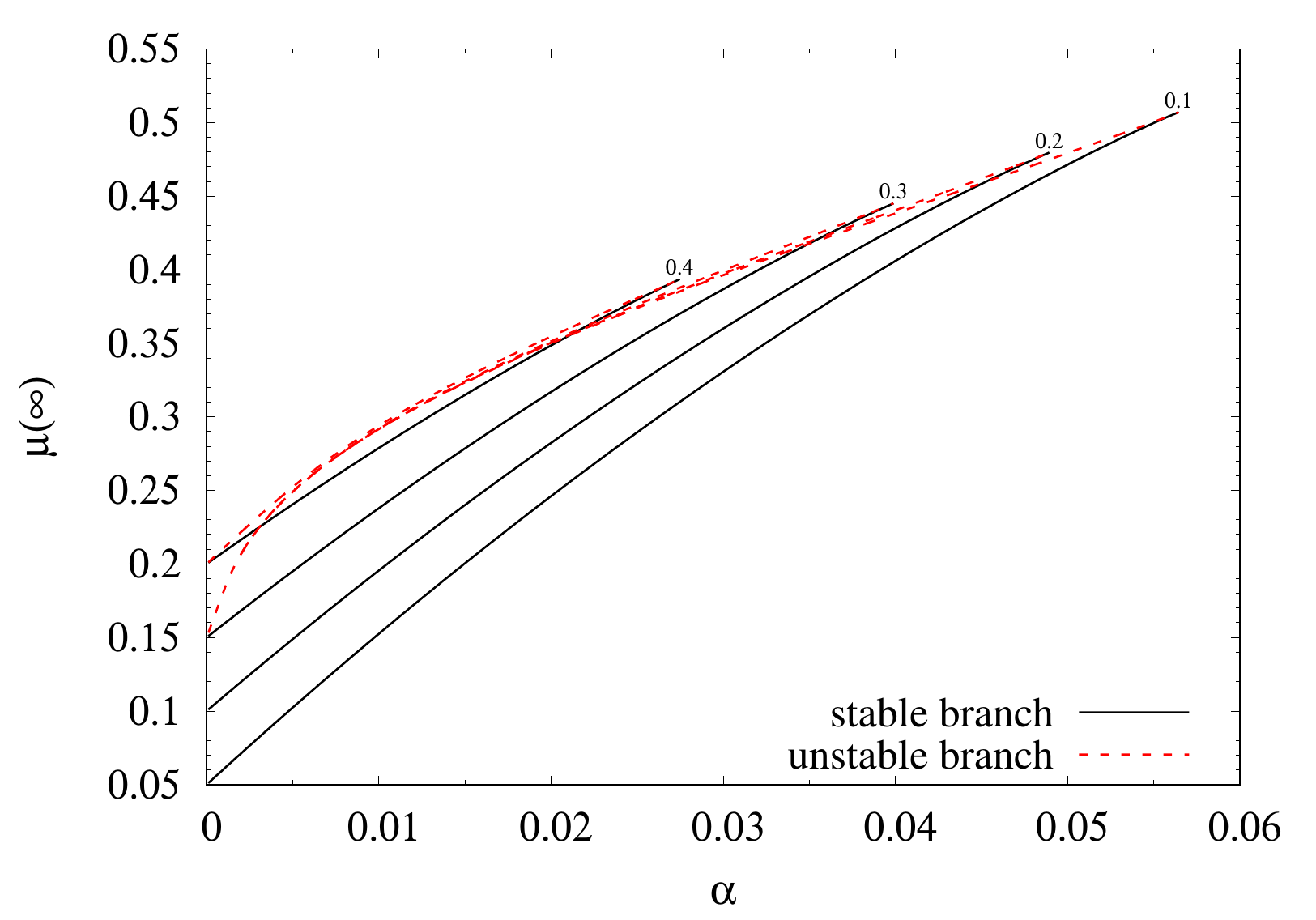}}
\subfloat[]{\includegraphics[width=0.49\linewidth]{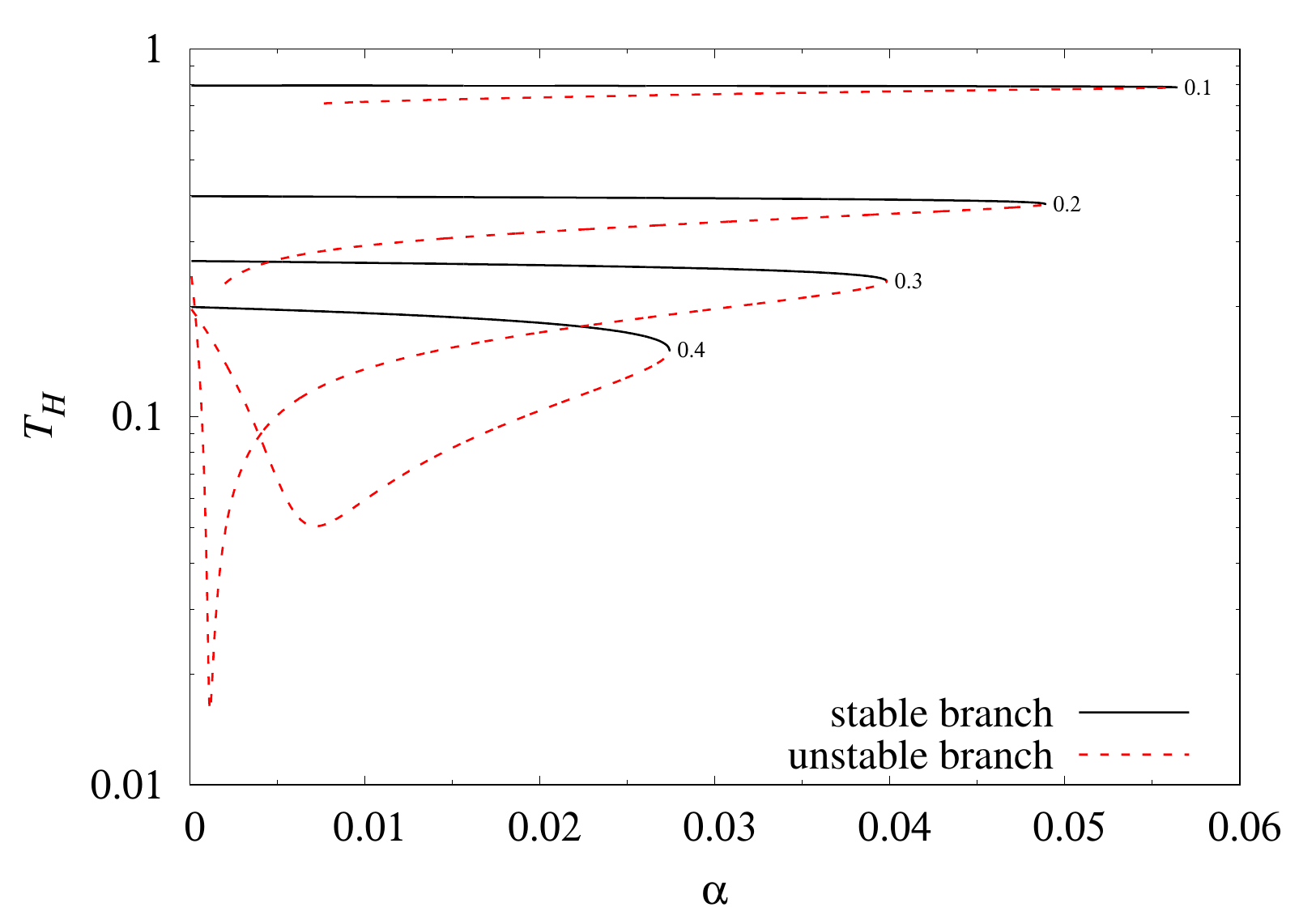}}}
\caption{Stable (black solid lines) and unstable (red dashed lines)
  branches of solutions in the $2+8$ model with $\gamma=\frac13$: (a)
  the value of the profile function at the horizon, $f_h$; (b) the 
  derivative of the profile function at the horizon, $f_\rho(\rho_h)$;
  (c) the ADM mass, $\mu(\infty)$; (d) the Hawking temperature $T_H$,
  all as functions of the gravitational coupling, $\alpha$. 
  The numbers on the figures indicate the different values of
  $\rho_h=0.1,0.2,0.3,0.4$.
}
\label{fig:m28_alpha_gamma1third}
\end{center}
\end{figure}

\begin{figure}[!thp]
\begin{center}
\mbox{\subfloat[]{\includegraphics[width=0.49\linewidth]{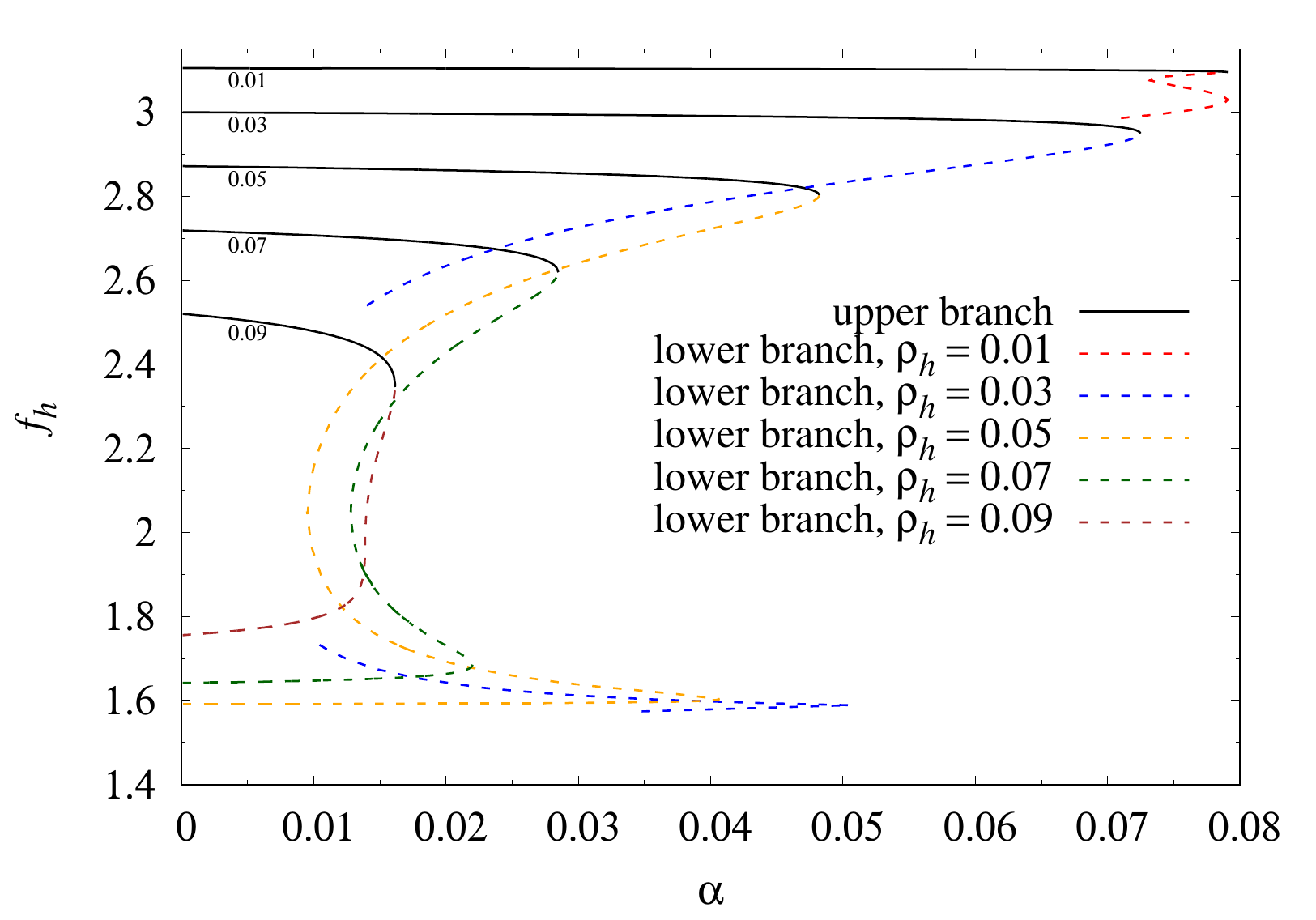}}
\subfloat[]{\includegraphics[width=0.49\linewidth]{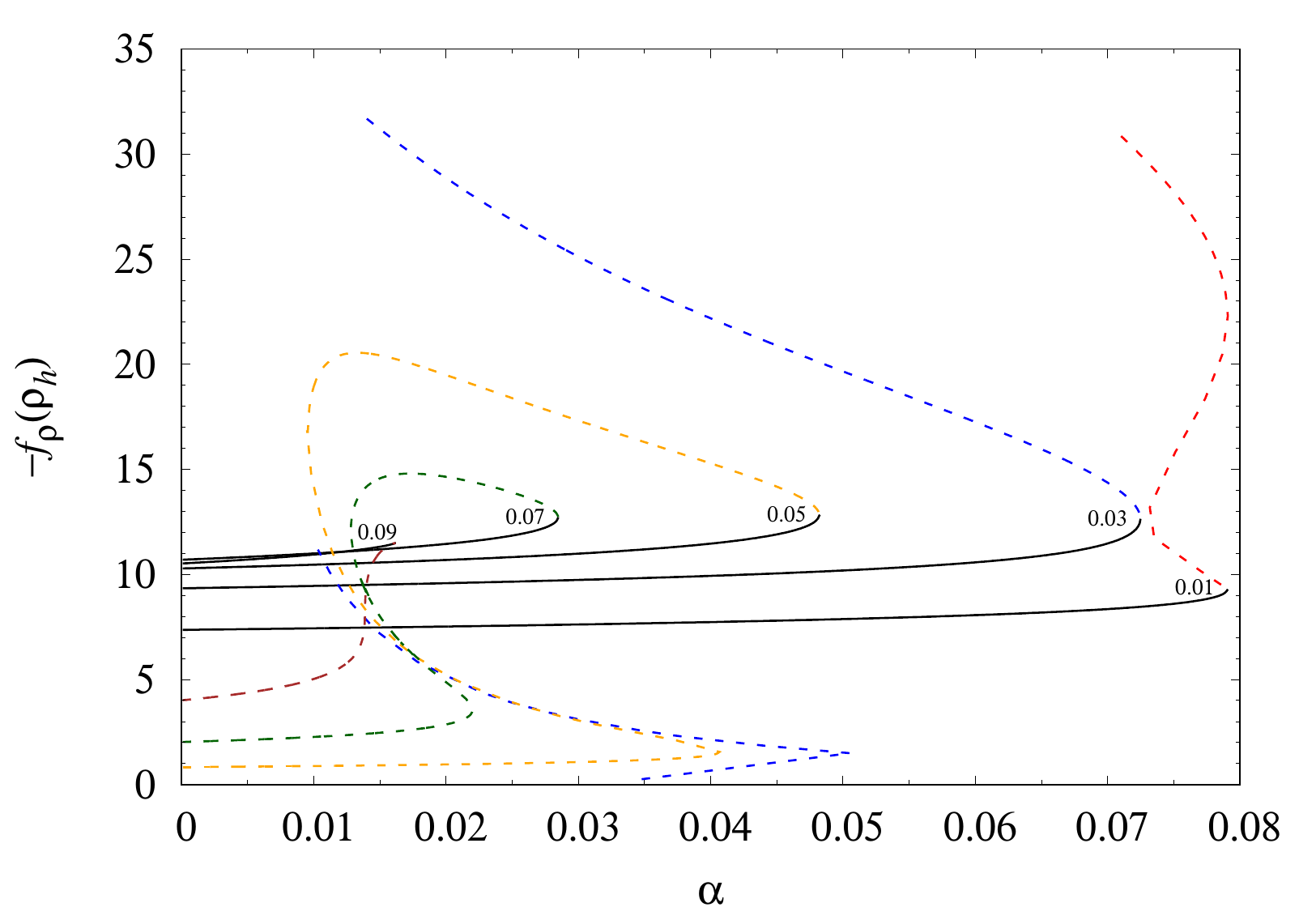}}}
\mbox{\subfloat[]{\includegraphics[width=0.49\linewidth]{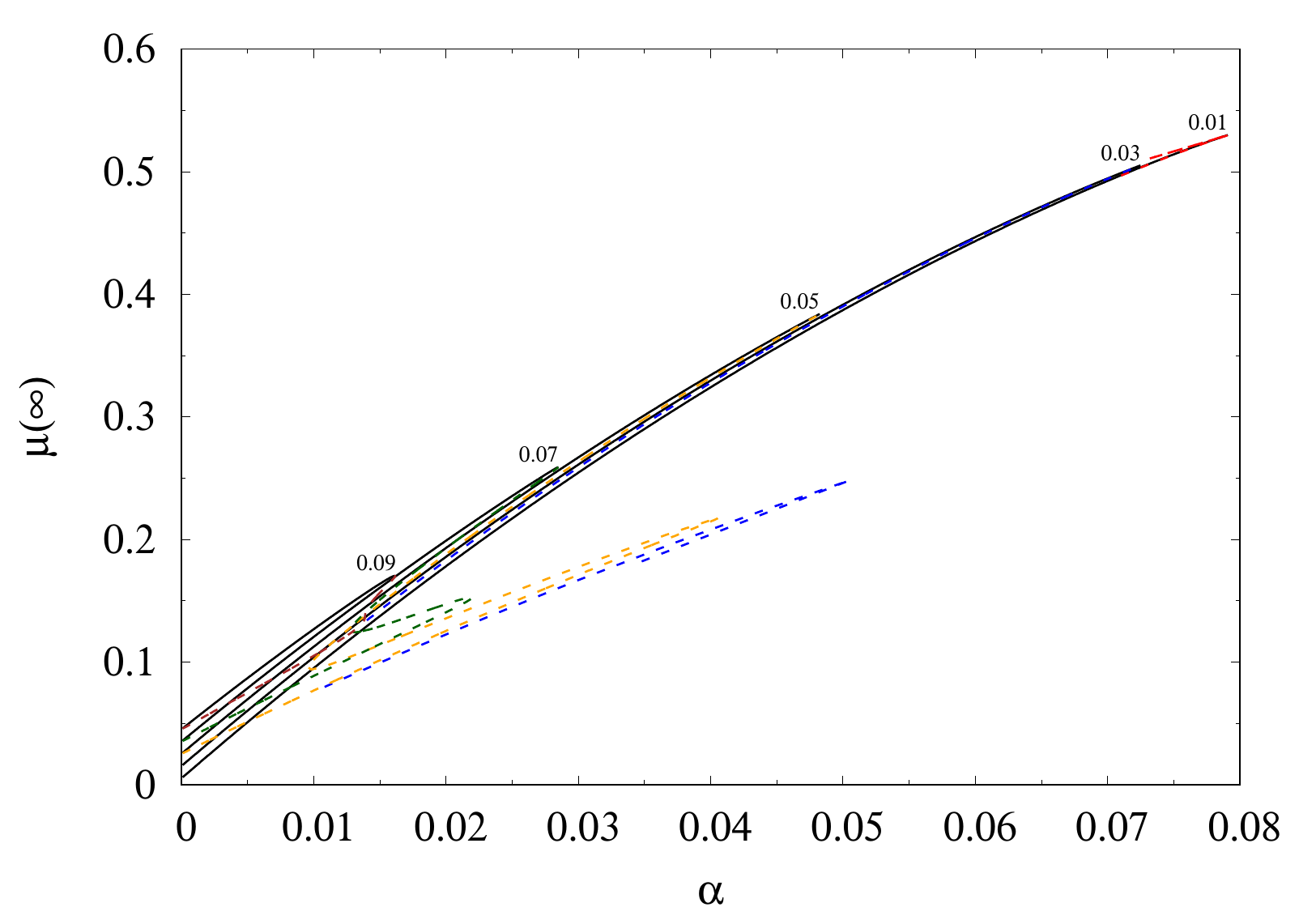}}
\subfloat[]{\includegraphics[width=0.49\linewidth]{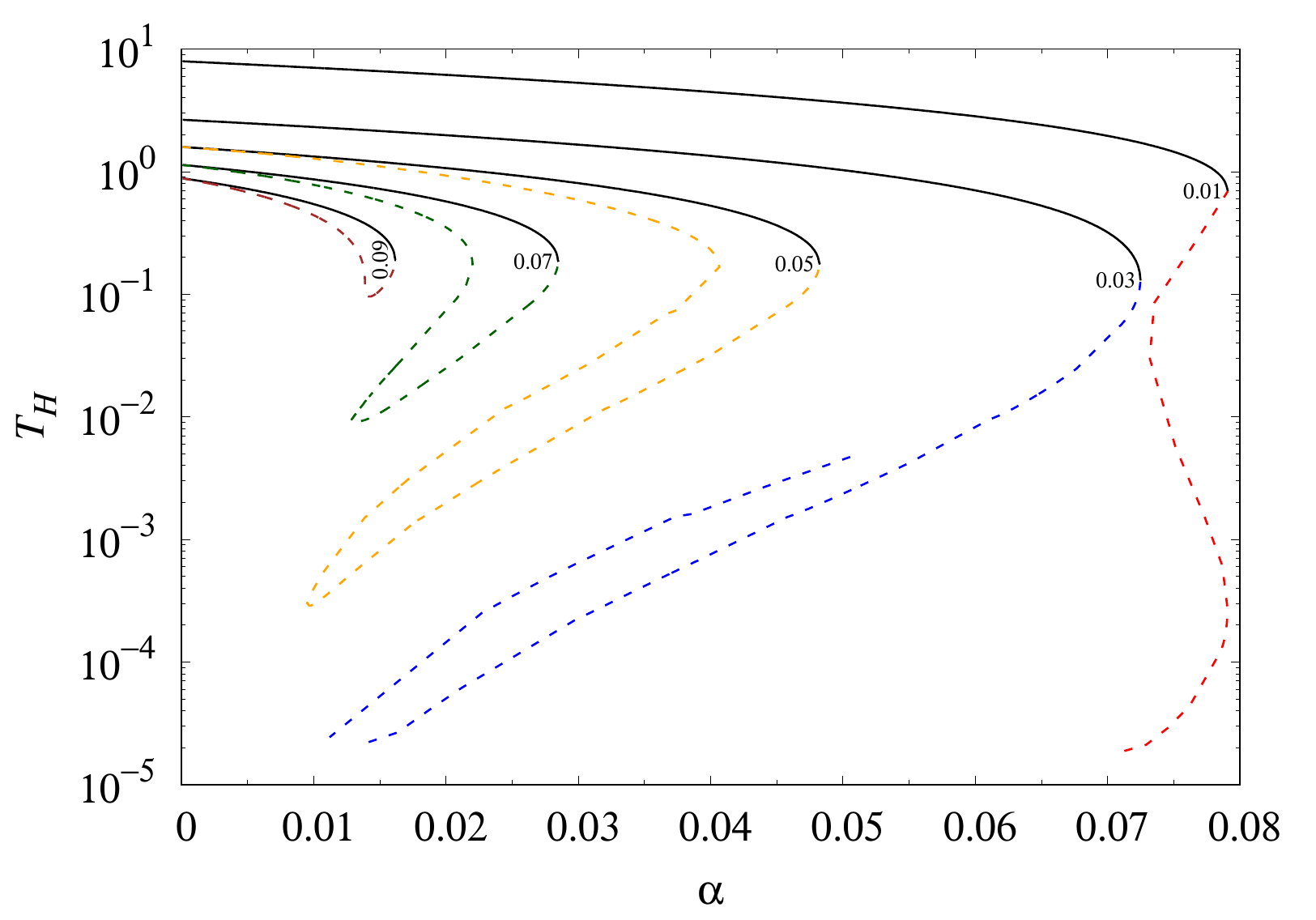}}}
\caption{Upper (black solid lines) and lower (red dashed lines)
  branches of solutions in the $2+12$ model: (a)
  the value of the profile function at the horizon, $f_h$; (b) the 
  derivative of the profile function at the horizon, $f_\rho(\rho_h)$;
  (c) the ADM mass, $\mu(\infty)$; (d) the Hawking temperature $T_H$,
  all as functions of the gravitational coupling, $\alpha$. 
  The numbers on the figures indicate the different values of
  $\rho_h=0.01,0.03,\ldots,0.09$.
  The lower branches are colored differently for each horizon radius,
  see the legend in panel (a). 
}
\label{fig:m212_alpha}
\end{center}
\end{figure}

In Fig.~\ref{fig:m28_alpha_gamma1}, the standard quantities are shown
for the $\gamma=1$ case of the $2+8$ model as functions of the
gravitational coupling $\alpha$.
The different branches correspond to different horizon radii and as
expected, the branch with the smallest horizon radius has the largest
value of the profile function, $f_h$. As the horizon radius is
increased, the branches move downwards in $f_h$,
Fig.~\ref{fig:m28_alpha_gamma1}(a), as expected, since the branches
will move downwards in $f_h$ to meet their respective bifurcation
point. 
The upper branches are called stable here, as they everywhere have
lower ADM mass than the corresponding lower branches, see
Fig.~\ref{fig:m28_alpha_gamma1}(c).
We have not carried out a linear perturbation analysis for this case,
as we expect the lower branches to be unstable, if not linearly, then
at best metastable.
Although the ADM mass increases roughly linearly with the
gravitational coupling, $\alpha$ (see
Fig.~\ref{fig:m28_alpha_gamma1}(c)), the Hawking temperature remains
almost constant as $\alpha$ is varied, for the stable branches.
We made an extensive search for the unstable or lower branch for 
$\rho_h=0.1$, but were unable to find any solutions -- both for large
and small values of $f_h$.
Finally, let us mention that we have found that the unstable branches
for $\rho_h=0.4,0.5$ continue all the way to $\alpha\to 0$.

For completeness, we have made the same plots for the $\gamma=\frac13$
case in Fig.~\ref{fig:m28_alpha_gamma1third}.
Because these plots are quite similar to those of
Fig.\ref{fig:m28_alpha_gamma1}, let us just mention the differences.
Indeed the quantitative behavior is the same, but we have been able to
find an unstable branch with horizon radius $\rho_h=0.1$.
Both the unstable branches $\rho_h=0.1$ and $\rho_h=0.2$ end at a
finite horizon radius, while those with $\rho_h=0.3,0.4$ continue back
in the limit of $\alpha\to 0$.
Again the ADM masses suggest that the lower branches are unstable, see
Fig.~\ref{fig:m28_alpha_gamma1third}(c).

Finally, we will consider the last model, i.e.~the $2+12$ model, which
only has the gravitational coupling, $\alpha$, as a parameter (after
rescaling of the length and energy units).
The result is shown in Fig.~\ref{fig:m212_alpha}.
This model possesses a more complicated branch structure than the two
flavors of the $2+8$ model.
The upper branches behave as expected; they start from above in $f_h$
with small horizon radius and move downwards as $\rho_h$ is increased.
They have little dependence on $\alpha$, expect near their bifurcation
point.
The lower branches, however are far more complicated.
The lower branch with $\rho_h=0.9$ (brown dashed line) is the only one
depicted which is single valued in $\alpha$.
The lower branches with $\rho_h=0.7$ (green dashed line) and
$\rho_h=0.5$ (yellow dashed line) are still continuous, but not single
valued in $\alpha$. 
Surprisingly, the lower part of the $\rho_h=0.5$ lower branch exists
up till quite large $\alpha\simeq 0.04$, before it sharply turns back,
see Fig.~\ref{fig:m212_alpha}(a).
That part of the unstable branch has in fact a smaller ADM mass than
the upper branch, see the lower yellow dashed line in
Fig.~\ref{fig:m212_alpha}(c).
As the horizon radius is decreased, the lower branches become
discontinuous.
Indeed, the $\rho_h=0.3$ lower branch has an upper part connected to
the upper branch and a disconnected lower part, see
Fig.~\ref{fig:m212_alpha}(a).
Finally, the $\rho_h=0.1$ lower branch only exists very close to the
bifurcation point, i.e.~for quite large values of $\alpha$.
We did not find a disconnected lower part for this value of the
horizon radius.

As mentioned above, the ADM masses interestingly show that the lower
parts of the lower branches have lower ADM masses than their
corresponding upper branches, which would make them the stable
branches, Fig.~\ref{fig:m212_alpha}(c).
Note, however, that these lower parts do not exist all the way up to
the bifurcation point, so for values of $\alpha$ close the bifurcation
point, the upper branch would be the stable one.
We have not carried out a linear perturbation analysis for this case,
see comments in the next section.

It is interesting to see what happens to the derivative of the profile
function at the horizon for the lower parts of the lower branches,
which according to the ADM masses are the stable ones, see
Fig.~\ref{fig:m212_alpha}(b).
Let us consider the $\rho_h=0.5$ lower branch (yellow dashed
line). The unstable part of the lower branch has a higher derivative
at the horizon and about where this branch becomes the stable one, the
derivative drops below that of the upper branch, see
Fig.~\ref{fig:m212_alpha}(b).
The lower parts of the lower branches, which are the stable ones for
those values of $\alpha$, do in fact all have smaller derivatives of 
the profile function at the horizon.

Finally, we will show the Hawking temperature for this model.
This calculation turned out to be the hardest one in the paper and we
needed to use over 1 billion lattice points for our integrator to get
convergence for the temperature.
The interesting result from this calculation is that the lower parts
(in $f_h$) of the lower branches, which are the stable ones for the
corresponding values of $\alpha$, have higher Hawking temperature
compared with the upper parts of the same branches, see
Fig.~\ref{fig:m212_alpha}(d).

\section{Discussion and conclusion}\label{sec:discussion}

In this paper, we investigated a number of Skyrme-like models with
terms containing six to twelve derivatives.
The higher-derivative terms considered here are not the most generic
ones, but the so-called minimal terms constructed in
Ref.~\cite{Gudnason:2017opo}.
The main motivation was to get a better understanding of the criteria
for when a Schwarzschild-type BH can support scalar hair.
Indeed in Refs.~\cite{Gudnason:2016kuu,Adam:2016vzf}, it was shown
that although the Skyrme term can support BH hair, the sextic
BPS-Skyrme term cannot. 
In this paper we have checked 3 further models, i.e.~a model with a kinetic
term and a $2n$-th order term, $n=4,5,6$.
The $2+8$ model is a one-parameter family of models and it turned out
that it can support BH hair as long as the model does not purely
consist of the BPS-Skyrme term times the standard kinetic term.
One of the possibilities that are stable, is the Skyrme-term squared.
The $2+10$ model has turned out not to be able to sustain BH hair.
Finally, the $2+12$ model is basically the kinetic term and the
BPS-Skyrme term squared and surprisingly it does possess a stable BH
hair.
The BH hair comes in two branches, one upper branch (in the profile
function at the horizon, $f_h$) which is typically stable, and one
lower branch (in $f_h$) which is typically unstable (see below,
however). 

A feature already seen in the generalized Skyrme model coupled
to Einstein gravity, is that the unstable branches, for sizable
coupling to the BPS-Skyrme term, end at a finite BH horizon radius
simply because the temperature approaches zero.
This can be viewed as the BH approaching an extremal BH state or more
pragmatically as the derivative of the field profile at the BH horizon
blowing up.
This feature has turned out to be shared by the other models that do not
possess BH hair in the limit where the Skyrme term is turned off; in
particular, those are the $2+4+6$, the $2+4+8$ ($\gamma=0$), and
$2+4+10$ models.

For the standard Skyrme model coupled to Einstein gravity, the upper
and the lower branches of solutions correspond to the stable and
unstable solutions.
This was checked in Refs.~\cite{Heusler:1991xx,Heusler:1992av} with a
linear perturbation analysis which showed that the lower branch
contains a single negative frequency of the perturbation modes of the
linearized system.
This system contains implicitly both the linear perturbations of the
metric as well as of the Skyrme field radial profile. The metric
perturbations are then eliminated, yielding a single field master
equation.
Evidence for the instability of the lower branches was already seen by
calculating their respective ADM masses, for which the lower branches
always possessed the higher ADM masses compared to the upper (stable)
branches. 
However, in the $2+4+12$ model the situation turned out to be somewhat 
more intricate.
Indeed, when the Skyrme term is slowly turned off, the lower branch
switched to become the one with a lower ADM mass.
For an intermediate range of the Skyrme-term coefficient, the lower
branch possesses a lower ADM mass compared to the upper branch, for a
finite range from the bifurcation point down to a critical radius
where the ADM masses cross over (details have been shown in
Fig.~\ref{fig:m2412_rh_detail1}). 
Finally, in the limit of a vanishing Skyrme-term coefficient, the
lower branches of solutions remain the ones with the lowest ADM
masses.
Those lower branches, however, still terminate at a finite horizon
radius.
This behavior is mapped out in the phase diagram of
Fig.~\ref{fig:m2412_phasediagram}.
To this end, we have carried out a linear perturbation analysis of the
BH hair system analogous to that of
Refs.~\cite{Heusler:1991xx,Heusler:1992av}.
It turned out, however, that the problem at hand is more complicated
due to the higher nonlinearity of the problem, which causes the kernel
and the weight function of the resulting Sturm-Liouville problem to
differ.
Consequently, the master equation for the perturbations of the
$2+4+12$ model cannot be written as a Schr\"odinger equation and the
full Sturm-Liouville problem needed to be solved. 
The result of this analysis was consistent with the naive conclusion
from the ADM masses, namely, the lower branches become the stable ones
in the limit of the Skyrme term being turned off in the $2+4+12$
model.
This result is surprising.
As a double check we have tried turning off the twelfth-order term and thus
returning to the standard Skyrme model, and indeed, the lowest
eigenvalue of the perturbations returned to the standard behavior.
That is, the upper branch has only positive eigenvalues and the
lower branch has a single negative eigenvalue.
Once both the Skyrme term and the twelfth-order term are turned on
with sizable (order one) coefficients, the eigenvalue possesses a
hybrid behavior; for small horizon radii the lowest eigenvalue is
negative, but it turns positive at a finite radius and crosses over
that of the upper branch until they meet at the bifurcation point.

Since the models under study in this paper are highly nonlinear
problems, the linearized perturbation analysis may not suffice.
Indeed, as a future investigation, full nonlinear stability should be
considered seriously in order to understand the situation of the
$2+4+12$ model in the limit of a small or vanishing Skyrme-term
coefficient.

A peculiar observation about the limit of the Skyrme-term coefficient
being turned off in the $2+4+12$ model, is that the explanation for
the lower branches terminating at a finite horizon radius until now
was that the Hawking temperature would go to zero, or equivalently the
derivative of the profile function at the horizon would blow up.
In this case, however, the lower branches still terminate at a finite
radius, but the reason switches from being the derivative of the
profile function blowing up (i.e.~equal to the Hawking temperature
dropping to zero) to the derivative going to zero.
That also has as consequence that the solutions cease to exist, but the
Hawking temperature remains finite. 

To complicate the situation with the termination of the lower
branches, in the $2+4+8$ model (for any value of $\beta$, the Skyrme-term
coefficient) the lower branches terminate at a finite horizon radius,
but with a finite Hawking temperature and an apparently finite first
derivative of the profile function at the horizon.
Further studies are needed to conclude the fate of the small
horizon-radius limit for these models.

Finally, in the $2+12$ model which has BH hair stabilized by a
twelfth-order term and no Skyrme term, a certain range of the
gravitational coupling exists for which there are not 2 but 4
solutions with the same horizon radius, $\rho_h$.
According to their ADM masses, the lower part of the lower branch is
the stable one.
The upper branch may possess metastability and the upper part of the
lower branch may be either unstable or metastable.

Throughout this paper, we have turned off a potential in order to keep
the analysis as clean as possible. Although we have not checked, we
think that most, if not all, results will be qualitatively similar if
a potential would be turned on; in particular a standard pion mass
term.
The absence of BH hair in the $2+4+6$ model without a mass term or
other potential is thus confirmed; Ref.~\cite{Adam:2016vzf} carried
out all their numerical calculations with the pion mass term present.
Ref.~\cite{Adam:2016vzf} also gives a physical explanation for the
lack of BH hair in the model with only the BPS-Skyrme term, which
claims that the pressure becomes negative at the horizon due to the
potential.
This explanation cannot cover the case considered here, where we have
excluded the potential altogether.
It may be that the zero pressure from the BPS-Skyrme term is not
sufficient for preventing a collapse of the hair.
Further studies are needed for a conclusion on this issue.

Finally, a question tightly related to the above mentioned issue, is
whether the stress-energy tensor of the model (excluding the kinetic
term) can be rewritten in the form of a perfect fluid and whether this
may be a criteria for whether a BH can possess stable hair or not. 
In Ref.~\cite{Adam:2015rna}, it was shown that the BPS-Skyrme model
can be rewritten as a non-barotropic perfect fluid using the Eulerian
formulation of a relativistic fluid.
The fluid element velocity was identified with the baryon charge
current.
Not all models considered here can be written in terms of just the
baryon charge current or baryon charge density.
Although some can, others cannot and some are hybrids of the latter
two options. 

It is known in non-gravitating cases that the $2+4+6$ model with
potential terms different from the standard pion mass term admit
non-spherical
Skyrmions \cite{Gudnason:2014jga,Gudnason:2014nba,Gudnason:2014hsa,Gudnason:2015nxa,Gudnason:2016yix}. 
It is an open question whether such potentials can give rise to
non-spherical BH Skyrmions, other than axially symmetric ones. 

An obvious generalization of this study is to consider more general
terms with the same number of derivatives; i.e.~non-minimal
Lagrangians. The first class of more general Lagrangians will contain
only a second-order equation of motion (albeit nonlinear in derivative
terms), but with more than four derivatives in the same direction, for
the sixth, eighth, tenth and twelfth order terms.
A further generalization is to include more than one derivative acting
on the same field, yielding a higher-order equation of motion.
We avoided such a complication here and in
Ref.~\cite{Gudnason:2017opo}, because in general it could give rise to
an Ostrogradsky instability.
A first interesting question would be, if the sixth-order Lagrangian
was relaxed to contain more general terms than the BPS-Skyrme term,
would it be able to sustain stable BH hair.
We will leave this question for future work.

\subsection*{Acknowledgments}

The work of S.~B.~G.~is supported by the National Natural Science 
Foundation of China (Grant No.~11675223).
The work of M.~N.~is supported in part by a Grant-in-Aid for
Scientific Research on Innovative Areas ``Topological Materials
Science'' (KAKENHI Grant No.~15H05855) from the Ministry of
Education, Culture, Sports, Science (MEXT) of Japan. The work of
M.~N.~is also supported in part by the Japan Society for the Promotion
of Science (JSPS) Grant-in-Aid for Scientific Research (KAKENHI Grant 
No.~16H03984) and by the MEXT-Supported Program for the Strategic
Research Foundation at Private Universities ``Topological Science''
(Grant No.~S1511006).
The TSC-computer of the ”Topological Science” project in Keio
university was used for some of the numerical calculations.

\end{document}